\documentclass[twocolumn,prb,showpacs,floatfix,amsmath,amsfonts]{revtex4-1}
\usepackage[latin9]{inputenc}
\usepackage{verbatim}
\usepackage{amsmath}
\usepackage{amssymb}
\usepackage{graphicx}

\makeatletter
\pdfoutput=1
\usepackage{hyperref}
\usepackage{dcolumn}

\makeatother

\begin{document}

\title{Magnetic-field-assisted transmission of THz waves through a graphene
layer combined with a periodically perforated metallic film}

\author{Yu. V. Bludov\email{bludov@fisica.uminho.pt}, M. I. Vasilevskiy,
N. M. R. Peres }

\affiliation{Departmento de  F\'{i}sica and Centro de  F\'{i}sica, Universidade do
Minho, PT-4710-057, Braga, Portugal;\\
QuantaLab, University of Minho, PT-4710-057, Braga, Portugal}

\begin{abstract}
We consider a graphene sheet encapsulated in a two-dimensional metallic grating and a substrate ($\mathrm{Al_{2}O_{3}}$) and subjected to an external magnetic field (in Faraday configuration). The grating consists of a thin perfectly
conducting metal film perforated with a 2D periodic array of square holes. 
According to our calculations, significant changes in the spectra of the Faraday rotation angle of the transmitted wave and of the magnetic circular dichroism should be expected in this situation compared  to bare graphene. We explain this enhancement by the excitation of graphene magnetoplasmons that accompanies the transmission of the electromagnetic wave through the structure. The results can be interesting for applications in THz photonics, such as switchable rotating polarizer and optical isolator.
\end{abstract}
\maketitle

\section{Introduction}

One of the prominent features of the magnetoactive structures is a
strong dependence of the electromagnetic wave characteristics upon
the direction of propagation. An illustrative example is the
Faraday effect \cite{mo-Faraday1846-PTRSL}, where the direction of
the rotation of the plane of polarization is different for the forward-
and backward-proparating waves in magnetoactive media along the direction
of static external magnetic field (Faraday geometry). Similar phenomenon
\textendash{} nonreciprocal phase shift\cite{mo-nrps-Gabriel1965-mtt,mo-nrps-Auracher1975-OptComm}
\textendash{} takes place in asymmetric structures \cite{mo-rev-Dotsch2005-josab}(e.g.,
waveguides) when the direction of propagation is perpendicular
to the magnetic field (Voigt geometry). These phenomena consitute the
basis for the operation principles of a variety of microwave
photonic devices, such as optical isolators\cite{mo-oi-Aplet1964-ao,mo-oi-Jalas2013-nature},
circulators \cite{mo-circ-Ribbens1965-ao,mo-circ-Takei2010-japj},
and switches \cite{mo-switch-Tien1974-jap,mo-switch-Ishida2016-jjap}
(for a review on practical application of magnetooptical materials
see \cite{mo-rev-Shoji2014-stam,mo-rev-Shoji2016-jo}).

The general tendency to the minituarization of the photonic components
resulted in the creation of magnetoptical devices, which key building
blocks are photonic crystals\cite{pc-Qiu2011-oe,pc-circ-Smigaj2010-ol,pc-Dmitriev2012-ao,pc-circ-Wang2005-ol,pc-Zhang2013-olt,pc-Fan2011-josab,pc-Wang2009-nature,pc-circ-Fan2012-OptComm}
or electromagnetic metasurfaces \cite{mm-oi-Chen2015-oe,mm-oi-Chen2015-ao}. Yet,
the diffraction limit of electromagnetic waves imposes one of the
fundamental obstacles for further minituarization and growth of integration
of photonic devices in opto-electronics circuitry. One of the possible
ways to circumvent this limit is to build the photonic components,
whose operation principles are based on the surface electromagnetic
waves instead of on their free space counterparts. To be specific,
incorporating metallic structures into a photonic platform allows
one to create circuitry, operating on \emph{surface plasmon-polaritons} -- a special kind of the electromagnetic waves, whose energy
is localized near the metal-dielectric interface and whose wavelength
is considerably smaller than that of the free space wave with the
same frequency \cite{spp-rev-Ozbay2006-science,spp-rev-Stockman2011-pt}.

Combining plasmonics with magnetooptics \cite{spp-mf-rev-Temnov2012-NatPhot,spp-mf-rev-Armelles2013-aom}gives
a variety of advantages\cite{spp-mo-rev-Maksymov2015-nanomat}. First,
magnetic field allows to achieve the dynamical tunability (control
of parameters in real time) of plasmonic structure\cite{spp-Chiu1972-ncb,spp-mo-Temnov2010-NatPhot,spp-mf-Fei2013-apl,spp-mo-Fan2016-oe}.
Secondly, the aforementioned phenomenon of the nonreciprocal propagation
allows to create plasmonic analogues of different magnetooptical devices
\cite{spp-nrps-Firby2015-optica,spp-circ-Davoyan2013-njp,spp-oi-Firby2016-ol}.
Third, the resonant excitation of the magnetoplasmons and their coupling
to the transmitted and reflected bulk waves results in a significant
enhancement of the Faraday and magnetooptical Kerr effects \cite{spp-fr-Chin2013-NatComm,spp-fr-Caballero2015-oe,spp-mo-Belotelov2011-NatNanotech,spp-mf-Kleilkamp2013-prx}.
However, introducing the magnetooptical materials into the plasmonic
circuitry enhances significantly the losses in the system, thus reducing
the free path of magnetoplasmons (compared to the conventional
plasmonic structures). As a result, nowadays there is a huge demand
for novel magnetoplasmonic materials.

In the context discussed above, graphene emerges as a promising candidate,
operating the THz and mid-IR spectral range. This material possesses
several properties, which can be advantageous for the magnetooptics
and magnetoplasmonics. To begin with, surface plasmon-polaritons in graphene
\cite{spp-gr-rev-Xiao2016-fp,spp-gr-rev-deAbajo2014-ACSPhot,spp-gr-rev-Chen2017-nanophot,spp-gr-rev-Bludov2013-ijmfb,spp-gr-rev-Low2014-ACSNano}are
characterized by both large lifetime and high degree of field confinement
\cite{spp-gr-rev-Koppens2011-nl,spp-gr-Nikitin2011-prb}. Simultaneously, graphene is a magnetoactive
material: being subjected to an external static magnetic field (perpendicular
to its surface), graphene exhibits some unusual magnetic properties,
like the Hall effect at room temperture\cite{gr-qhe-Novoselov2005-nature,gr-qhe-Zhang2005-nature}.
Furthermore, its conductivity (and, consequently, its transmittance
and the reflectance\cite{gr-mf-oi-Tamagnone2014-NatPhot,gr-mf-oi-Tamagnone2016-NatComm,gr-mf-Shimano2013-NatComm,gr-mf-Crassee2010-NatPhys,gr-fr-Sounas2013-apl}
as well as the dispersion properties of magnetoplasmons \cite{spp-mf-gr-Berman2008-prb,spp-mf-gr-Roldan2009-prb,spp-mf-gr-Crassee2012-nl,spp-mf-gr-Yan2012-nl})
can be effectively tuned by changing the applied magnetic field. The
possibility to achieve the magnetoplasmon-mediated enhancement of
the magnetooptical phenomena was demonstrated in various graphene-based
structures, such as a periodic array of graphene ribbons \cite{spp-gr-mo-Tymchenko2013-ACSNano},
a graphene monolayer patterned with the periodical antidot array \cite{spp-gr-mo-Poumirol2017-NatComm},
an array of graphene-covered nanowires \cite{spp-gr-mo-Kuzmin2016-nl},
or a monolayer graphene metasurface \cite{spp-gr-mo-Fallahi2012-apl}.

In this paper we study the interaction of THz electromagnetic wave
with a graphene monolayer cladded by a semi-infinite substrate
and a periodically perforated metallic film of finite thickness. This
metallic film is assumed made of a perfect metal and containing a
two-dimensional (2D) periodic array of square holes. The structure is subjected 
to an external magnetic field directed perpendicularly to the surface (Faraday geometry). 
We demonstrate that if the two-dimensional metallic grating is sparse enough, 
it screens the incident long-wavelength electromagnetic wave. 
As a result, in the low-frequency range the
Faraday rotation in the graphene covered with the perforated metal
film is less than that in the graphene layer alone. At high frequencies, the sparse metallic grating almost
does not influence the propagation of the electromagnetic wave. 
At the same time, in the intermetiate frequency range the diffraction
of the incident electromagnetic wave on the grating results
in the excitation of graphene magnetoplasmons. This process increases
considerably the Faraday rotation angle of the polarization vector for a linearly polarized wave.
For a circularly polarized propagating wave, the presence of the perforated
metallic film leads to inversion of the sign of the magnetic circular dichroism (MCD)
in the low-frequency range.

The paper is organized as follows. In Sec.\ref{sec:main-equations}
we obtain the principal equations governing the process of incident
wave diffraction on graphene combined with the periodical array of
holes in the metal film. Section \ref{sec:Faraday-rotation} is devoted
to a detailed discussion of how the parameters of this structure influence
the Faraday rotation angle of the transmitted electromagnetic wave.
In Sec.\,\ref{sec:Circular-dichroism} we investigate the magnetic
circular dichroism in this structure. The conclusions are presented
in Sec.\ref{sec:conclusions}.

\section{Diffraction of plane electromagnetic wave on metal film with periodic array of  
holes}

\label{sec:main-equations}
\begin{figure}
\includegraphics[width=8.5cm]{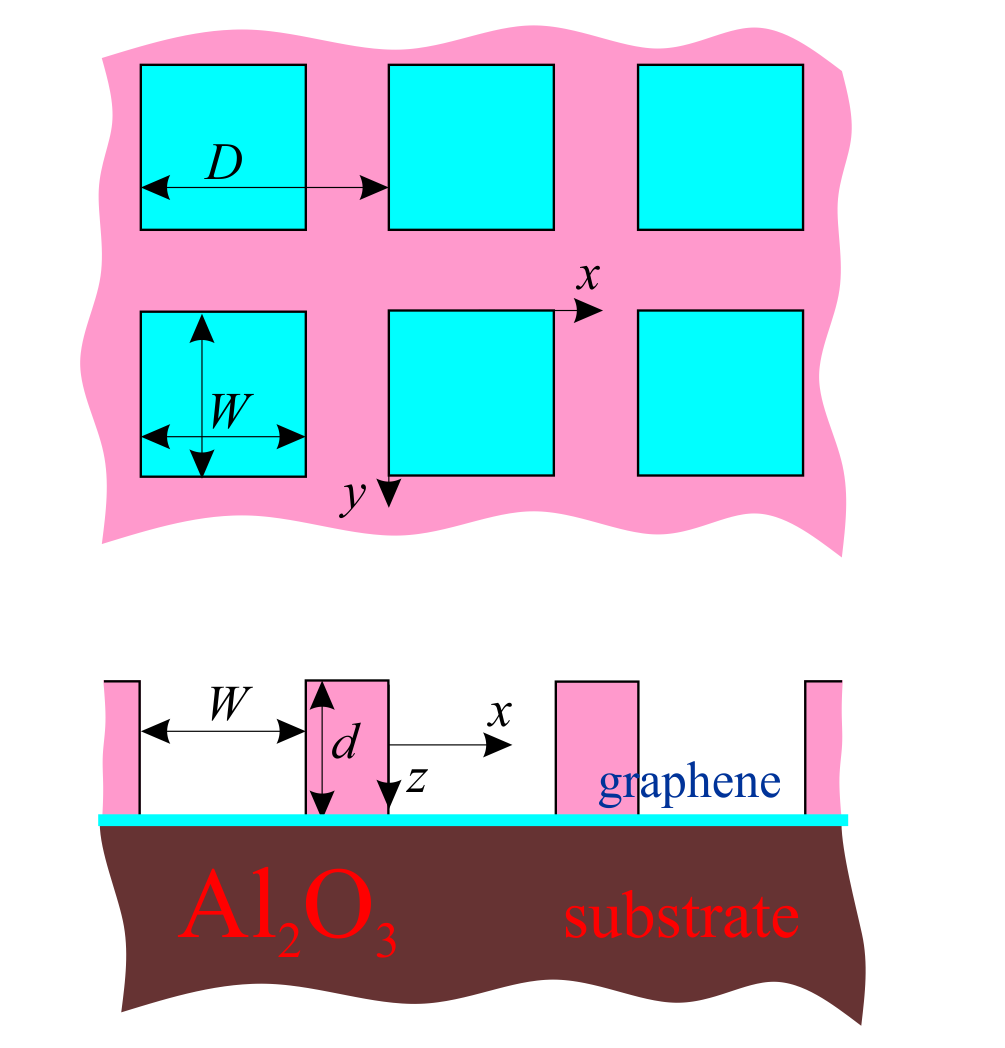}

\caption{Two-dimensional periodic array of holes in metallic film deposited
on top of a graphene layer covering a dielectric substrate.}
\label{fig:geometry}
\end{figure}

We consider a perfectly conducting metal
film of thickness $d$ (whose film surfaces are situated at planes
$z=\pm d/2$, see Fig.\ref{fig:geometry}), containing a periodic array of square holes, each of width $W$, arranged at $lD<x<lD+W$,
$l^{\prime}D<y<l^{\prime}D+W$ and forming a square lattice. Here $D$
is the period of the square lattice and $l$, $l^{\prime}$ are the hole indices.
We also assume that this metallic grating is deposited on top of a
graphene monolayer, arranged at the plane $z=d/2$. The graphene monolayer
is deposited on top of a semi-infinite dielectric substrate ($Al_2O_3$) occupying the half-space
$z>d/2$ and characterized by the dielectric function $\varepsilon$.
The plane wave impinges on the metal grating from air (the half-space $z<-d/2$) at normal incidence. 

The periodicity of the structure in the directions $x$ and $y$,
as well as the normal incidence of the external wave impose the requirement
for the solution of Maxwell equations to be also periodic. In half-spaces
$z<-d/2$ and $z>d/2$ (vacuum and substrate) electromagnetic fields
can be represented in the form of Fourier series with respect to the
lattice vector $g=2\pi/D$, namely 
\begin{eqnarray}
\left(\begin{array}{c}
H_{x}^{(1)}(\mathbf{r},z)\\
H_{y}^{(1)}(\mathbf{r},z)\\
E_{x}^{(1)}(\mathbf{r},z)\\
E_{y}^{(1)}(\mathbf{r},z)
\end{array}\right) & = & \left(\begin{array}{c}
\hat{{\cal I}}\\
i\hat{\sigma}_{y}
\end{array}\right)\left(\begin{array}{c}
H_{x}^{(i)}\\
H_{y}^{(i)}
\end{array}\right)\nonumber \\
 &  & \times\exp\left[ip_{0,0}^{(1)}\left(z+d/2\right)\right]\label{eq:mat-vacuum}\\
 &  & +\sum_{s,j=-\infty}^{\infty}\left(\begin{array}{c}
\hat{\cal{I}}\\
-{\cal \hat{Q}}_{s,j}^{(1)}
\end{array}\right)\left(\begin{array}{c}
H_{x||s,j}^{(r)}\\
H_{y||s,j}^{(r)}
\end{array}\right)\nonumber \\
 &  & \times\exp\left[i\mathbf{k}_{s,j}\mathbf{r}-ip_{s,j}^{(1)}\left(z+d/2\right)\right],\nonumber \\
\left(\begin{array}{c}
H_{x}^{(3)}(\mathbf{r},z)\\
H_{y}^{(3)}(\mathbf{r},z)\\
E_{x}^{(3)}(\mathbf{r},z)\\
E_{y}^{(3)}(\mathbf{r},z)
\end{array}\right) & = & \sum_{s,j=-\infty}^{\infty}\left(\begin{array}{c}
\hat{{\cal I}}\\
\hat{{\cal Q}}_{s,j}^{(3)}
\end{array}\right)\left(\begin{array}{c}
H_{x||s,j}^{(t)}\\
H_{y||s,j}^{(t)}
\end{array}\right)\label{eq:mat-substrate}\\
 &  & \times\exp\left[i\mathbf{k}_{s,j}\mathbf{r}+ip_{s,j}^{(3)}\left(z-d/2\right)\right],\nonumber 
\end{eqnarray}
where $s,j$ stand for the indices of spatial harmonics, and $\mathbf{k}_{s,j}=\left(sg,\,jg\right)$
is the 2D wavevector in the transverse plane $\mathbf{r}=\left(x,y\right)$.
In Eqs. (\ref{eq:mat-vacuum}) and (\ref{eq:mat-substrate}) the electromagnetic
field time-dependence is implicitly assumed as $\vec{E},\vec{H}\sim\exp\left(-i\omega t\right)$.
Also 
\begin{eqnarray*}
\hat{{\cal I}}=\left(\begin{array}{cc}
1 & 0\\
0 & 1
\end{array}\right) 
\end{eqnarray*}
is the 2$\times$2 unit matrix and 
\begin{eqnarray*}
\hat{\sigma}_{y}=\left(\begin{array}{cc}
0 & -i\\
i & 0
\end{array}\right)
\end{eqnarray*}
is the Pauli matrix, 
\begin{eqnarray*}
{\cal \hat{Q}}_{s,j}^{(1)}=\left(\begin{array}{cc}
\frac{c}{\omega}\frac{sjg^{2}}{p_{s,j}^{(1)}} & \frac{c}{\omega}\frac{\left(p_{s,0}^{(1)}\right)^{2}}{p_{s,j}^{(1)}}\\
-\frac{c}{\omega}\frac{\left(p_{0,j}^{(1)}\right)^{2}}{p_{s,j}^{(1)}} & -\frac{c}{\omega}\frac{sjg^{2}}{p_{s,j}^{(1)}}
\end{array}\right),\\
\hat{{\cal Q}}_{s,j}^{(3)}=\left(\begin{array}{cc}
\frac{c}{\omega\varepsilon}\frac{sjg^{2}}{p_{s,j}^{(3)}} & \frac{c}{\omega\varepsilon}\frac{\left(p_{s,0}^{(3)}\right)^{2}}{p_{s,j}^{(3)}}\\
-\frac{c}{\omega\varepsilon}\frac{\left(p_{0,j}^{(3)}\right)^{2}}{p_{s,j}^{(3)}} & -\frac{c}{\omega\varepsilon}\frac{sjg^{2}}{p_{s,j}^{(3)}}
\end{array}\right),
\end{eqnarray*}
$p_{s,j}^{\left(1\right)}=\sqrt{(\omega/c)^{2}-\left|\mathbf{k}_{s,j}\right|^{2}}$
and $p_{s,j}^{\left(3\right)}=\sqrt{(\omega/c)^{2}\varepsilon-\left|\mathbf{k}_{s,j}\right|^{2}}$
are $z$-components of the wavevector, corresponding to $\left(s,j\right)$
harmonics in the vacuum and substrate, respectively.

Equations (\ref{eq:mat-vacuum}) and (\ref{eq:mat-substrate}) can
be interpreted in the following manner. In the half-space $z<-d/2$,
occupied by air, the total electromagnetic field is composed
of the electromagnetic fields, which correspond to incident {[}first
term in Eq.(\ref{eq:mat-vacuum}){]} and reflected {[}second term
in Eq.(\ref{eq:mat-vacuum}){]} waves. At the same time, in the substrate
(half-space $z>d/2$) electromagnetic field {[}see Eq.
(\ref{eq:mat-substrate}){]} contains the transmitted wave only. If
the sign of the wavevector's $z$-component $p_{s,j}^{\left(n\right)}$
(where $n=1,3$) satisfies the conditions ${\rm Re}(p_{s,j}^{\left(n\right)})\ge0$,
${\rm Im}(p_{s,j}^{\left(n\right)})\ge0$, the transverse incident
wave propagates along $z$-axis in the positive direction and is characterized
by the wavevector $p_{0,0}^{\left(1\right)}=\omega/c$ and the magnetic
field amplitude $\mathbf{H}^{(i)}=\left(H_{x}^{(i)},\,H_{y}^{(i)},\,0\right)$.
The reflected and transmitted waves contain an infinite number of
spatial harmonics. The reflected wave harmonic with index $\left(s,j\right)$
{[}characterized by the amplitude $\mathbf{H}_{s,j}^{(r)}=\left(H_{x||s,j}^{(r)},\,H_{y||s,j}^{(r)},\,H_{z||s,j}^{(r)}\right)$
and 3D wavevector $\left(\mathbf{k}_{s,j},\,-p_{s,j}^{\left(1\right)}\right)${]}
in the negative direction of $z$-axis can be either propagating wave
(in this case $|\mathbf{k}_{s,j}|<\omega/c$ and $p_{s,j}^{\left(1\right)}$
will be purely real and positive) or evanescent (in this case $|\mathbf{k}_{s,j}|>\omega/c$
and $p_{s,j}^{\left(1\right)}$ will be purely imaginary and positive).
Owing to the fact that substrate's dielectric constant $\varepsilon$
is complex, each of the transmitted wave harmonics $s,j$ {[}with
amplitude $\mathbf{H}_{s,j}^{(t)}=\left(H_{x||s,j}^{(t)},\,H_{y||s,j}^{(t)},\,H_{z||s,j}^{(t)}\right)$
and 3D wavevector $\left(\mathbf{k}_{s,j},\,p_{s,j}^{\left(3\right)}\right)${]}
is of mixed type: along $z$ axis it propagates in its positive direction
with exponential factor $\sim\exp\left[i{\rm Re}(p_{s,j}^{\left(3\right)})z\right]$,
but having a decaying amplitude $\sim\exp\left[-{\rm Im}(p_{s,j}^{\left(3\right)})z\right]$. 

Inside the holes in the metallic film, for $lD<x<lD+W$, $l^{\prime}D<y<l^{\prime}D+W$,$-d/2\le z\le d/2$,
transverse components of the magnetic and electric fields can be represented
in matrix form as (see Appendix \ref{sec:Append-em-field} for details)\begin{widetext}
\begin{eqnarray}
\left(\begin{array}{c}
H_{x}^{(2)}(\mathbf{r},z)\\
H_{y}^{(2)}(\mathbf{r},z)
\end{array}\right) & = & \sum_{m,n=0}^{\infty}{\cal \hat{P}}_{m,n}\left(x-lD,\,y-l^{\prime}D\right)\hat{U}_{m,n}\left[\left(\begin{array}{c}
A_{m,n}^{(s)}\delta_{m,0}^{\prime}\delta_{0,n}^{\prime}\\
B_{m,n}^{(s)}
\end{array}\right)\sin\left(\mu_{m,n}\left(z+\frac{d}{2}\right)\right)\right.\nonumber \\
 &  & \left.+\left(\begin{array}{c}
A_{m,n}^{(c)}\delta_{m,0}^{\prime}\delta_{0,n}^{\prime}\\
B_{m,n}^{(c)}
\end{array}\right)\cos\left(\mu_{m,n}\left(z-\frac{d}{2}\right)\right)\right],\label{eq:H-res-mat-mod}\\
\left(\begin{array}{c}
E_{x}^{(2)}(\mathbf{r},z)\\
E_{y}^{(2)}(\mathbf{r},z)
\end{array}\right) & = & \sum_{m,n=0}^{\infty}{\cal \hat{P}}_{m,n}^{\prime}\left(x-lD,\,y-l^{\prime}D\right)\hat{U}_{m,n}\left[\left(\begin{array}{c}
B_{m,n}^{(s)}\\
A_{m,n}^{(s)}\delta_{m,0}^{\prime}\delta_{0,n}^{\prime}
\end{array}\right)\cos\left(\mu_{m,n}\left(z+\frac{d}{2}\right)\right)\right.\nonumber \\
 &  & \left.-\left(\begin{array}{c}
B_{m,n}^{(c)}\\
A_{m,n}^{(c)}\delta_{m,0}^{\prime}\delta_{0,n}^{\prime}
\end{array}\right)\sin\left(\mu_{m,n}\left(z-\frac{d}{2}\right)\right)\right],\label{eq:E-res-mat-mod}
\end{eqnarray}
\end{widetext}
where the matrices 
\begin{eqnarray*}
{\cal \hat{P}}_{m,n}\left(\xi,\eta\right)=\\
\left(\begin{array}{cc}
\sin\left(\frac{m\pi}{W}\xi\right)\cos\left(\frac{n\pi}{W}\eta\right) & 0\\
0 & \cos\left(\frac{m\pi}{W}\xi\right)\sin\left(\frac{n\pi}{W}\eta\right)
\end{array}\right),\\
{\cal \hat{P}}_{m,n}^{\prime}\left(\xi,\eta\right)=\hat{\sigma}_{x}{\cal \hat{P}}_{m,n}\left(\xi,\eta\right)\hat{\sigma}_{x}=\\
\left(\begin{array}{cc}
\cos\left(\frac{m\pi}{W}\xi\right)\sin\left(\frac{n\pi}{W}\eta\right) & 0\\
0 & \sin\left(\frac{m\pi}{W}\xi\right)\cos\left(\frac{n\pi}{W}\eta\right)
\end{array}\right),\\
\hat{U}_{m,n}=\frac{W}{\left(m^{2}+n^{2}\right)\pi}\left(\begin{array}{cc}
-\frac{i\omega}{c}n & m\mu_{m,n}\\
\frac{i\omega}{c}m & n\mu_{m,n}
\end{array}\right),
\end{eqnarray*}
$m$ and $n$ are the mode indices, $A_{m,n}^{(c)}$, $A_{m,n}^{(s)}$
are the amplitudes of E-modes, while $B_{m,n}^{(c)}$, $B_{m,n}^{(s)}$ are the H-modes amplitudes. In general, E-modes inside the square
hole exist for nonzero mode indices $m,n\ge1$ \textendash{} this
is taken into account in Eqs. (\ref{eq:H-res-mat-mod}) and (\ref{eq:E-res-mat-mod})
introducing the factor $\delta_{m,0}^{\prime}\delta_{0,n}^{\prime}$ in front of the E-mode coefficients (where $\delta_{m,n}^{\prime}=1-\delta_{m,n}$
and $\delta_{m,n}$ is the Kronecker delta). On the other hand, the H-mode
index ($m$ or $n$) can be zero \textendash{} excepting the case when
both indices $m=n=0$, and this term is implicitly excluded from
the summation in Eqs. (\ref{eq:H-res-mat-mod}) and (\ref{eq:E-res-mat-mod}).
Notice, that Eqs. (\ref{eq:H-res-mat-mod}) and (\ref{eq:E-res-mat-mod})
satisfy the boundary conditions, namely, zero tangential components of the
electric field and zero normal components of the magnetic field on
the metal surfaces, $E_{x}^{(2)}(x,l^{\prime}D,z)=H_{y}^{(2)}(x,l^{\prime}D,z)=0$,
$E_{x}^{(2)}(x,l^{\prime}D+W,z)=H_{y}^{(2)}(x,l^{\prime}D+W,z)=0$,
$E_{y}^{(2)}(lD,y,z)=H_{x}^{(2)}(lD,y,z)=0$, $E_{y}^{(2)}(lD+W,y,z)=H_{x}^{(2)}(lD+W,y,z)=0$.
Along with this, the periodicity of the solutions of Eqs. (\ref{eq:H-res-mat-mod})
and (\ref{eq:E-res-mat-mod}) is achieved through the independence
of the mode amplitudes upon the hole indices $l,l^{\prime}$. 

Applying the boundary conditions (continuity of the tangential components
of the electric and magnetic fields) at the surface of the metal film
$z=-d/2$, where graphene is absent, results in the following system
of equations: \begin{widetext}
\begin{eqnarray}
\left(\begin{array}{c}
H_{x}^{(i)}\\
H_{y}^{(i)}
\end{array}\right)+\sum_{s,j=-\infty}^{\infty}\exp\left[i\mathbf{k}_{s,j}\mathbf{r}\right]\left(\begin{array}{c}
H_{x||s,j}^{(r)}\\
H_{y||s,j}^{(r)}
\end{array}\right)=\label{eq:cond-H1}\\
=\sum_{m,n=0}^{\infty}{\cal \hat{P}}_{m,n}\left(x-lD,\,y-l^{\prime}D\right)\hat{U}_{m,n}\left(\begin{array}{c}
A_{m,n}^{(c)}\delta_{m,0}^{\prime}\delta_{0,n}^{\prime}\\
B_{m,n}^{(c)}
\end{array}\right)\cos\left(\mu_{m,n}d\right),\begin{array}{c}
\quad lD\le x\le lD+W,\\
\quad l^{\prime}D\le y\le l^{\prime}D+W,
\end{array}\nonumber \\
i\hat{\sigma}_{y}\left(\begin{array}{c}
H_{x}^{(i)}\\
H_{y}^{(i)}
\end{array}\right)-\sum_{s,j=-\infty}^{\infty}\exp\left[i\mathbf{k}_{s,j}\mathbf{r}\right]\hat{{\cal Q}}_{s,j}^{(1)}\left(\begin{array}{c}
H_{x||s,j}^{(r)}\\
H_{y||s,j}^{(r)}
\end{array}\right)=\nonumber \\
=\left\{ \begin{array}{cc}
\sum_{m,n=0}^{\infty}{\cal \hat{P}}_{m,n}^{\prime}\left(x-lD,\,y-l^{\prime}D\ \right)\hat{U}_{m,n}\left(\begin{array}{c}
B_{m,n}^{(s)}+B_{m,n}^{(c)}\sin\left(\mu_{m,n}d\right)\\
\left[A_{m,n}^{(s)}+A_{m,n}^{(c)}\sin\left(\mu_{m,n}d\right)\right]\delta_{m,0}^{\prime}\delta_{0,n}^{\prime}
\end{array}\right), & \begin{array}{c}
\quad lD\le x\le lD+W,\\
\quad l^{\prime}D\le y\le l^{\prime}D+W,
\end{array}\\
0, & \quad\mathrm{otherwise}
\end{array}\right.\label{eq:cond-E1}
\end{eqnarray}
\end{widetext}Boundary conditions across the graphene sheet (at $z=d/2$)
imply the continuity of the electric field tangential components and
discontinuity of the tangential components of the magnetic field due
to the presence of the currents in graphene, namely:
\begin{eqnarray}
H_{x}^{(3)}(\mathbf{r},d/2)-H_{x}^{(2)}(\mathbf{r},d/2)\nonumber \\
=\left(4\pi/c\right)\left[\Sigma_{yx}E_{x}^{(3)}(\mathbf{r},d/2)\right.\label{eq:bc-Hx}\\
\left.+\Sigma_{yy}E_{y}^{(3)}(\mathbf{r},d/2)\right],\nonumber \\
H_{y}^{(3)}(\mathbf{r},d/2)-H_{y}^{(2)}(\mathbf{r},d/2)\nonumber \\
=-\left(4\pi/c\right)\left[\Sigma_{xx}E_{x}^{(3)}(\mathbf{r},d/2)\right.\label{eq:bc-Hy}\\
\left.+\Sigma_{xy}E_{y}^{(3)}(\mathbf{r},d/2)\right],\nonumber \\
E_{x}^{(3)}(\mathbf{r},d/2)=E_{x}^{(2)}(\mathbf{r},d/2),\label{eq:bc-Ex}\\
E_{y}^{(3)}(\mathbf{r},d/2)=E_{y}^{(2)}(\mathbf{r},d/2),\label{eq:bc-Ey}
\end{eqnarray}
where 
\begin{eqnarray*}
\Sigma_{xx}=\Sigma_{yy}=i\frac{2\Sigma_{0}\hbar^{2}v_{F}^{2}}{l_{B}^{2}}\sum_{m=-\infty}^{\infty}\left[\frac{n_{F}(E_{|m|+1})-n_{F}(E_{m})}{E_{|m|+1}-E_{m}}\right.\\
\times\left\{ \frac{1}{E_{|m|+1}-E_{m}-\hbar\omega-i\gamma}+\frac{1}{E_{m}-E_{|m|+1}-\hbar\omega-i\gamma}\right\} \\
+\frac{n_{F}(E_{-(|m|+1)})-n_{F}(E_{m})}{E_{-(|m|+1)}-E_{m}}\left\{ \frac{1}{E_{-(|m|+1)}-E_{m}-\hbar\omega-i\gamma}\right.\\
\left.\left.+\frac{1}{E_{m}-E_{-(|m|+1)}-\hbar\omega-i\gamma}\right\} \right]\left(1+\delta_{m,0}\right),\\
\Sigma_{xy}=-\Sigma_{yx}=\frac{2\Sigma_{0}\hbar^{2}v_{F}^{2}}{\pi l_{B}^{2}}\sum_{m=-\infty}^{\infty}\left[\frac{n_{F}(E_{|m|+1})-n_{F}(E_{m})}{E_{|m|+1}-E_{m}}\right.\\
\times\left\{ \frac{1}{E_{|m|+1}-E_{m}-\hbar\omega-i\gamma}-\frac{1}{E_{m}-E_{|m|+1}-\hbar\omega-i\gamma}\right\} \\
+\frac{n_{F}(E_{-(|m|+1)})-n_{F}(E_{m})}{E_{-(|m|+1)}-E_{m}}\left\{ \frac{1}{E_{-(|m|+1)}-E_{m}-\hbar\omega-i\gamma}\right.\\
\left.\left.-\frac{1}{E_{m}-E_{-(|m|+1)}-\hbar\omega-i\gamma}\right\} \right]\left(1+\delta_{m,0}\right)
\end{eqnarray*}
are the components of the graphene conductivity tensor in magnetic field\cite{gr-mf-cond-Ferreira2011-pra},
\begin{eqnarray*}
\hat{\Sigma}=\left(\begin{array}{cc}
\Sigma_{xx} & \Sigma_{xy}\\
\Sigma_{yx} & \Sigma_{yy}
\end{array}\right).
\end{eqnarray*}
In the above relations $E_{m}=\mathrm{sign}\left(m\right)\hbar v_{F}\sqrt{2|m|}/l_{B}$ are
energy levels of the graphene in external magnetic field, $v_{F}\approx10^{6}\thinspace$m/s
is the Fermi velocity in graphene, $l_{B}=(\hbar/eB)^{1/2}$ is the
magnetic length, $\Sigma_{0}=e^2/4\hbar$ is the so-called ac universal conductivity
of graphene,
$$
n_{F}\left(E\right)={\left [ \exp\left(\frac{E-\mu}{kT}\right)+1\right ]^{-1}}
$$
is the Fermi-Dirac distribution function, $\mu$ is the Fermi energy,
and $\gamma$ is the electron relaxation rate in graphene. 

Boundary conditions
\eqref{eq:bc-Hx}\textendash \eqref{eq:bc-Ey} can also be written
in the matrix form as
\begin{eqnarray}
\left(\begin{array}{c}
H_{x}^{(2)}(\mathbf{r},d/2)\\
H_{y}^{(2)}(\mathbf{r},d/2)\\
E_{x}^{(2)}(\mathbf{r},d/2)\\
E_{y}^{(2)}(\mathbf{r},d/2)
\end{array}\right)=\hat{{\cal Q}}_{g}\left(\begin{array}{c}
H_{x}^{(3)}(\mathbf{r},d/2)\\
H_{y}^{(3)}(\mathbf{r},d/2)\\
E_{x}^{(3)}(\mathbf{r},d/2)\\
E_{y}^{(3)}(\mathbf{r},d/2)
\end{array}\right)\label{eq:bc}
\end{eqnarray}
with the $4\times 4$ matrix
\begin{eqnarray*}
\hat{{\cal Q}}_{g}=\left(\begin{array}{cc}
\hat{{\cal I}} & \hat{{\cal G}}\\
0 & \hat{{\cal I}}
\end{array}\right),\\
\hat{{\cal G}}=-i\frac{4\pi}{c}\hat{\sigma}_{y}\hat{\Sigma}=\frac{4\pi}{c}\left(\begin{array}{cc}
-\Sigma_{yx} & -\Sigma_{yy}\\
\Sigma_{xx} & \Sigma_{xy}
\end{array}\right).
\end{eqnarray*}

Combining Eqs. (\ref{eq:bc}) with (\ref{eq:mat-substrate}), it
is possible to obtain the following expressions for the electromagnetic
field at the metal film surface ($z=d/2$),
\begin{eqnarray*}
\left(\begin{array}{c}
H_{x}^{(2)}(\mathbf{r},d/2)\\
H_{y}^{(2)}(\mathbf{r},d/2)\\
E_{x}^{(2)}(\mathbf{r},d/2)\\
E_{y}^{(2)}(\mathbf{r},d/2)
\end{array}\right)=\sum_{s,j=-\infty}^{\infty}\exp\left[i\mathbf{k}_{s,j}\mathbf{r}\right]\\
\times\left(\begin{array}{c}
\hat{{\cal I}}+\hat{{\cal G}}\hat{{\cal Q}}_{s,j}^{(3)}\\
{\cal \hat{Q}}_{s,j}^{(3)}
\end{array}\right)\left(\begin{array}{c}
H_{x||s,j}^{(t)}\\
H_{y||s,j}^{(t)}
\end{array}\right)\; .
\end{eqnarray*}

Similar to Eqs.(\ref{eq:cond-H1}) and (\ref{eq:cond-E1}) we can
write: \begin{widetext}
\begin{eqnarray}
\sum_{s,j=-\infty}^{\infty}\exp\left[i\mathbf{k}_{s,j}\mathbf{r}\right]\left(\hat{{\cal I}}+\hat{{\cal G}}\hat{{\cal Q}}_{s,j}^{(3)}\right)\left(\begin{array}{c}
H_{x||s,j}^{(t)}\\
H_{y||s,j}^{(t)}
\end{array}\right)=\label{eq:cond-H3}\\
=\sum_{m,n=0}^{\infty}{\cal \hat{P}}_{m,n}\left(x-lD,\,y-l^{\prime}D\right)\hat{U}_{m,n}\left(\begin{array}{c}
\left[A_{m,n}^{(s)}\sin\left(\mu_{m,n}d\right)+A_{m,n}^{(c)}\right]\delta_{m,0}^{\prime}\delta_{0,n}^{\prime}\\
B_{m,n}^{(s)}\sin\left(\mu_{m,n}d\right)+B_{m,n}^{(c)}
\end{array}\right),\begin{array}{c}
\quad lD\le x\le lD+W,\\
\quad l^{\prime}D\le y\le l^{\prime}D+W,
\end{array}\; ;\nonumber \\
\sum_{s,j=-\infty}^{\infty}\exp\left[i\mathbf{k}_{s,j}\mathbf{r}\right]\hat{{\cal Q}}_{s,j}^{(3)}\left(\begin{array}{c}
H_{x||s,j}^{(t)}\\
H_{y||s,j}^{(t)}
\end{array}\right)=\nonumber \\
=\left\{ \begin{array}{cc}
\sum_{m,n=0}^{\infty}{\cal \hat{P}}_{m,n}^{\prime}\left(x-lD,\,y-l^{\prime}D\ \right)\hat{U}_{m,n}\left(\begin{array}{c}
B_{m,n}^{(s)}\\
A_{m,n}^{(s)}\delta_{m,0}^{\prime}\delta_{0,n}^{\prime}
\end{array}\right)\cos\left(\mu_{m,n}d\right), & \begin{array}{c}
\quad lD\le x\le lD+W,\\
\quad l^{\prime}D\le y\le l^{\prime}D+W,
\end{array}\\
0, & \quad\mathrm{otherwise}
\end{array}\right.\; . \label{eq:cond-E3}
\end{eqnarray}
Multiplying Eqs. (\ref{eq:cond-H1}) and (\ref{eq:cond-H3})
by ${\cal \hat{P}}_{m^{\prime},n^{\prime}}\left(x-lD,\,y-l^{\prime}D\right)$
and integrating over the area of the hole $lD\le x\le lD+W$, $l^{\prime}D\le y\le l^{\prime}D+W$,
\end{widetext}
we obtain
\begin{eqnarray}
{\cal \hat{P}}_{m^{\prime},n^{\prime}||0,0}\left(\begin{array}{c}
H_{x}^{(i)}\\
H_{y}^{(i)}
\end{array}\right)+\sum_{s,j=-\infty}^{\infty}{\cal \hat{P}}_{m^{\prime},n^{\prime}||s,j}\left(\begin{array}{c}
H_{x||s,j}^{(r)}\\
H_{y||s,j}^{(r)}
\end{array}\right)\nonumber \\
=\left(\begin{array}{cc}
\delta_{m^{\prime},0}^{\prime}\left(1+\delta_{0,n^{\prime}}\right) & 0\\
0 & \delta_{0,n^{\prime}}^{\prime}\left(1+\delta_{m^{\prime},0}\right)
\end{array}\right)\label{eq:final-H1}\\
\times\frac{W^{2}}{4}\hat{U}_{m^{\prime},n^{\prime}}\left(\begin{array}{c}
A_{m^{\prime},n^{\prime}}^{(c)}\delta_{m^{\prime},0}^{\prime}\delta_{0,n^{\prime}}^{\prime}\\
B_{m^{\prime},n^{\prime}}^{(c)}
\end{array}\right)\cos\left(\mu_{m^{\prime},n^{\prime}}d\right)\; ;\nonumber \\
\sum_{s,j=-\infty}^{\infty}{\cal \hat{P}}_{m^{\prime},n^{\prime}||s,j}\left(\hat{I}+\hat{{\cal G}}\hat{{\cal Q}}_{s,j}^{(3)}\right)\left(\begin{array}{c}
H_{x||s,j}^{(t)}\\
H_{y||s,j}^{(t)}
\end{array}\right)\nonumber \\
=\left(\begin{array}{cc}
\delta_{m^{\prime},0}^{\prime}\left(1+\delta_{0,n^{\prime}}\right) & 0\\
0 & \delta_{0,n^{\prime}}^{\prime}\left(1+\delta_{m^{\prime},0}\right)
\end{array}\right)\frac{W^{2}}{4}\hat{U}_{m^{\prime},n^{\prime}}\nonumber \\
\label{eq:final-H3}\\
\times\left(\begin{array}{c}
\left[A_{m^{\prime},n^{\prime}}^{(s)}\sin\left(\mu_{m^{\prime},n^{\prime}}d\right)+A_{m^{\prime},n^{\prime}}^{(c)}\right]\delta_{m^{\prime},0}^{\prime}\delta_{0,n^{\prime}}^{\prime}\\
B_{m^{\prime},n^{\prime}}^{(s)}\sin\left(\mu_{m^{\prime},n^{\prime}}d\right)+B_{m^{\prime},n^{\prime}}^{(c)}
\end{array}\right),\nonumber 
\end{eqnarray}
where
\begin{eqnarray*}
{\cal \hat{P}}_{m^{\prime},n^{\prime}||s,j}=\int_{0}^{W}dx\int_{0}^{W}dy{\cal \hat{P}}_{m,n}\left(\mathbf{r}\right)\exp\left[i\mathbf{k}_{s,j}\mathbf{r}\right]\\
=\left(\begin{array}{cc}
S_{m^{\prime}||s}C_{n^{\prime}||j} & 0\\
0 & C_{m^{\prime}||s}S_{n^{\prime}||j}
\end{array}\right),
\end{eqnarray*}
and $C_{m||s}=\int_{0}^{W}\cos\left(\frac{m\pi}{W}x\right)\exp(isgx)dx$,
$S_{m||s}=\int_{0}^{W}\sin\left(\frac{m\pi}{W}x\right)\exp(isgx)dx$.
While obtaining Eqs. (\ref{eq:final-H1}) and (\ref{eq:final-H3}),
we have used the orthogonality of trigonometric functions for $m,m^{\prime},n,n^{\prime}\ge0$,
\begin{eqnarray*}
\int_{0}^{W}\sin\left(\frac{m\pi}{W}\xi\right)\sin\left(\frac{m^{\prime}\pi}{W}\xi\right)d\xi=\frac{W}{2}\delta_{m,m^{\prime}}\delta_{m^{\prime},0}^{\prime},\\
\int_{0}^{W}\cos\left(\frac{n\pi}{W}\xi\right)\cos\left(\frac{n^{\prime}\pi}{W}\xi\right)d\xi=\frac{W}{2}\left(1+\delta_{n^{\prime},0}\right)\delta_{n,n^{\prime}}.
\end{eqnarray*}
Also we multiply Eqs. (\ref{eq:cond-E1}) and (\ref{eq:cond-E3})
by $\exp\left[-i\mathbf{k}_{s^{\prime},j^{\prime}}\mathbf{r}\right]$,
and integrate over the area of one period of the structure $lD\le x\le\left(l+1\right)D$,
$l^{\prime}D\le y\le\left(l^{\prime}+1\right)D$, and after taking
into account orthogonality of the plane waves in the unit cell,
\begin{eqnarray*}
\int_{0}^{D}dx\int_{0}^{D}dy\exp\left[i\left(\mathbf{k}_{s,j}-\mathbf{k}_{s^{\prime},j^{\prime}}\right)\mathbf{r}\right]=D^{2}\delta_{s,s^{\prime}}\delta_{j,j^{\prime}}\; ,
\end{eqnarray*}
we obtain
\begin{eqnarray}
i\hat{\sigma}_{y}\left(\begin{array}{c}
H_{x}^{(i)}\\
H_{y}^{(i)}
\end{array}\right)D^{2}\delta_{s^{\prime},0}\delta_{j^{\prime},0}-D^{2}\hat{{\cal Q}}_{s^{\prime},j^{\prime}}^{(1)}\left(\begin{array}{c}
H_{x||s,j}^{(r)}\\
H_{y||s,j}^{(r)}
\end{array}\right)\nonumber \\
=\sum_{m,n=0}^{\infty}\overline{{\cal \hat{P}}_{n,m||j^{\prime},s^{\prime}}}\hat{U}_{m,n}\nonumber \\
\label{eq:final-E1}\\
\times\left(\begin{array}{c}
B_{m,n}^{(s)}+B_{m,n}^{(c)}\sin\left(\mu_{m,n}d\right)\\
\left[A_{m,n}^{(s)}+A_{m,n}^{(c)}\sin\left(\mu_{m,n}d\right)\right]\delta_{m,0}^{\prime}\delta_{0,n}^{\prime}
\end{array}\right),\nonumber \\
D^{2}{\cal \hat{Q}}_{s^{\prime},j^{\prime}}^{(3)}\left(\begin{array}{c}
H_{x||s^{\prime},j^{\prime}}^{(t)}\\
H_{y||s^{\prime},j^{\prime}}^{(t)}
\end{array}\right)=\sum_{m,n=0}^{\infty}\overline{{\cal \hat{P}}_{n,m||j^{\prime},s^{\prime}}}\hat{U}_{m,n}\label{eq:final-E3}\\
\times\left(\begin{array}{c}
B_{m,n}^{(s)}\\
A_{m,n}^{(s)}\delta_{m,0}^{\prime}\delta_{0,n}^{\prime}
\end{array}\right)\cos\left(\mu_{m,n}d\right),\nonumber 
\end{eqnarray}
where 
\begin{eqnarray*}
\overline{{\cal \hat{P}}_{n,m||j^{\prime},s^{\prime}}}=\int_{0}^{W}dx\int_{0}^{W}dy{\cal \hat{P}}_{m,n}^{\prime}\left(\mathbf{r}\right)\exp\left[-i\mathbf{k}_{s,j}\mathbf{r}\right]\\
=\left(\begin{array}{cc}
\overline{C_{m||s^{\prime}}}\overline{S_{n||j^{\prime}}} & 0\\
0 & \overline{S_{m||s^{\prime}}}\overline{C_{n||j^{\prime}}}
\end{array}\right),
\end{eqnarray*}
and the overbar stands for the complex conjugation. Thus, solving
the linear system of Eqs. (\ref{eq:final-H1}) \textendash{} (\ref{eq:final-E3}),
it is possible to obtain the amplitudes of the excited modes inside
the holes as well as amplitudes of the reflected and transmitted harmonics.
From these it is possible to calculate the total reflectance $R$
and transmittace $T$ of the structure as {[}see Applendix
\ref{sec:Append-em-field} for details{]} 
\begin{eqnarray}
R & = & -\left[\left|H_{x}^{(i)}\right|^{2}+\left|H_{y}^{(i)}\right|^{2}\right]^{-1}\label{eq:R-final}\\
 &  & \times\sum_{i,j=-\infty}^{\infty}\mathrm{Re}\left\{ \left(\begin{array}{c}
H_{x||s,j}^{(r)}\\
H_{y||s,j}^{(r)}
\end{array}\right)^{\dagger}i\hat{\sigma}_{y}\hat{{\cal Q}}_{s,j}^{(1)}\left(\begin{array}{c}
H_{x||s,j}^{(r)}\\
H_{y||s,j}^{(r)}
\end{array}\right)\right\} ,\nonumber \\
T & = & -\left[\left|H_{x}^{(i)}\right|^{2}+\left|H_{y}^{(i)}\right|^{2}\right]^{-1}\label{eq:T-final}\\
 &  & \times\sum_{i,j=-\infty}^{\infty}\mathrm{Re}\left\{ \left(\begin{array}{c}
H_{x||s,j}^{(t)}\\
H_{y||s,j}^{(t)}
\end{array}\right)^{\dagger}i\hat{\sigma}_{y}\hat{{\cal Q}}_{s,j}^{(3)}\left(\begin{array}{c}
H_{x||s,j}^{(t)}\\
H_{y||s,j}^{(t)}
\end{array}\right)\right\} .\nonumber 
\end{eqnarray}
It is common to characterize the transmission of graphene-based structures in terms of the so called extinction relative to bare graphene with zero Fermi energy (designated by the abbreviation CNP that stands for "charge neutrality point")\cite{spp-gr-mo-Poumirol2017-NatComm}, $1-T/T_{CNP}$, where both transmittances can be calculated using the above equation. We shall present this quantity calculated in the subsequent sections.

\section{Faraday rotation}
\label{sec:Faraday-rotation}

One of the principal goals of the present work is to investigate the influence of the magnetoplasmon resonance
on the Faraday rotation of an electromagnetic wave traversing the graphene layer. In order to clarify the role
of magnetoplasmons in the Faraday rotation, we start this section by briefly considering
the dispersion properties of magnetoplasmons.

\subsection{Magnetoplasmons in graphene}

\begin{figure}
\includegraphics[width=8.5cm]{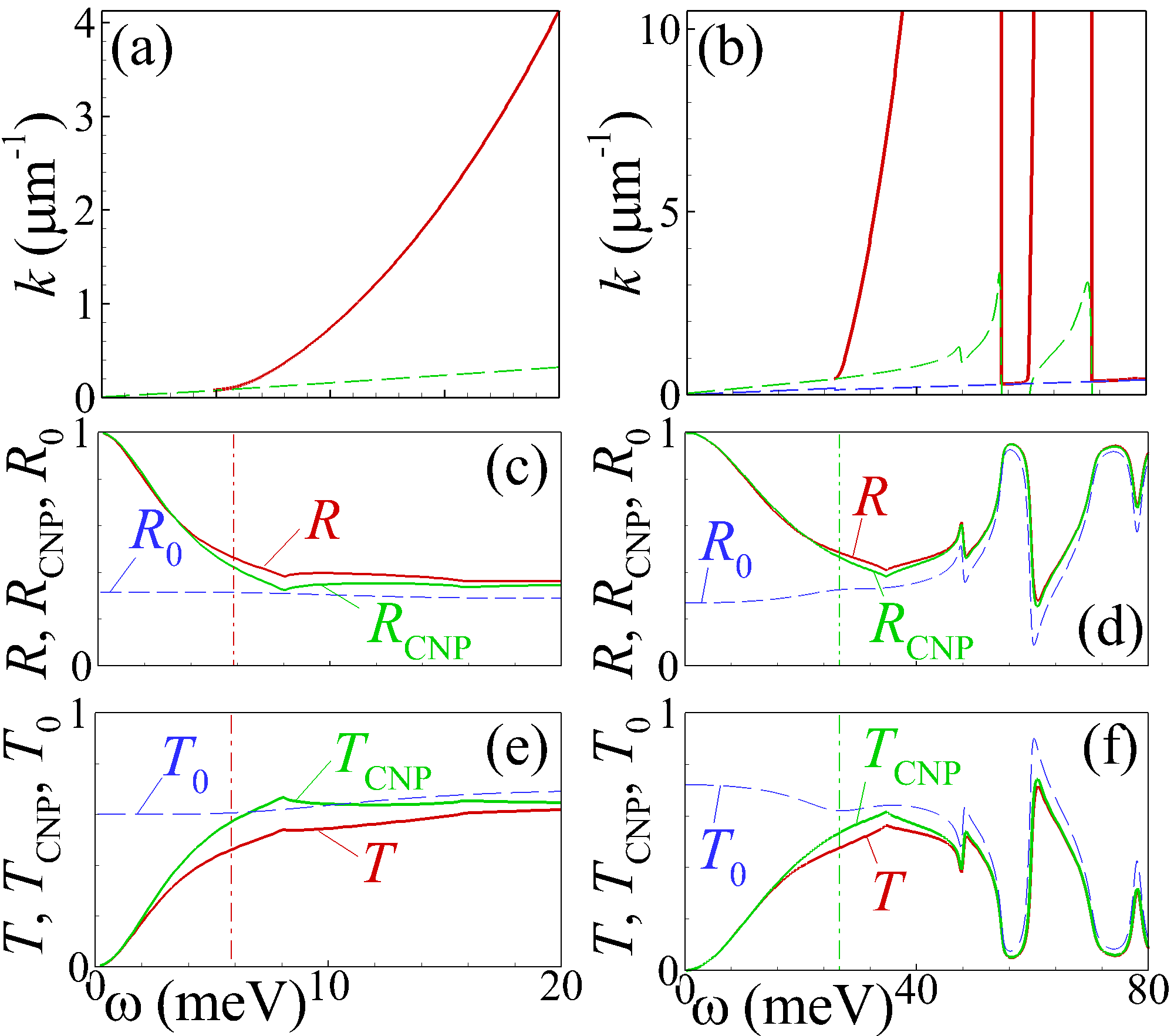}

\caption{(a,b) Dispersion relation of magnetoplasmons (solid red lines) in
graphene with the Fermi energy $\mu=0.175\,\mathrm{eV}$ placed into
magnetic field with magnitude $B=1.5\,\mathrm{T}$ {[}panel (a){]}
or $B=7\,\mathrm{T}$ {[}panel (b){]}. Light lines in vacuum $k=\omega/c$
and in the substrate $k=\omega\sqrt{\varepsilon}/c$ are depicted
by blue and green dashed lines, respectively; (c)\textendash (f) Frequency
dependence of the reflectance {[}panels (c) and (d){]} and transmittance
{[}panels (e) and (f){]} of doped ($\mu=0.175\,\mathrm{eV}$, solid
red lines) or undoped ($\mu=0\,\mathrm{eV}$, green solid lines) graphene
placed into the magnetic field and cladded between metal film with
square grating of thickness $d=50\,\mathrm{nm}$ and semi-infinite $\mathrm{Al}_{2}\mathrm{O}_{3}$
substrate as well as doped ($\mu=0.175\,\mathrm{eV}$) graphene without
grating ($d=0\,\mathrm{nm}$, dashed blue lines). Vertical dash-and-dotted lines depict the frequencies of the magnetoplasmon resonance $\omega_{mp}$ for wavevector $k=2\pi/D$. Other parameters
of the structure are: $B=1.5\,\mathrm{T}$, $D=50\,\mu\mathrm{m}$
{[}panels (c) and (e){]}, $B=7\,\mathrm{T}$, $D=10\,\mu\mathrm{m}$
{[}panels (d) and (f){]}, $W=0.9\,D$, $\gamma=7.54\,\mathrm{meV}$. The subscript 0 refers to doped graphene without grating and CNP stands for "charge neutrality point" and means undoped graphene with grating.}
\label{fig:eigen-sp}
\end{figure}

The dispersion relation of magnetoplasmons $k\left(\omega\right)$
can be obtained from the following equation (see Applendix \ref{subsec:dr-magnetoplasmons}
for details):
\begin{eqnarray}
\left[p^{(3)}(k)+\frac{4\pi\omega}{c^{2}}\sigma_{xx}+p^{(1)}(k)\right]\times\nonumber \\
\times\left[\frac{\varepsilon}{p^{(3)}(k)}+\frac{1}{p^{(1)}(k)}+\frac{4\pi}{\omega}\sigma_{xx}\right]+\label{eq:dr-magnetoplasmons}\\
+\left(\frac{4\pi}{c}\sigma_{xy}\right)^{2}=0,\nonumber 
\end{eqnarray}
where $p^{\left(1\right)}\left(k\right)=\sqrt{(\omega/c)^{2}-k^{2}}$,
$p^{\left(3\right)}\left(k\right)=\sqrt{(\omega/c)^{2}\varepsilon-k^{2}}$
and, similar to the previous section we require ${\rm Re}\left\{ p^{\left(n\right)}\left(k\right)\right\} \ge0$,
${\rm Im}\left\{ p^{\left(n\right)}\left(k\right)\right\} \ge0$ ($n=1,3$).
If we suppose that both the substrate and the graphene layer are lossless
{[}$\mathrm{Im}\left(\varepsilon\right)\equiv0$, $\gamma\equiv0${]},
then the dispersion relation (\ref{eq:dr-magnetoplasmons}) possesses
a solution in terms of purely real frequency $\omega$ and wavevector
$k$. The magnetoplasmon dispersion is depicted in Figs. \ref{fig:eigen-sp}(a)
and \ref{fig:eigen-sp}(b) for low and high magnetic fields, respectively.
When the external perpendicular magnetic field is applied to graphene,
the magnetoplasmon spectrum contains a low-frequency gap and the magnetoplasmons
exist at frequencies higher than a threshold frequency $\omega_{th}$
(which is approximately equal to the cyclotron frequency, $\omega>\omega_{th}\approx\omega_{c}=e^{2}v_{F}B/\mu$)
and for the wavevectors larger than the corresponding threshold wavevector $k_{th}$.
At the threshold frequency, which is equal to $4.85\,\mathrm{meV}$
in Fig. \ref{fig:eigen-sp}(a) and $25.6\,\mathrm{meV}$ in Fig. \ref{fig:eigen-sp}(b),
the magnetoplasmon spectrum splits off from the substrate light line, $k=\omega\sqrt{\varepsilon}/c$, depicted by green dashed lines. Consequently, the value of the wavevector threshold is $k_{th}=\omega_{th}\sqrt{\varepsilon}/c$).
For higher frequencies, the magnetoplasmon spectrum lies above the
light line in the substrate and it results in purely imaginary $p^{\left(3\right)}\left(k\right)$
and $p^{\left(1\right)}\left(k\right)$, which determine the localization
of the electromagnetic field of the magnetoplasmon close to the graphene
layer. Notice that the real part of the substrate dielectric constant is positive {[}$\mathrm{Re}\left(\varepsilon\right)>0${]},
in this low-frequency range $\omega\gtrsim\omega_{th}$. The physical reason for the existence of magnetoplasmons is the
coupling of the electromagnetic wave with excitations of free charge-carriers
in graphene. 

It is important to notice that in the frequency ranges $55\,\mathrm{meV}\lesssim\omega\lesssim60\:\mathrm{meV}$
and $\omega\gtrsim70.6\,\mathrm{meV}$ {[}inside the limits of the
horizontal axis of Fig.\ref{fig:eigen-sp}(b){]} the real part of
the substrate dielectric constant is negative {[}$\mathrm{Re}\left(\varepsilon\right)<0${]}
owing to the excitations of optical phonons.\footnote{The substrate dielectric function includes optical phonon response,
\begin{equation}  
	\label{DF}
	\varepsilon _3 (\omega )=\varepsilon _{\infty}+ \sum_{n=1}^{4}\frac {f_n\omega _{TO,n}^2} {\omega _{TO,n}^2-\omega ^2-i\omega \Gamma_{TO,n}},
\end{equation}
where $\varepsilon _{\infty}=3.2$, $\omega_{TO,1}=47.7\,$meV, $\omega_{TO,2}=54.8\,$meV, $\omega_{TO,3}=70.5\,$meV, $\omega_{TO,4}=78.7\,$meV are phonon frequencies, $\Gamma_{TO,1}=0.72\,$meV, $\Gamma_{TO,2}=0.54\,$meV, $\Gamma_{TO,3}=1.41\,$meV, $\Gamma_{TO,4}=1.57\,$meV are the phonon dampings, and $f_1=0.3$, $f_2=2.7$, $f_3=3.0$, $f_4=0.3$ are the weighting coefficients.} Consequently, in these frequency
ranges the physical reason for the existence of surface waves is somewhat
different since the substrate-air interface is able to sustain surface modes
due to coupling of the electromagnetic wave to the substrate phonons 
rather than because of the interaction with the free electron oscillations in graphene
(in the following these modes will be referred to as surface phonon-polaritons).
In this case the polariton dispersion curve lies above the light line in vacuum
$k=\omega/c$ {[}blue dashed lines in Fig.\ref{fig:eigen-sp}(b){]}.

\subsection{Magnetoplasmon-enhanced Faraday rotation}

The mismatch between the wavevectors of magnetoplasmons and the light line
(either in the substrate or in air) determines the impossibility
to excite magnetoplasmons by a propagating electromagnetic wave, falling
directly onto the graphene layer. One way to overcome this wavevector
mismatch is to use the electromagnetic wave diffraction on some kind
of periodic structure added to the graphene layer. In this case the electromagnetic wave diffraction on the periodic
structure gives rise to a variety of harmonics {[}in Eqs.
(\ref{eq:mat-vacuum}) and (\ref{eq:mat-substrate}) their 2D wavevectors
are $\mathrm{\mathbf{k}}_{s,j}=\left(sg,jg\right)${]}. If at a certain
frequency $\omega_{mp}$ the wavevector of one of them coincides with
that of the magnetoplasmon spectrum \eqref{eq:dr-magnetoplasmons}, then
the energy of the external electromagnetic wave can be effectively
transferred into the energy of the excited magnetoplasmon. 
\begin{figure}
\includegraphics[width=8.5cm]{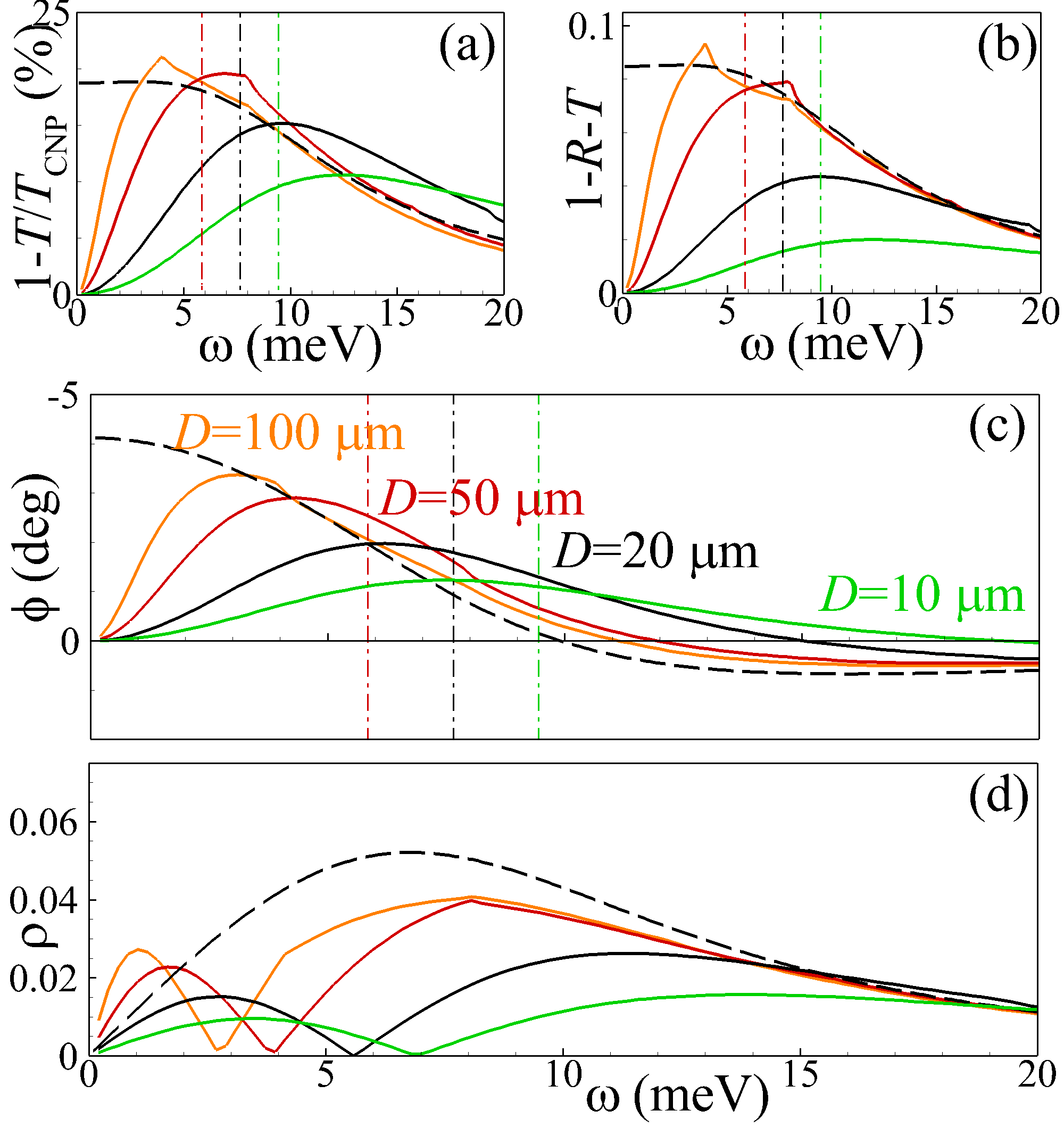}

\caption{Frequency dependence of the extinction {[}panel (a){]}, the absorbance
{[}panel (b){]}, Faraday rotation angle {[}panel (c){]} and transmitted
wave ellipticity {[}panel (d){]} of the doped graphene monolayer placed
into the magnetic field with $\mathrm{B}=1.5\,T$ and cladded between
the semi-infinite $\mathrm{Al}_{2}\mathrm{O}_{3}$ substrate and metal
film with hole grating of period $D=100\,\mu\mathrm{m}$ (solid orange
lines), $D=50\,\mu\mathrm{m}$ (solid red lines), $D=20\,\mu\mathrm{m}$
(solid black lines), or $D=10\,\mu\mathrm{m}$ (solid green lines).
Other parameters are the same as in Fig.\ref{fig:eigen-sp}. In panels
(a), (b) and (c) vertical dash-and-dotted lines depict the frequencies
of the magnetoplasmon resonance for wavevector $k=2\pi/D$ with $D=50\,\mu\mathrm{m}$
(red lines), $D=20\,\mu\mathrm{m}$ (black lines), and $D=10\,\mu\mathrm{m}$
(green lines). In all panels the black dashed lines correspond
to the extinction $1-T_{0}/T_{0,CNP}$, absorbance $1-R_{0}-T_{0}$,
Faraday rotation angle $\phi_{0}$ and ellipticity $\rho_{0}$ of
the transmitted wave in the case of bare graphene. }
\label{fig:var_Al2O3_B_eq1.5}
\end{figure}

It can be understood from Figs. \ref{fig:eigen-sp}(c) \textendash{}
\ref{fig:eigen-sp}(f), that the presence of the perforated metal
film on top of the graphene layer modifies significantly the reflectance
and transmittance of the structure both in the case of doped (solid
red lines) and undoped (solid green lines) graphene. The incident
wave is considered to be linearly polarized with the magnetic field along the $x$-axis (i.e. $H_{x}^{(i)}\ne0$, $H_{y}^{(i)}\equiv0$).
For the parameters of Figs. \ref{fig:eigen-sp}(c) and \ref{fig:eigen-sp}(e)
the lattice vector is equal to $g=2\pi/D\approx0.126\,\mu\mathrm{m}^{-1}$,
for which the predicted frequency of the magnetoplasmon resonance
{[}for $k=g$ see Fig. \ref{fig:eigen-sp}(a){]} is $\omega_{mp}\approx5.85\,\mathrm{meV}$.
Similarly, for the parameters of Figs. \ref{fig:eigen-sp}(d) and
\ref{fig:eigen-sp}(f) the lattice vector is $g\approx0.628\,\mu\mathrm{m}^{-1}$
and the magnetoplasmon resonance frequency is $\omega_{mp}\approx26.75\,\mathrm{meV}$
{[}which can be obtained from Fig. \ref{fig:eigen-sp}(b){]}. In the
low frequency range, $\omega\ll\omega_{mp}$, the wavelength of the
incident electromagnetic wave exceeds significantly the period of
the grating, $\lambda=2\pi c/\omega\gg D$. As a result, the
perforated metal film screens the incident wave almost like a continuous
metal as evidenced by the enhanced reflectance of the
structure, $R\lesssim1$ {[}see Figs. \ref{fig:eigen-sp}(c) and \ref{fig:eigen-sp}(d){]}
and its suppressed transmittance, $T\gtrsim0$ {[}see Figs. \ref{fig:eigen-sp}(e)
and \ref{fig:eigen-sp}(f){]}, as compared to the reflectance, $R_{0}$
and transmittance, $T_{0}$ of bare graphene 
{[}dashed blue lines in Figs. \ref{fig:eigen-sp}(c)\textendash \ref{fig:eigen-sp}(f){]}
\footnote{The expressions for $R_{0}$ and $T_{0}$ as well as the details of derivation
can be found in Appendix \ref{subsec:R0-T0}.}. In the high-frequency range, $\omega\gg\omega_{mp}$, the situation
is opposite, namely, the wavelength of the incident electromagnetic wave is considerably
shorter than the period of structure, $\lambda\ll D$, and, of course,
also smaller than the hole width, $\lambda\ll W$. In this case the
metal film almost does not influence the propagation of the electromagnetic
wave, hence the reflectance, $R$, and the transmittace, $T$, of graphene
with perforated metal film almost conside with those of bare graphene ($R_{0}$ and
$T_{0}$).
\begin{figure}
\includegraphics[width=8.5cm]{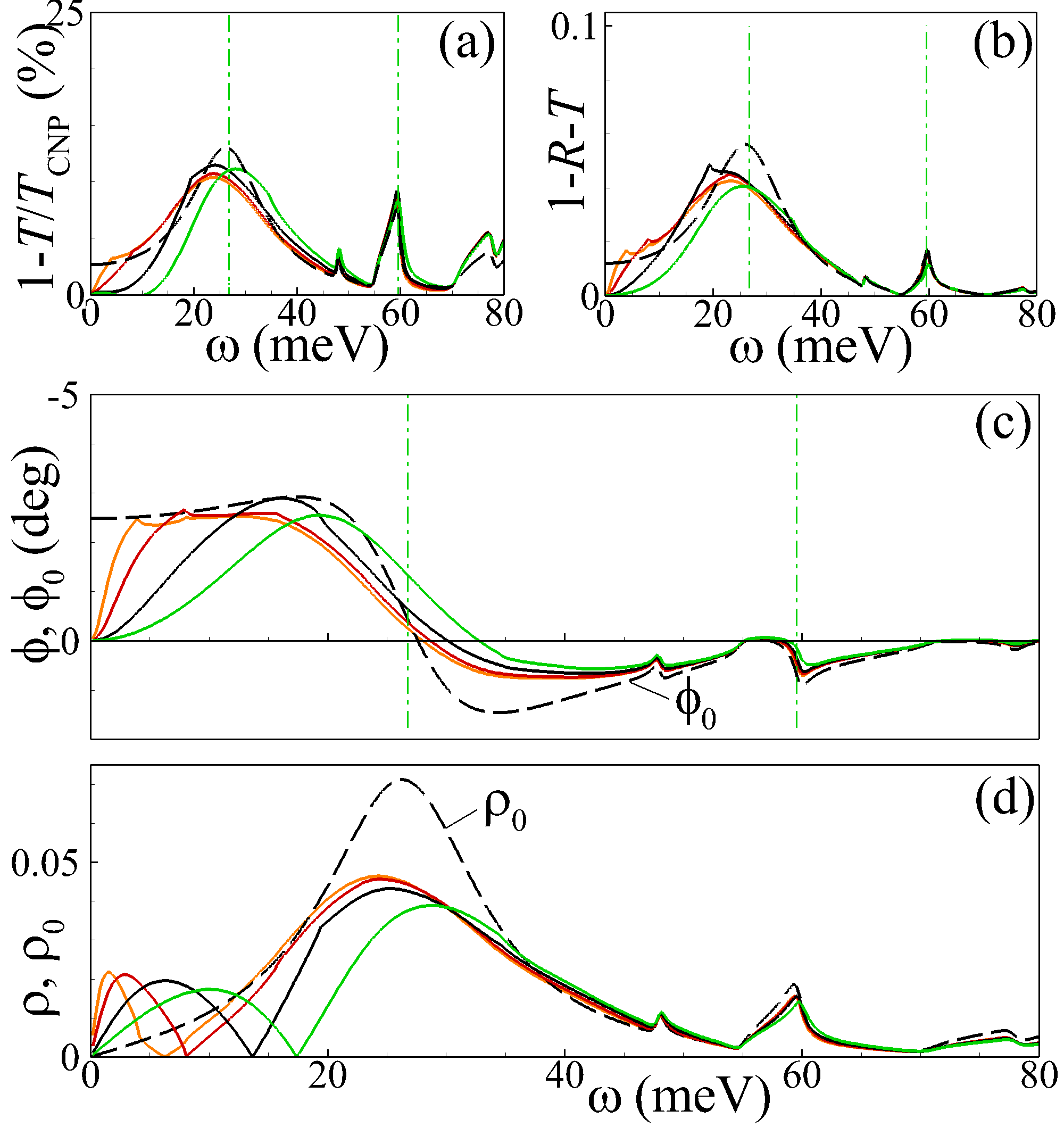}

\caption{The same as in Fig.\ref{fig:var_Al2O3_B_eq1.5}, but for the magnetic
field $\mathrm{B}=7\,T$. }
\label{fig:var_Al2O3_B_eq7}
\end{figure}

The most interesting situation takes place in the intermediate frequency
range, when the frequency of the incident wave is close to the magnetoplasmon
resonance frequency $\omega\sim\omega_{mp}$. As it can be seen from
Figs. \ref{fig:eigen-sp}(e) and \ref{fig:eigen-sp}(f), it is in this frequency
range that the maximal deviation between the transmittance of the
perforated metal film with doped graphene ($T$) and that
of the same structure with undoped graphene ($T_{CNP}$)  takes place owing to the
the excitation of magnetoplasmons. This phenomenon is shown in Figs.\,\ref{fig:var_Al2O3_B_eq1.5}(a)
and \ref{fig:var_Al2O3_B_eq7}(a). Near the frequency of magnetoplasmon
resonance, $\omega_{mp}$ (depicted by vertical dash-and-dotted lines),
the extinction attains its local maximum, as does the
absorbance, $1-R-T$ {[}see Figs. \ref{fig:var_Al2O3_B_eq1.5}(b) and
\ref{fig:var_Al2O3_B_eq7}(b){]}. Interestingly, both the extinction and the absorbance attain their maxima in the
structures that do not allow for the excitation of magnetoplasmons.
Such a situation can take place in two kinds of structures: (i) graphene
layer without perforated metallic film (dashed black lines in Figs.\,\eqref{fig:var_Al2O3_B_eq1.5}
and \eqref{fig:var_Al2O3_B_eq7}); (ii) graphene layer with grating but when the lattice vector of the latter, $g=2\pi/D$,
is below the threshold $k_{th}$ (that is, when the period $D$ is too large, so that the magnetoplasmon resonance eigenfrequency does not exist) -- see the corresponding spectra in Figs. \ref{fig:eigen-sp}(a) and \ref{fig:eigen-sp}(b){]}.
Examples of such situation can be found for structures with $D=100\,\mu m$ (orange lines) in Figs.\,\ref{fig:var_Al2O3_B_eq1.5}(a)
and \ref{fig:var_Al2O3_B_eq1.5}(b), as well as for $D=100\,\mu m$ (orange
lines), $D=50\,\mu m$ (red lines), and $D=20\,\mu m$ (black lines)
in Figs. \ref{fig:var_Al2O3_B_eq7}(a) and \ref{fig:var_Al2O3_B_eq7}(b).
Nevertheless, in these cases the extinction and absorbance attain
their maxima in the vicinity of the cyclotron frequency $\omega_{c}$
owing to the resonant interaction between the electromagnetic wave
and graphene's charge carriers, rotating in the perpendicular magnetic
field (in more details this phenomenon will be described in Sec.\,\ref{sec:Circular-dichroism}).
It should be noticed, that at low magnetic field {[}Figs. \ref{fig:var_Al2O3_B_eq1.5}(a)
and \ref{fig:var_Al2O3_B_eq1.5}(b){]} there is a certain descrepancy
between the predicted magnetoplasmon resonance and the maxima of absorbance
and extinction. The reason for this seems to be the following: 
the relaxation rate of free carriers in graphene used in the calculation was $\gamma=7.54\,\mathrm{meV}$,
i.e., its value is comparable with the predicted frequencies of magnetoplasmon
resonance $\omega_{mp}$. As a result, the magnetoplasmon oscillations
are overdamped in this case. A possible solution of this
problem is to shift the magnetoplasmon resonance to the higher-frequency
range, e.g., by increasing the external magnetic field strength
$B$. In this case the agreement between the predicted magnetoplasmon
resonance frequency and the maxima of the absorbance and the extinction
is considerably better {[}see Figs. \ref{fig:var_Al2O3_B_eq7}(a)
and \ref{fig:var_Al2O3_B_eq7}(b){]}. At the same time, the maxima in the extinction and absorbance spectra at $\omega_{mp}\approx59.5\,\mathrm{meV}$
appear owing to the excitation of the aforementioned surface phonon-polaritons.
\begin{figure}
\includegraphics[width=8.5cm]{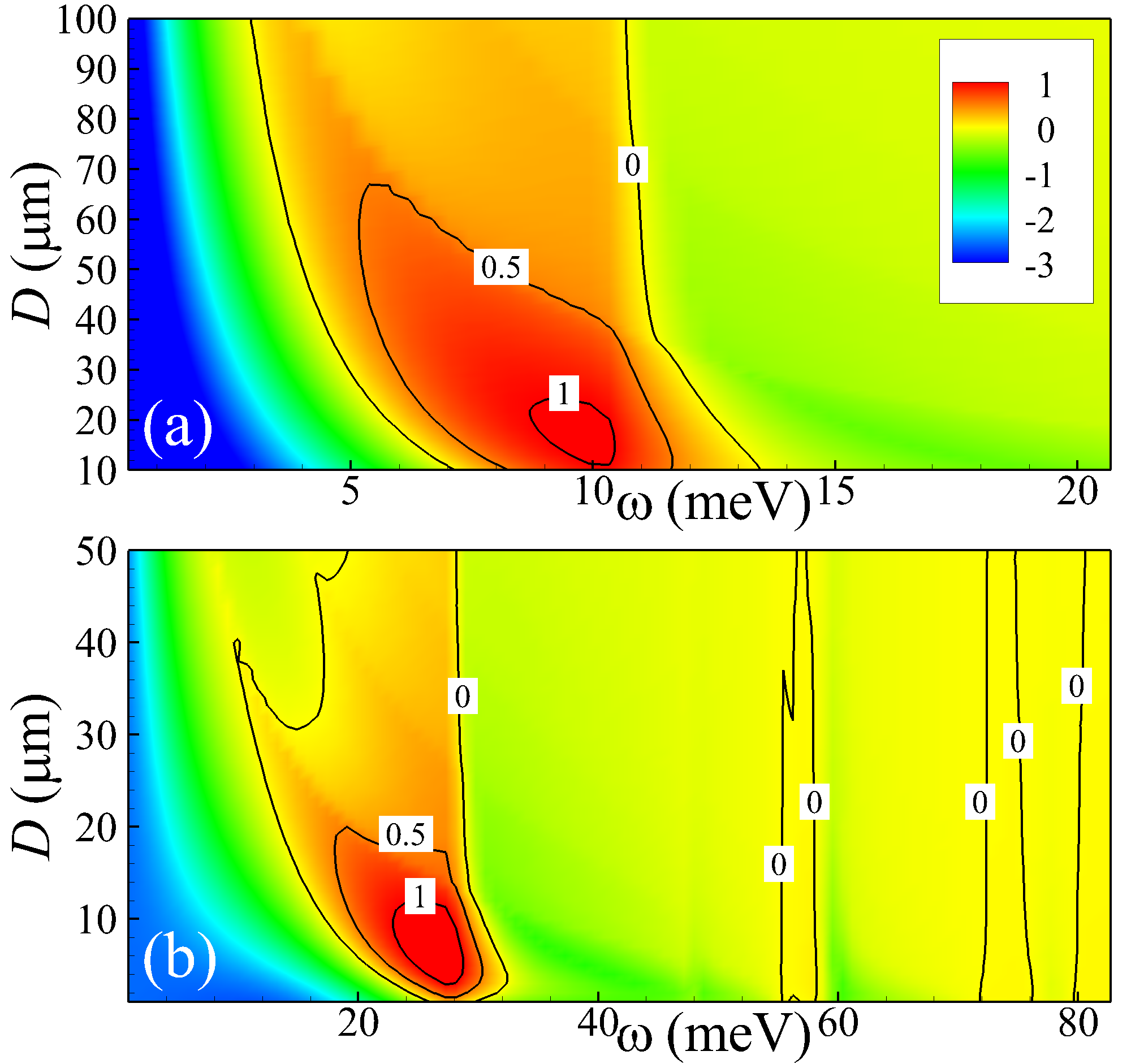}

\caption{Faraday rotation angle difference $\Delta \phi$ (in degrees, depicted
by color map) \emph{versus} frequency $\omega$ and period of the
hole array $D$ for the graphene layer placed into perpendicular
magnetic field with $B=1.5\,T$ {[}panel (a){]} or $B=7\,T$ {[}panel
(b){]} and cladded by the semi-infinite $\mathrm{Al}_{2}\mathrm{O}_{3}$
substrate and the metal film with the periodc array of holes, which
widths are $W=0.9\,D$. Other parameters are the same as those in
Fig.\ref{fig:eigen-sp}.}
\label{fig:del-phi(w,D)}
\end{figure}

How does the excitation of the graphene
magnetoplasmons, which manifests itself by the increase of extinction and absorbance,
influences the Faraday rotation angle of the transmitted wave? To
answer this question we compare the Faraday rotation angle $\phi$
of the electromagnetic wave transmitted through the perforated metal
film on top of the doped graphene layer {[}solid lines in Figs. \ref{fig:var_Al2O3_B_eq1.5}(c)
and \ref{fig:var_Al2O3_B_eq7}(c){]} with that
($\phi_{0}$) of the electromagnetic wave passing through bare
graphene with the same parameters but without the grating
{[}dashed lines in Figs. \ref{fig:var_Al2O3_B_eq1.5}(c) and \ref{fig:var_Al2O3_B_eq7}(c){]}.
The formulae permitting to calculate the Faraday rotation angle from
zero-harmonics amplitudes $H_{x||0,0}^{(t)}$$H_{y||0,0}^{(t)}$ can
be found, e.g., in Ref.\cite{Born99}. It is evident that in the low-frequency
region, $\omega\ll\omega_{mp}$, the rotation angle $\phi$
is considerably below $\phi_{0}$,
i.e., the presence of the perforated metal film hampers the Faraday
rotation effect. This fact is a consequence of the above-mentioned attenuated
transmittance of the perforated metal film, shown in Figs. \ref{fig:eigen-sp}(c)
and \ref{fig:eigen-sp}(d). Notice that the decrease of the Faraday
rotation angle, $\phi$ (compared to $\phi_{0}$), in the low-frequency
range is accompanied by the enhancement of the ellipticity, $\rho$,
of the transmitted wave polarization (again, compared to the ellipticity
$\rho_{0}$ of the polarization of the electromagnetic wave transmitted
through the bare graphene) -- see Figs. \ref{fig:var_Al2O3_B_eq1.5}(d)
and \ref{fig:var_Al2O3_B_eq7}(d). In the high-frequency range,
where $\omega\gg\omega_{mp}$, the aforementioned neglible interaction
between the electromagnetic wave and perforated metal film {[}see
Figs. \ref{fig:eigen-sp}(c) and \ref{fig:eigen-sp}(d){]} reveals
into the fact that $\phi\approx\phi_{0}$ and $\rho\approx\rho_{0}$.
Nevertheless, in the intermediate frequency range, $\omega\sim\omega_{mp}$,
when the magnetoplasmon resonance eigenfrequency does exist {[}red,
black, and green solid lines in Figs.\,\ref{fig:var_Al2O3_B_eq1.5}(c)
and \ref{fig:var_Al2O3_B_eq1.5}(d), as well as green solid lines
in Figs. \ref{fig:var_Al2O3_B_eq7}(c) and \ref{fig:var_Al2O3_B_eq7}(d){]},
the ellipticity $\rho<\rho_{0}$ (so, the transmitted wave polarization
becomes closer to the linearly polarized wave), but the Faraday rotation
angle satisfies the condition $\phi>\phi_{0}$. In other words, excitation
of graphene magnetoplasmons owing to the presence of the perforated
metal film on top of graphene increases the Faraday rotation angle.
In the situation, where the magnetoplasmon resonance eigenfrequency
does not exist {[}$2\pi/D<k_{th}$, examples are orange solid lines
in Figs. \ref{fig:var_Al2O3_B_eq1.5}(c) and \ref{fig:var_Al2O3_B_eq1.5}(d),
as well as orange, red, and black solid lines in Figs. \ref{fig:var_Al2O3_B_eq7}(c)
and \ref{fig:var_Al2O3_B_eq7}(d){]} the Faraday rotation angle $\phi$
is almost the same as $\phi_{0}$ (or even less), even though the
ellipticity $\rho$ can be smaller than $\rho_{0}$. 

More detailed information of how the presence of the perforated metal
film on top of graphene changes the Faraday rotation angle can be
extracted from Fig. \ref{fig:del-phi(w,D)}, which demonstrates the
dependence of the Faraday angle difference, $\Delta\phi=\left|\phi\right|-\left|\phi_{0}\right|$,
upon the frequency $\omega$ and the period of the hole grating $D$.
It is clearly seen that in the intermediate frequency range {[}$4\,\mathrm{meV}\lesssim\omega\lesssim12\,\mathrm{meV}$
in Fig.\ref{fig:del-phi(w,D)}(a) and $15\,\mathrm{meV}\lesssim\omega\lesssim30\,\mathrm{meV}$
in Fig.\ref{fig:del-phi(w,D)}(b){]} the Faraday angle difference
is positive, $\Delta\phi>0$, that is, the presence of the perforated metal
film increases the absolute value of the Faraday rotation angle as
compared to the case without the grating. Beyond this frequency
range the Faraday angle difference is mainly negative, $\Delta\phi<0$.\footnote{Excepting narrow frequency windows at $55\,\mathrm{meV}\lesssim\omega\lesssim60\,\mathrm{meV}$,
and $72\,\mathrm{meV}\lesssim\omega\lesssim75\,\mathrm{meV}$, where,
nevertheless, an increase of the Faraday rotation angle is negligibly small. } In other words, both in the low- and in the high-frequency ranges
presence of the perforated metal film on top of graphene suppresses
the Faraday rotation of transmitted wave. At the same time, inside
the intermediate frequency range there exist some optimal values
of the hole grating period $D\approx17\,\mu\mathrm{m}$ {[}in Fig.\ref{fig:del-phi(w,D)}(a){]}
and $D\approx8\,\mu\mathrm{m}$ {[}in Fig.\ref{fig:del-phi(w,D)}(b){]},
for which the Faraday angle difference is maximal and exceeds $1\textdegree$.

\section{Magnetic circular dichroism\label{sec:Circular-dichroism}}

\begin{figure}
\includegraphics[width=8.5cm]{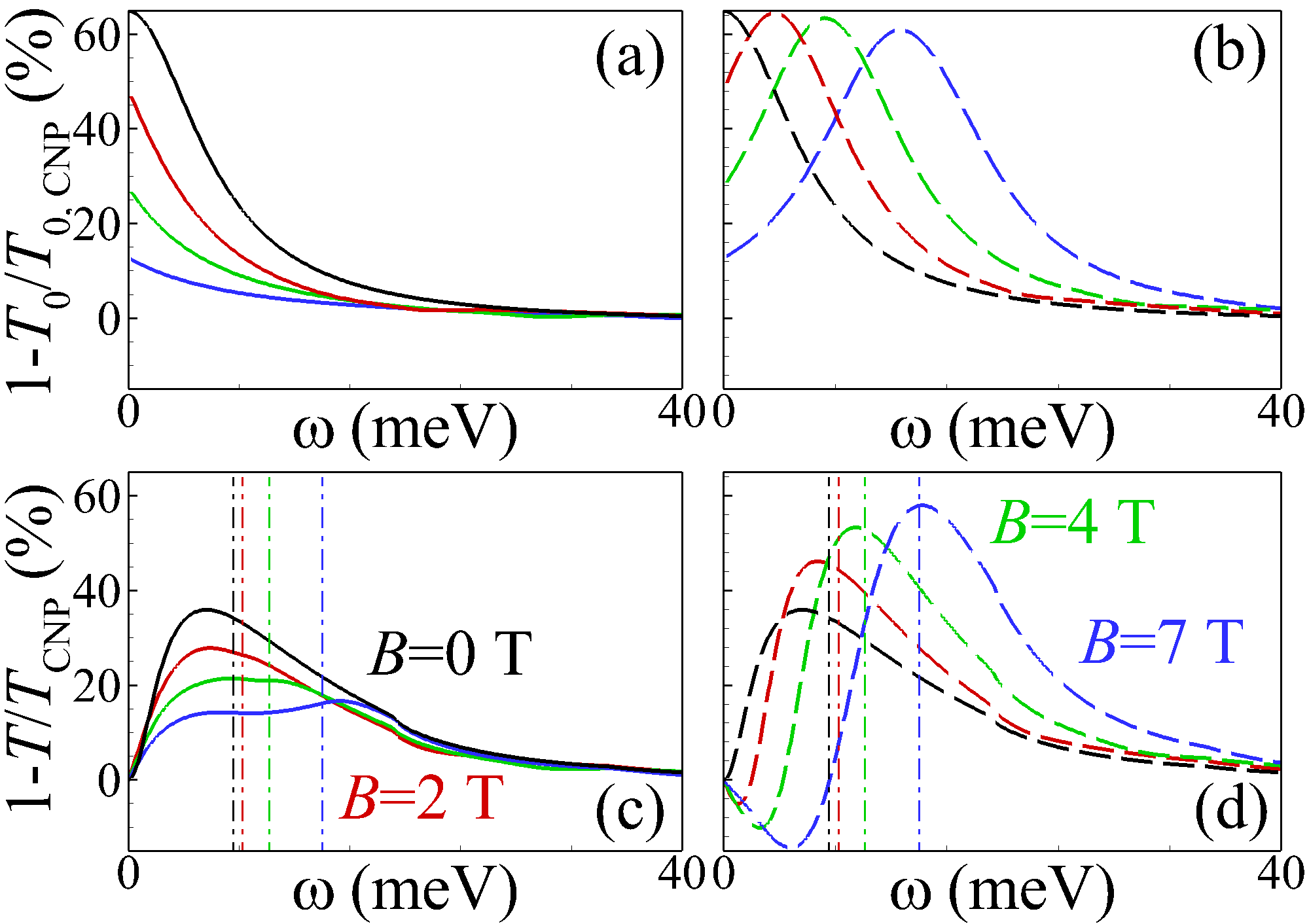}

\caption{(a,b) Frequency dependence of the extinction of the clockwise {[}panel
(a){]} and anticlockwise {[}panel (b){]} circularly polarized electromagnetic
wave transmitted through graphene with Fermi energy $\mu=-0.358\,$eV
and relaxation rate $\gamma=3.68\,$meV on semi-infinite $\mathrm{Al}_{2}\mathrm{O}_{3}$ substrate,
sujected to perpendicular magnetic field of strength $B=0\,$T (black lines), $B=2\,$T (red lines), $B=4\,$T (green
lines) and $B=7\,$T (blue lines); (c,d) The extinction (as a function
of frequency) of the clockwise {[}panel (c){]} and anticlockwise {[}panel
(d){]} circularly polarized electromagnetic wave impining on the graphene
cladded by semi-infinite $\mathrm{Al}_{2}\mathrm{O}_{3}$
substrate and perforated metal film with parameters $D=20\,\mu$m,
$W=18\,\mu$m, $d=50\,$nm. The parameters of graphene, the magnetic
field values and the meaning of the different curves are the same as those in panels
(a) and (b).}
\label{fig:Ext_LR}
\end{figure}
The term dichroism means the property shown by certain materials of having different absorption coefficients for light polarized in different directions.\cite{Born99}
If the amplitudes of the transmitted clockwise and anticlockwise polarized waves will not be equal in the presence of a magnetic field applied along their direction of propagation, 
the material possesses the magnetic circular dichroism (MCD). This is different from the Faraday rotation effect that exists in non-absorbing media. Applying a magnetic field to the dielectric
causes the material to exhibit circular birefringence, i.e. the propagation velocities of the clockwise and anticlockwise polarized waves become unequal. The Faraday rotation angle is then proportional to the difference in 
the refractive indices for two circular polarizations.\cite{Basu} Of course, the two effects are interconnected, especially in an intrinsically absorbing plasmonic structure such as the one considered here, but still we can attempt to make a distinction between them.
Let us point it out that the system under study has no dichroism in the absence of magnetic field.

In order to clarify the influence of magnetoplasmons on the MCD of the structure, we consider first the extinction
of the clockwise and anticlockwise circularly polarized electromagnetic
waves {[}depicted in Figs.\,\ref{fig:Ext_LR}(a) and \ref{fig:Ext_LR}(b),
respectively{]}, transmitted through the graphene layer without 2D grating on top of it. A clockwise polarized 
incident wave is characterized by the $\pi/2$ phase shift
between the $x$ and $y$ components of its magnetic field ($H_{y}^{(i)}=iH_{x}^{(i)}$),
while their amplitudes are equal. The phase shift between the magnetic field components in the anticlockwise
circularly polarized incident wave is $-\pi/2$, i.e. $H_{y}^{(i)}=-iH_{x}^{(i)}$.

As follows from the comparison of Figs.\,\ref{fig:Ext_LR}(a)
and \ref{fig:Ext_LR}(b), at zero magnetic field (black lines) the
extinction values of clockwise and anticlockwise circularly polarized electromagnetic
waves are equal. When graphene is doped with holes (negative chemical
potential $\mu<0$), application of external magnetic field results
in the circular MCD: for the same frequency the extinction
of the anticlockwise-polarized wave {[}Fig.\,\ref{fig:Ext_LR}(b){]}
exceeds that of the clockwise circularly polarized wave {[}Fig.\,\ref{fig:Ext_LR}(a){]}.
The frequency dependence of the extinction, being a monotonically
decreasing function in the case of clockwise-polarized wave, in the
case of anticlockwise-polarized wave exhibits its maximum at the cyclotron
frequency $\omega_{c}$. As a consequence, the growth of the magnetic
field leads to the blue-shift of the maximum of the anticlockwise-polarized
wave extinction spectrum {[}Fig.\,\ref{fig:Ext_LR}(b){]} and to a monotonic
decrease of the clockwise-polarized wave extinction {[}see Fig.\,\ref{fig:Ext_LR}(a){]}.
\begin{figure*}
\includegraphics[width=17cm]{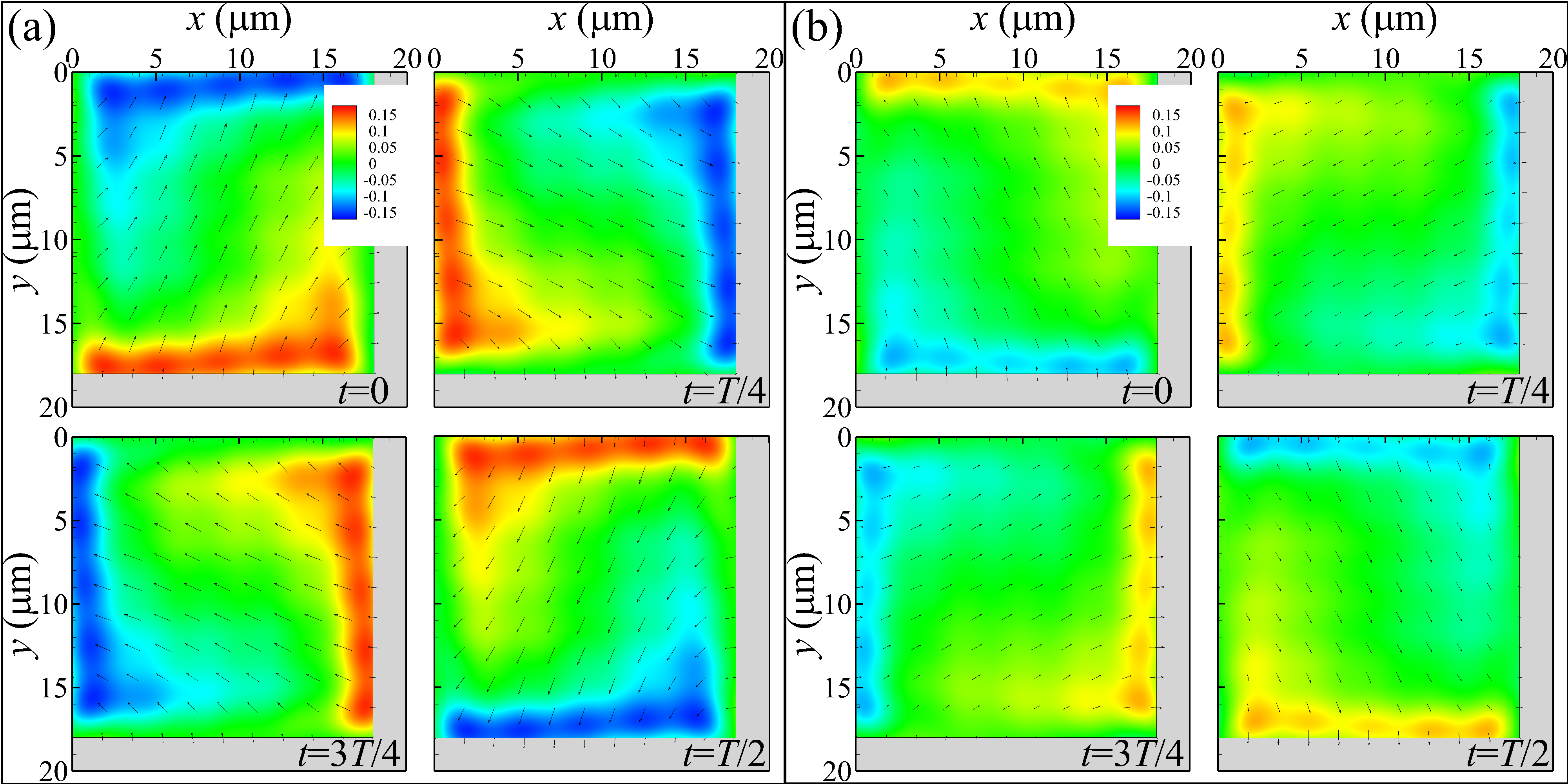}

\caption{Spatial distribution (over a unit cell of the periodically perforated
metal film) of the electric field components $E_{z}^{(3)}(\mathbf{r},d/2)$
(depicted by color map) and $E_{x}^{(3)}(\mathbf{r},d/2)$, $E_{y}^{(3)}(\mathbf{r},d/2)$
(depicted by vectors) on graphene, subjected to the magnetic field
$B=7\thinspace$T for the clockwise {[}box (a){]} and anticlockwise
{[}box (b){]} circularly polarized incident waves with frequency $\omega=14\,$meV
at time moments $t=0$ (upper left panels in each box), after quarter-period
$t=\pi/2\omega$ (upper right panels), after half-period $t=\pi/2\omega$
(lower right panels) and after three-quarters of the period $t=3\pi/2\omega$
(lower left panels). Other parameters of the structure are the same
as those in Fig.\,\ref{fig:Ext_LR}.}
\label{fig:fields}
\end{figure*}

Adding the perforated metal film on top of graphene changes considerably
the extinction spectrum. Thus, in the low-frequency region both clockwise
and anticlockwise circularly polarized waves {[}depicted in Figs.\,\ref{fig:Ext_LR}(c)
and \ref{fig:Ext_LR}(d), correspondingly{]} are characterized by
diminished extinction owing to the aforementioned long-wavelength
screening. This result is completely different from the extinction
spectra of the perforated graphene, considered in Ref.\,\cite{spp-gr-mo-Poumirol2017-NatComm}:
in the low-frequency region perforated graphene exhibits enhanced
extinction. Along with this, the extinction values for electromagnetic waves
of both polarizations reach their maxima in the vicinity of the
magnetoplasmon resonance frequency $\omega_{mp}$ {[}which frequencies
are calculated from Eq.\,\eqref{eq:dr-magnetoplasmons} and depicted
in Figs.\,\ref{fig:Ext_LR}(c) and \ref{fig:Ext_LR}(d) by vertical
dash-and-dotted lines{]}. For the clockwise polarization {[}Fig.\,\ref{fig:Ext_LR}(c){]}
such a behaviour is quite different from that of bare graphene
 {[}where the extinction is a monotonically
decreasing function of frequency, see Fig.\,\ref{fig:Ext_LR}(a){]}.
It is worth noting that in the case of clockwise polarization the
graphene with perforated metal film is characterized by the magnetoplasmon-enhanced
extinction. Indeed, in the vicinity of magnetoplasmon resonance frequency
the extinction of such structure exceeds that of the graphene without
metal film. For the anticlockwise polarization {[}Fig.\,\ref{fig:Ext_LR}(d){]}
the resonant excitation of magnetoplasmons results in the blue-shift
of the extinction maximum, compared to the case of bare graphene {[}Fig.\,\ref{fig:Ext_LR}(b){]}. Noteworthy,
in the low-frequency range the anticlockwise polarized wave is characterized
by a negative extinction, which becomes more pronounced at high magnetic
field {[}compare red and blue lines in Fig.\,\ref{fig:Ext_LR}(d),
which correspond to $B=2\thinspace$T and $B=7\thinspace$T, respectively{]}. 

At the same time, for both polarizations high magnetic field provides
better correspondence between the frequency of maximal extinction
and the frequency of the magnetoplasmon resonance. It can be
seen from the comparison of black, red, green and blue solid lines
maxima in Figs.\,\ref{fig:Ext_LR}(c) and \ref{fig:Ext_LR}(d) with
the positions of the vertical dash-and-dotted lines of respective
colors. Indeed, for $B=7\thinspace$T (blue solid and dash-and-dotted line)
the difference is neglible. The respective spatial distributions of
the electric field on graphene in the vicinity of the magnetoplasmon
resonance frequency are shown in Fig.\,\ref{fig:fields} for clockwise
and anticlockwise circularly polarized incident waves {[}Figs.\,\ref{fig:fields}(a)
and \ref{fig:fields}(b), respectively{]}. For both polarizations, $z$ component of the electric field
(depicted by color map) has a maximum and a minimum near the opposite
edges of the square hole at the magnetoplasmon resonance. In other words, the distribution of charge
carriers in graphene is dipolar. During one period of the electromagnetic
wave, $T=2\pi/\omega$, the dipolar distribution rotates along the hole perimeter 
in the same direction as the incident wave's polarization vector, clockwise
{[}Fig.\,\ref{fig:fields}(a){]} or anticlockwise {[}Fig.\,\ref{fig:fields}(b){]}.

\begin{figure}
\includegraphics[width=8.5cm]{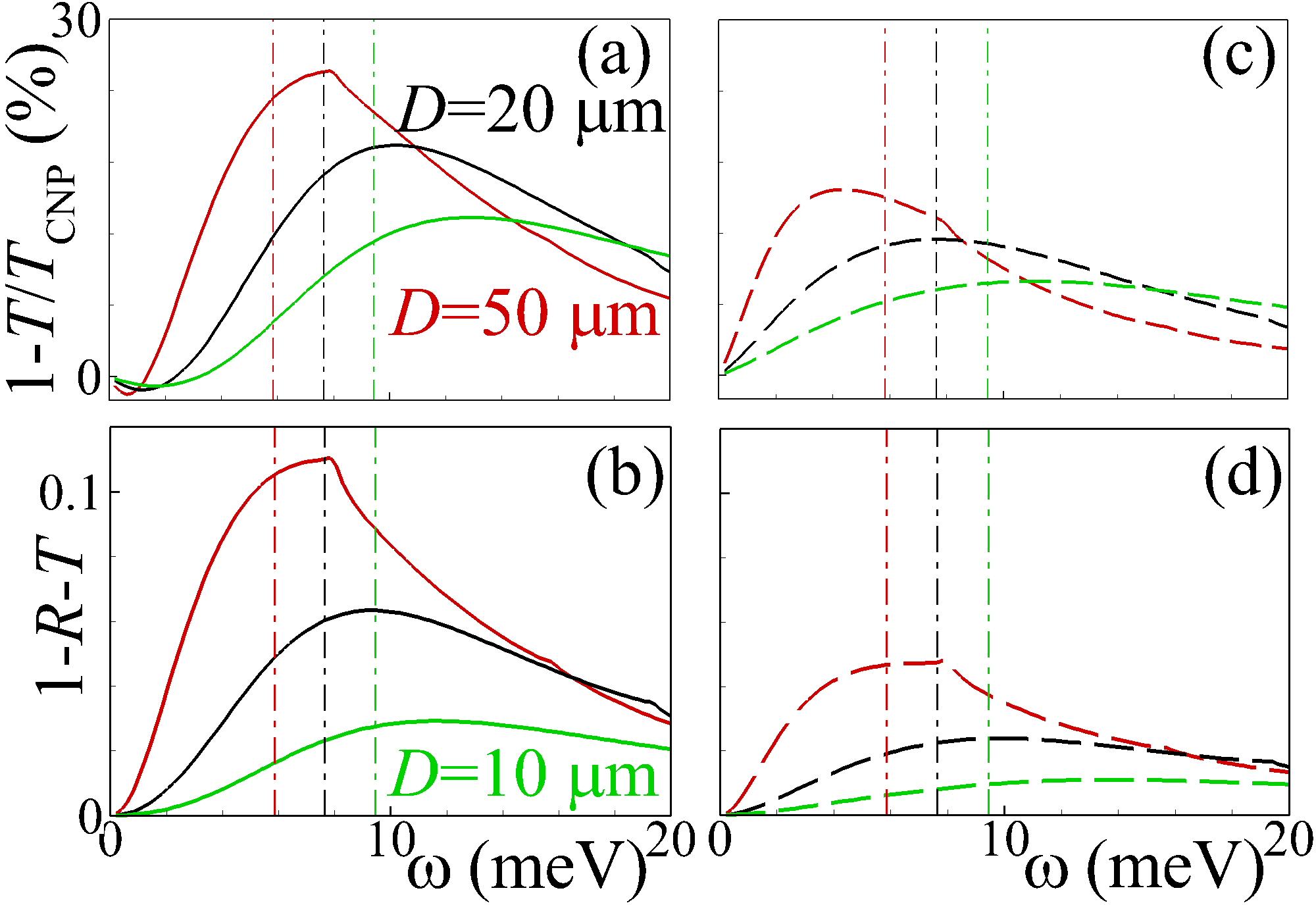}

\caption{Frequency dependence of the extinction {[}upper row, panels (a) and
(c){]}, and absorbance {[}lower row, panels (b) and (d){]} of the
clockwise {[}left column, panels (a) and (b){]} and anticlockwise
{[}right column, panels (a) and (b){]} circularly polarized electromagnetic
wave impingin on graphene cladded by a semi-infinite
$\mathrm{Al}_{2}\mathrm{O}_{3}$ substrate and a perforated metal film (2D grating), subjected to a erpendicular magnetic field. 
The parameters of the structure and the meaning of the different curves are
the same as in Fig.\ref{fig:var_Al2O3_B_eq1.5}.}
\label{fig:var_Al2O3_LR_B_eq1.5}
\end{figure}

When the graphene is doped with electrons (positive chemical potential,
$\mu>0$, Fig.\,\ref{fig:var_Al2O3_LR_B_eq1.5}), the extinction
of the clockwise polarized wave {[}Fig.\,\ref{fig:var_Al2O3_LR_B_eq1.5}(a){]}
at high-frequency range is larger than that of the anticlockwise-polarized
wave {[}Fig.\,\ref{fig:var_Al2O3_LR_B_eq1.5}(c){]}. The situation
is totally opposite to the case of graphene doped with holes
{[}compare Figs.\,\ref{fig:var_Al2O3_LR_B_eq1.5}(a) and \ref{fig:Ext_LR}(c)
as well as Figs.\,\ref{fig:var_Al2O3_LR_B_eq1.5}(c) and \ref{fig:Ext_LR}(d){]}.
Moreover, if compared to the case of the linearly polarized incident
wave with the same parameters {[}shown in Fig.\,\ref{fig:var_Al2O3_B_eq1.5}{]},
the one with clockwise circular polarization both exhibit a stronger
extinction {[}compare Figs.\,\ref{fig:var_Al2O3_LR_B_eq1.5}(a) and
\ref{fig:var_Al2O3_B_eq1.5}(a){]} and stronger absorbance {[}compare
Figs.\,\ref{fig:var_Al2O3_LR_B_eq1.5}(b) and \ref{fig:var_Al2O3_B_eq1.5}(b){]}.
At the same time, the case of anticlockwise circular polarization
is quite different: here the extinction and the absorbance are considerably
lower than those of linearly polarized incident wave {[}compare Figs.\,\ref{fig:var_Al2O3_LR_B_eq1.5}(c)
and \ref{fig:var_Al2O3_B_eq1.5}(a) as well as Figs.\,\ref{fig:var_Al2O3_LR_B_eq1.5}(d)
and \ref{fig:var_Al2O3_B_eq1.5}(b){]}. The dependence
of the magnetic circular dichoism upon the parameters of the perforated
metal film is described in Appendix \ref{sec:Coef-MCD}.

\section{Conclusions}

\label{sec:conclusions}To conclude, we calculated the spectral dependence of the 
Faraday rotation and MCD of an electromagnetic wave transmitted through
a graphene layer subjected to an external perpendicular magnetic field. The calculations show that
these effects are strongly influenced  by adding a periodically perforated metallic film (a 2D grating) on top of graphene. 
We have demonstrated that, if the width of the perforation holes is close to the array period (i.e. the grating looks like a thin metallic net),  the incident
electromagnetic wave is strongly screened by this structure in the low-frequency range. 
It results in a decrease of the Faraday rotation angle of the transmitted wave, in comparison with bare graphene. 
In contrast, if the wave frequency is close to that of the magnetoplasmon resonance supported by the structure with 2D grating, one can expect an increase of the
Faraday rotation angle, which is a result of the magnetoplasmon-mediated transmission. 
The maximum of the Faraday rotation angle is shifted to higher frequency when the period of the grating decreases. An important advantage of the graphene-based structure with 2D grating is that it introduces lower ellipticity to the transmitted linear-polarized wave and for some frequencies it can vanish, which means maintaining the linear polarization (see Fig. \ref{fig:var_Al2O3_B_eq1.5}). It can be potentially interesting for making a switchable rotating polariser in the THz range by combining a stack of graphene layers with a thin metal net on top of each of them in order to get high rotation angles. 

The effect introduced by the grating is even more substantial for the MCD, which is quantified by the difference in the extinction between left-hand and right-hand circular polarized waves\footnote{Alternatively, it can be quantified in terms of the coefficient of circular dichroism introduced in Appendix \ref{sec:Coef-MCD}.}  (Fig. \ref{fig:var_Al2O3_LR_B_eq1.5}).
We notice that the dichroism changes its sign depending on the wave frequency and the crossover point in the spectrum can be tuned by adjusting magnetic field and also the graphene Fermi energy that determine the magnetoplasmon dispersion curve.
This opens the way to design an electrically switchable optical isolator based on the MCD effect by utilizing non-reciprocal losses.\cite{spp-gr-mo-Poumirol2017-NatComm}

One possible extension of this work can be the study of other (non-square)
types of hole arrays (e.g. triangular, rectangular, etc.), which are
nonsymmetric with respect to the polarization plane of the incident wave.
In this case the transmittance and the Faraday rotation angle will depend upon the direction of polarization of the incident wave with respect to the translation vectors of the 
2D grating. Another
possible further development of the present work may consist in taking into account
the dependence of the substrate dielectric constant upon the magnetic
field, which can result in the additional anisotropy and, as a consequence, in
a dependence of the transmittance upon the direction of the external
magnetic field. It can also be interesting to consider
a thin spacer between the graphene and the metal grating that could work as a resonator and can eventually enhance the magnetoplasmon's amplitude and the effect induced by them in the structure. Also this type of structure allows one to use graphene magnetoplasmons for the controlling of spoof plasmons in the metal grating \cite{spp-gr-spoof-Dias2017-acsphotonics} as well as the enhanced optical transmission \cite{spp-eot-Ebessen1998-nat}.

\acknowledgments

The authors thank Alexey Kuzmenko from the University of Geneva, Switzerland and Luis Mart\'{i}n-Moreno from the University of Zaragoza, Spain for the careful reading, valuable discussions and comments. Funding from the European Commission within the project "Graphene-Driven Revolutions in ICT and Beyond" (ref. no. 696656) and the
Portuguese Foundation for Science and Technology (FCT) in
the framework of the Strategic Funding UID/FIS/04650/
2013 is gratefully acknowledged.

\appendix

\section{Explicit form of the electromagnetic fields}

\label{sec:Append-em-field}Assuming electromagnetic field time-dependence
as $\mathbf{E},\mathbf{H}\sim\exp\left(-i\omega t\right)$, we represent
the Maxwell equations as
\begin{eqnarray}
\mathrm{rot}\,\mathbf{E}^{(m)}=\frac{i\omega}{c}\mathbf{H}^{(m)},\label{eq:max-rot-E}\\
\mathrm{rot}\,\mathbf{H}^{(m)}=-\frac{i\omega}{c}\varepsilon_{m}\mathbf{E}^{(m)},\label{eq:max-rot-H}\\
\mathrm{div}\,\mathbf{E}^{(m)}=0,\label{eq:max-div-E}\\
\mathrm{div}\,\mathbf{H}^{(m)}=0,\label{eq:max-div-H}
\end{eqnarray}
where $\omega$ is the frequency, $c$ is the velocity of
light in vacuum. The superscripts $m=1,2,3$ correspond to the spatial
domains $z<-d/2$, $-d/2<z<d/2$, and $z>d/2$, respectively. Consequently, the
dielectric constants are: $\varepsilon_{1}=\varepsilon_{2}=1$, $\varepsilon_{3}=\varepsilon$.
After substituting Eq. (\ref{eq:max-rot-E}) into (\ref{eq:max-rot-H}),
and using Eq. (\ref{eq:max-div-E}), one obtains: 
\begin{eqnarray}
\left[\frac{\partial^{2}}{\partial x^{2}}+\frac{\partial^{2}}{\partial y^{2}}+\frac{\partial^{2}}{\partial z^{2}}+\frac{\omega^{2}}{c^{2}}\varepsilon_{m}\right]\mathbf{E}^{(m)}=0.\label{eq:wave-eq-E}
\end{eqnarray}
In a similar manner, substitution of Eq. (\ref{eq:max-rot-H}) into
(\ref{eq:max-rot-E}) gives {[}taking into account Eq. (\ref{eq:max-div-H}){]}:
\begin{eqnarray}
\left[\frac{\partial^{2}}{\partial x^{2}}+\frac{\partial^{2}}{\partial y^{2}}+\frac{\partial^{2}}{\partial z^{2}}+\frac{\omega^{2}}{c^{2}}\varepsilon_{m}\right]\mathbf{H}^{(m)}=0.\label{eq:wave-eq-H}
\end{eqnarray}
Incide the holes in the metal film (spatial domain $-d/2<z<d/2$,
$m=2$), we can consider $z$ components of the electromagnetic field,
$E_{z}^{(2)}$ and $H_{z}^{(2)}$, as independent variables, thus
\begin{eqnarray}
\frac{\partial^{2}E_{z}^{(2)}}{\partial x^{2}}+\frac{\partial^{2}E_{z}^{(2)}}{\partial y^{2}}+\frac{\partial^{2}E_{z}^{(2)}}{\partial z^{2}}+\frac{\omega^{2}}{c^{2}}E_{z}^{(2)}=0,\label{eq:wave-eq-Ez2}\\
\frac{\partial^{2}H_{z}^{(2)}}{\partial x^{2}}+\frac{\partial^{2}H_{z}^{(2)}}{\partial y^{2}}+\frac{\partial^{2}H_{z}^{(2)}}{\partial z^{2}}+\frac{\omega^{2}}{c^{2}}H_{z}^{(2)}=0.\label{eq:wave-eq-Hz2}
\end{eqnarray}
As a result, we can divide eigenfunctions into two types: (i)
E-waves with $E_{z}\ne0$, $H_{z}\equiv0$ and (ii) H-waves with $H_{z}\ne0$,
$E_{z}\equiv0$. For E-waves after matching boundary conditions at
the hole edges inside the film $E_{z}^{(2,E)}(lD,y,z)=E_{z}^{(2,E)}(lD+W,y,z)=0$,
$E_{z}^{(2,E)}(x,l^{\prime}D,z)=E_{z}^{(2,E)}(x,l^{\prime}D+W,z)=0$,
we obtain the solution of Eq.(\ref{eq:wave-eq-Ez2}) for the electric
field's $z$ component inside the film in the form
\begin{eqnarray}
E_{z}^{(2,E)}(\mathbf{r},z) & = & \sum_{m,n=1}^{\infty}\sin\left(\frac{m\pi}{W}\left(x-lD\right)\right)\nonumber \\
 &  & \times\sin\left(\frac{n\pi}{W}\left(y-l^{\prime}D\right)\right)\label{eq:Ez-E-res}\\
 &  & \times\left\{ A_{m,n}^{(s)}\sin\left(\mu_{m,n}\left(z+\frac{d}{2}\right)\right)\right.\nonumber \\
 &  & \left.+A_{m,n}^{(c)}\cos\left(\mu_{m,n}\left(z-\frac{d}{2}\right)\right)\right\} ,\nonumber 
\end{eqnarray}
where $m,n\ge1$ are the mode indices; $A_{m,n}^{(s)}$ and $A_{m,n}^{(c)}$
are the amplitudes of the respective mode (sin- and cos-like wave,
correspondingly), $\mu_{m,n}=\sqrt{\left(\frac{\omega}{c}\right)^{2}-\left(\frac{m\pi}{W}\right)^{2}-\left(\frac{n\pi}{W}\right)^{2}}$.
Other components of the E-wave electromagnetic field can be represented
as 
\begin{eqnarray}
H_{x}^{(2,E)}(\mathbf{r},z) & = & -\frac{i\omega}{c}\sum_{m,n=1}^{\infty}\frac{1}{\left(\frac{m\pi}{W}\right)^{2}+\left(\frac{n\pi}{W}\right)^{2}}\frac{n\pi}{W}\nonumber \\
 &  & \times\sin\left(\frac{m\pi}{W}\left(x-lD\right)\right)\nonumber \\
 &  & \times\cos\left(\frac{n\pi}{W}\left(y-l^{\prime}D\right)\right)\label{eq:Hx-E-res}\\
 &  & \times\left\{ A_{m,n}^{(s)}\sin\left(\mu_{m,n}\left(z+\frac{d}{2}\right)\right)\right.\nonumber \\
 &  & \left.+A_{m,n}^{(c)}\cos\left(\mu_{m,n}\left(z-\frac{d}{2}\right)\right)\right\} ,\nonumber \\
H_{y}^{(2,E)}(\mathbf{r},z) & = & \frac{i\omega}{c}\sum_{m,n=1}^{\infty}\frac{1}{\left(\frac{m\pi}{W}\right)^{2}+\left(\frac{n\pi}{W}\right)^{2}}\frac{m\pi}{W}\nonumber \\
 &  & \times\cos\left(\frac{m\pi}{W}\left(x-lD\right)\right)\nonumber \\
 &  & \times\sin\left(\frac{n\pi}{W}\left(y-l^{\prime}D\right)\right)\label{eq:Hy-E-res}\\
 &  & \times\left\{ A_{m,n}^{(s)}\sin\left(\mu_{m,n}\left(z+\frac{d}{2}\right)\right)\right.\nonumber \\
 &  & \left.+A_{m,n}^{(c)}\cos\left(\mu_{m,n}\left(z-\frac{d}{2}\right)\right)\right\} ,\nonumber 
\end{eqnarray}
\begin{eqnarray}
E_{x}^{(2,E)}(\mathbf{r},z) & = & \sum_{m,n=1}^{\infty}\frac{\mu_{m,n}}{\left(\frac{m\pi}{W}\right)^{2}+\left(\frac{n\pi}{W}\right)^{2}}\frac{m\pi}{W}\nonumber \\
 &  & \times\cos\left(\frac{m\pi}{W}\left(x-lD\right)\right)\nonumber \\
 &  & \times\sin\left(\frac{n\pi}{W}\left(y-l^{\prime}D\right)\right)\label{eq:Ex-E-res}\\
 &  & \times\left\{ A_{m,n}^{(s)}\cos\left(\mu_{m,n}\left(z+\frac{d}{2}\right)\right)\right.\nonumber \\
 &  & \left.-A_{m,n}^{(c)}\sin\left(\mu_{m,n}\left(z-\frac{d}{2}\right)\right)\right\} ,\nonumber \\
E_{y}^{(2,E)}(\mathbf{r},z) & = & \sum_{m,n=1}^{\infty}\frac{\mu_{m,n}}{\left(\frac{m\pi}{W}\right)^{2}+\left(\frac{n\pi}{W}\right)^{2}}\frac{n\pi}{W}\nonumber \\
 &  & \times\sin\left(\frac{m\pi}{W}\left(x-lD\right)\right)\nonumber \\
 &  & \times\cos\left(\frac{n\pi}{W}\left(y-l^{\prime}D\right)\right)\label{eq:Ey-E-res}\\
 &  & \times\left\{ A_{m,n}^{(s)}\cos\left(\mu_{m,n}\left(z+\frac{d}{2}\right)\right)\right.\nonumber \\
 &  & \left.-A_{m,n}^{(c)}\sin\left(\mu_{m,n}\left(z-\frac{d}{2}\right)\right)\right\} .\nonumber 
\end{eqnarray}
Notice that Eqs. (\ref{eq:Ex-E-res})\textendash (\ref{eq:Hx-E-res})
also satisfy other natural boundary conditions: zero tangential component
of the electric field and zero normal component of the magnetic field
at hole surfaces, $E_{x}^{(2,E)}(x,l^{\prime}D,z)=E_{x}^{(2,E)}(x,l^{\prime}D+W,z)=0$,
$E_{y}^{(2,E)}(lD,y,z)=E_{y}^{(2,E)}(lD+W,y,z)=0$, $H_{x}^{(2,E)}(lD,y,z)=H_{x}^{(2,E)}(lD+W,y,z)=0$,
$H_{y}^{(2,E)}(x,l^{\prime}D,z)=H_{y}^{(2,E)}(x,l^{\prime}D+W,z)=0$.

For H-wave the solution of Eq. (\ref{eq:wave-eq-Hz2}) for the $z$
component of the magnetic field can be represented as
\begin{eqnarray}
H_{z}^{(2,H)}(\mathbf{r},z) & = & \sum_{m,n=0}^{\infty}\cos\left(\frac{m\pi}{W}\left(x-lD\right)\right)\nonumber \\
 &  & \times\cos\left(\frac{n\pi}{W}\left(y-l^{\prime}D\right)\right)\nonumber \\
 &  & \times\left\{ B_{m,n}^{(s)}\cos\left(\mu_{m,n}\left(z+\frac{d}{2}\right)\right)\right.\label{eq:Hz-H-res}\\
 &  & \left.-B_{m,n}^{(c)}\sin\left(\mu_{m,n}\left(z-\frac{d}{2}\right)\right)\right\} ,\nonumber 
\end{eqnarray}
while other components of electromagnetic field are
\begin{eqnarray}
H_{x}^{(2,H)}(\mathbf{r},z) & = & \sum_{m=1}^{\infty}\sum_{n=0}^{\infty}\frac{\mu_{m,n}}{\left(\frac{m\pi}{W}\right)^{2}+\left(\frac{n\pi}{W}\right)^{2}}\frac{m\pi}{W}\nonumber \\
 &  & \times\sin\left(\frac{m\pi}{W}\left(x-lD\right)\right)\nonumber \\
 &  & \times\cos\left(\frac{n\pi}{W}\left(y-l^{\prime}D\right)\right)\label{eq:Hx-H-res}\\
 &  & \times\left\{ B_{m,n}^{(s)}\sin\left(\mu_{m,n}\left(z+\frac{d}{2}\right)\right)\right.\nonumber \\
 &  & \left.+B_{m,n}^{(c)}\cos\left(\mu_{m,n}\left(z-\frac{d}{2}\right)\right)\right\} ,\nonumber \\
H_{y}^{(2,H)}(\mathbf{r},z) & = & \sum_{m=0}^{\infty}\sum_{n=1}^{\infty}\frac{\mu_{m,n}}{\left(\frac{m\pi}{W}\right)^{2}+\left(\frac{n\pi}{W}\right)^{2}}\frac{n\pi}{W}\nonumber \\
 &  & \times\cos\left(\frac{m\pi}{W}\left(x-lD\right)\right)\nonumber \\
 &  & \times\sin\left(\frac{n\pi}{W}\left(y-l^{\prime}D\right)\right)\label{eq:Hy-H-res}\\
 &  & \times\left\{ B_{m,n}^{(s)}\sin\left(\mu_{m,n}\left(z+\frac{d}{2}\right)\right)\right.\nonumber \\
 &  & \left.+B_{m,n}^{(c)}\cos\left(\mu_{m,n}\left(z-\frac{d}{2}\right)\right)\right\} ,\nonumber 
\end{eqnarray}
\begin{eqnarray}
E_{x}^{(2,H)}(\mathbf{r},z) & = & -\frac{i\omega}{c}\sum_{m=0}^{\infty}\sum_{n=1}^{\infty}\frac{1}{\left(\frac{m\pi}{W}\right)^{2}+\left(\frac{n\pi}{W}\right)^{2}}\frac{n\pi}{W}\nonumber \\
 &  & \times\cos\left(\frac{m\pi}{W}\left(x-lD\right)\right)\nonumber \\
 &  & \times\sin\left(\frac{n\pi}{W}\left(y-l^{\prime}D\right)\right)\label{eq:Ex-H-res}\\
 &  & \times\left\{ B_{m,n}^{(s)}\cos\left(\mu_{m,n}\left(z+\frac{d}{2}\right)\right)\right.\nonumber \\
 &  & \left.-B_{m,n}^{(c)}\sin\left(\mu_{m,n}\left(z-\frac{d}{2}\right)\right)\right\} ,\nonumber \\
E_{y}^{(2,H)}(\mathbf{r},z) & = & \frac{i\omega}{c}\sum_{m=1}^{\infty}\sum_{n=0}^{\infty}\frac{1}{\left(\frac{m\pi}{W}\right)^{2}+\left(\frac{n\pi}{W}\right)^{2}}\frac{m\pi}{W}\nonumber \\
 &  & \times\sin\left(\frac{m\pi}{W}\left(x-lD\right)\right)\nonumber \\
 &  & \times\cos\left(\frac{n\pi}{W}\left(y-l^{\prime}D\right)\right)\label{eq:Ey-H-res}\\
 &  & \times\left\{ B_{m,n}^{(s)}\cos\left(\mu_{m,n}\left(z+\frac{d}{2}\right)\right)\right.\nonumber \\
 &  & \left.-B_{m,n}^{(c)}\sin\left(\mu_{m,n}\left(z-\frac{d}{2}\right)\right)\right\} .\nonumber 
\end{eqnarray}
Here $B_{m,n}^{(s)}$ and $B_{m,n}^{(c)}$ are the amplitudes of sin-
and cos-like waves. It should be emphasized that in the case of H-wave
along with modes with nonzero indexes $m,n\ge1$, the existence of
modes with one zero and another nonzero index (like $m\ne0$, $n=0$
or $m=0$, $n\ne0$) is possible. Notice that Eqs. (\ref{eq:Hx-H-res})\textendash (\ref{eq:Ey-H-res})
also satisfy the above-mentioned boundary conditions at hole walls
$E_{x}^{(2,H)}(x,l^{\prime}D,z)=E_{x}^{(2,H)}(x,l^{\prime}D+W,z)=0$,
$E_{y}^{(2,H)}(lD,y,z)=E_{y}^{(2,H)}(lD+W,y,z)=0$, $H_{x}^{(2,H)}(lD,y,z)=H_{x}^{(2,H)}(lD+W,y,z)=0$,
$H_{y}^{(2,H)}(x,l^{\prime}D,z)=H_{y}^{(2,H)}(x,l^{\prime}D+W,z)=0$.
At the same time, there is no mode with $m=n=0$ (in the summation
in Eq. (\ref{eq:Hz-H-res}) this term is implicitly excluded), because
in this case it is impossible to satisfy above-mentioned boundary
conditions on hole walls. Using Eqs. (\ref{eq:Hx-E-res}), (\ref{eq:Hy-E-res}),
(\ref{eq:Hx-H-res}), (\ref{eq:Hy-H-res}), it is possible to obtain
Eq. (\ref{eq:H-res-mat-mod}), which represent the $x$ and $y$ components
of the total magnetic field $H_{x}^{(2)}(\mathbf{r},z)=H_{x}^{(2,E)}(\mathbf{r},z)+H_{x}^{(2,H)}(\mathbf{r},z)$,
$H_{y}^{(2)}(\mathbf{r},z)=H_{y}^{(2,E)}(\mathbf{r},z)+H_{y}^{(2,H)}(\mathbf{r},z)$
in the matrix form. Similarly, Eq. (\ref{eq:E-res-mat-mod}) for the
$x$ and $y$ components of the total electric field $E_{x}^{(2)}(\mathbf{r},z)=E_{x}^{(2,H)}(\mathbf{r},z)+E_{x}^{(2,E)}(\mathbf{r},z)$,
$E_{y}^{(2)}(\mathbf{r},z)=E_{y}^{(2,H)}(\mathbf{r},z)+E_{y}^{(2,E)}(\mathbf{r},z)$,
written in the matrix form, can be obtained from Eqs.(\ref{eq:Ex-E-res}),
(\ref{eq:Ey-E-res}), (\ref{eq:Ex-H-res}), (\ref{eq:Ey-H-res}).

Both inside the semi-infinite air ($z<-d/2$, $m=1$) and
the semi-infinite substrate ($z>d/2$, $m=3$) we choose transverse
($x$ and $y$) components of the magnetic field $H_{x}^{(1,3)}$,
$H_{y}^{(1,3)}$ as independent variables, thus 

\begin{eqnarray}
\left[\frac{\partial^{2}}{\partial x^{2}}+\frac{\partial^{2}}{\partial y^{2}}+\frac{\partial^{2}}{\partial z^{2}}+\frac{\omega^{2}}{c^{2}}\varepsilon_{m}\right]H_{x}^{(m)}=0,\label{eq:wave-eq-Hx13}\\
\left[\frac{\partial^{2}}{\partial x^{2}}+\frac{\partial^{2}}{\partial y^{2}}+\frac{\partial^{2}}{\partial z^{2}}+\frac{\omega^{2}}{c^{2}}\varepsilon_{m}\right]H_{y}^{(m)}=0.\label{eq:wave-eq-Hy13}
\end{eqnarray}
Also we represent solutions of wave equations (\ref{eq:wave-eq-Hx13})
and (\ref{eq:wave-eq-Hy13}) as two-dimensional Fourier series with
respect to the harmonics with wavevector in $\left(xy\right)$ plane
$\mathbf{k}_{s,j}$. As a matter of fact, electromagnetic field in
vacuum can be represented as 
\begin{eqnarray}
H_{x}^{(1)}(\mathbf{r},z) & = & H_{x}^{(i)}\exp\left[ip_{0,0}^{\left(1\right)}(z+d/2)\right]\nonumber \\
 &  & +\sum_{s,j=-\infty}^{\infty}H_{x||s,j}^{(r)}\label{eq:Hx1-vacuum}\\
 &  & \times\exp\left[i\mathbf{k}_{s,j}\mathbf{r}-ip_{s,j}^{(1)}(z+d/2)\right],\nonumber \\
H_{y}^{(1)}(\mathbf{r},z) & = & H_{y}^{(i)}\exp\left[ip_{0,0}^{\left(1\right)}(z+T/2)\right]\nonumber \\
 &  & +\sum_{s,j=-\infty}^{\infty}H_{y||s,j}^{(r)}\label{eq:Hy1-vacuum}\\
 &  & \times\exp\left[i\mathbf{k}_{s,j}\mathbf{r}-ip_{s,j}^{(1)}(z+d/2)\right],\nonumber \\
H_{z}^{(1)}(\mathbf{r},z) & = & \sum_{s,j=-\infty}^{\infty}\frac{sgH_{x||s,j}^{(r)}+jgH_{y||s,j}^{(r)}}{p_{s,j}^{(1)}}\label{eq:Hz1-vacuum}\\
 &  & \times\exp\left[i\mathbf{k}_{s,j}\mathbf{r}-ip_{s,j}^{(1)}(z+d/2)\right],\nonumber \\
E_{x}^{(1)}(\mathbf{r},z) & = & H_{y}^{(i)}\exp\left[ip_{0,0}^{\left(1\right)}(z+d/2)\right]\nonumber \\
 &  & -\frac{c}{\omega}\sum_{s,j=-\infty}^{\infty}\left\{ p_{s,j}^{(1)}H_{y||s,j}^{(r)}\right.\label{eq:Ex1-vacuum}\\
 &  & \left.+jg\frac{sgH_{x||s,j}^{(r)}+jgH_{y||s,j}^{(r)}}{p_{s,j}^{(1)}}\right\} \nonumber \\
 &  & \times\exp\left[i\mathbf{k}_{s,j}\mathbf{r}-ip_{s,j}^{(1)}(z+d/2)\right],\nonumber \\
E_{y}^{(1)}(\mathbf{r},z) & = & -H_{x}^{(i)}\exp\left[ip_{0,0}^{\left(1\right)}(z+d/2)\right]\nonumber \\
 &  & +\frac{c}{\omega}\sum_{s,j=-\infty}^{\infty}\left\{ p_{s,j}^{(1)}H_{x||s,j}^{(r)}\right.\label{eq:Ey1-vacuum}\\
 &  & \left.+sg\frac{sgH_{x||s,j}^{(r)}+jgH_{y||s,j}^{(r)}}{p_{s,j}^{(3)}}\right\} \nonumber \\
 &  & \times\exp\left[i\mathbf{k}_{s,j}\mathbf{r}+ip_{s,j}^{(3)}(z-d/2)\right],\nonumber \\
E_{z}^{(1)}(\mathbf{r},z) & = & -\frac{c}{\omega}\sum_{s,j=-\infty}^{\infty}\left\{ sgH_{y||s,j}^{(r)}-jgH_{x||s,j}^{(r)}\right\} \nonumber \\
 &  & \times\exp\left[i\mathbf{k}_{s,j}\mathbf{r}-ip_{s,j}^{(1)}(z+d/2)\right].
\end{eqnarray}
At the same time,  the electromagnetic field components in the substrate
are:
\begin{eqnarray}
H_{x}^{(3)}(\mathbf{r},z) & = & \sum_{s,j=-\infty}^{\infty}H_{x||s,j}^{(t)}\label{eq:Hx3-subsrtate}\\
 &  & \times\exp\left[i\mathbf{k}_{s,j}\mathbf{r}+ip_{s,j}^{(3)}(z-d/2)\right],\nonumber \\
H_{y}^{(3)}(\mathbf{r},z) & = & \sum_{s,j=-\infty}^{\infty}H_{y||s,j}^{(t)}\label{eq:Hy3-subsrtate}\\
 &  & \times\exp\left[i\mathbf{k}_{s,j}\mathbf{r}+ip_{s,j}^{(3)}(z-d/2)\right],\nonumber \\
H_{z}^{(3)}(\mathbf{r},z) & = & -\sum_{s,j=-\infty}^{\infty}\frac{sgH_{x||s,j}^{(t)}+jgH_{y||s,j}^{(t)}}{p_{s,j}^{(3)}}\label{eq:Hz3-subsrtate}\\
 &  & \times\exp\left[i\mathbf{k}_{s,j}\mathbf{r}+ip_{s,j}^{(3)}(z-d/2)\right],\nonumber \\
E_{x}^{(3)}(\mathbf{r},z) & = & \frac{c}{\omega\varepsilon}\sum_{s,j=-\infty}^{\infty}\left\{ p_{s,j}^{(3)}H_{y||s,j}^{(t)}\right.\nonumber \\
 &  & \left.+jg\frac{sgH_{x||s,j}^{(t)}+jgH_{y||s,j}^{(t)}}{p_{s,j}^{(3)}}\right\} \label{eq:Ex3-subsrtate}\\
 &  & \times\exp\left[i\mathbf{k}_{s,j}\mathbf{r}+ip_{s,j}^{(3)}(z-d/2)\right],\nonumber \\
E_{y}^{(3)}(\mathbf{r},z) & = & -\frac{c}{\omega\varepsilon}\sum_{s,j=-\infty}^{\infty}\left\{ p_{s,j}^{(3)}H_{x||s,j}^{(t)}\right.\nonumber \\
 &  & \left.+sg\frac{sgH_{x||s,j}^{(t)}+jgH_{y||s,j}^{(t)}}{p_{s,j}^{(3)}}\right\} \label{eq:Ey3-subsrtate}\\
 &  & \times\exp\left[i\mathbf{k}_{s,j}\mathbf{r}+ip_{s,j}^{(3)}(z-d/2)\right].\nonumber \\
E_{z}^{(3)}(\mathbf{r},z) & = & -\frac{c}{\omega\varepsilon}\sum_{s,j=-\infty}^{\infty}\left\{ sgH_{y||s,j}^{(t)}-jgH_{x||s,j}^{(t)}\right\} \nonumber \\
 &  & \times\exp\left[i\mathbf{k}_{s,j}\mathbf{r}+ip_{s,j}^{(3)}(z+d/2)\right].
\end{eqnarray}
Equations (\ref{eq:Hx1-vacuum}), (\ref{eq:Hy1-vacuum}), (\ref{eq:Ex1-vacuum}),
and (\ref{eq:Ey1-vacuum}) can be written in the matrix form {[}see
Eq. (\ref{eq:mat-vacuum}){]}, while Eqs. (\ref{eq:Hx3-subsrtate}),
(\ref{eq:Hy3-subsrtate}), (\ref{eq:Ex3-subsrtate}), and (\ref{eq:Ey3-subsrtate})
can be transformed into Eq. (\ref{eq:mat-substrate}).

In order to calculate the reflectance and transmittance we notice
that the flux density of energy is decribed by the Poynting vector $\mathbf{S}=(c/8\pi)\mathrm{Re}\left(\mathbf{E}\times\mathbf{H}\right)$.
Respectively, using Eqs. (\ref{eq:Ex3-subsrtate}) and (\ref{eq:Ey3-subsrtate}),
the $z$ component of the Poynting vector of the $(s,j)$-harmonics
in the substrate can be represented as 
\begin{eqnarray}
S_{z||s,j}^{(t)} & = & \frac{c^{2}}{8\pi\omega}\mathrm{Re}\left\{ \frac{\overline{H_{y||s,j}^{(t)}}}{\varepsilon}\left[p_{s,j}^{(3)}H_{y||s,j}^{(t)}\right.\right.\label{eq:St-z}\\
 &  & \left.+jg\frac{sgH_{x||s,j}^{(t)}+jgH_{y||s,j}^{(t)}}{p_{s,j}^{(3)}}\right]+\frac{\overline{H_{x||s,j}^{(t)}}}{\varepsilon}\nonumber \\
 &  & \left.\times\left[p_{s,j}^{(3)}H_{x||s,j}^{(t)}+sg\frac{sgH_{x||s,j}^{(t)}+jgH_{y||s,j}^{(t)}}{p_{s,j}^{(3)}}\right]\right\} .\nonumber 
\end{eqnarray}
In the matrix form this relation can be represented as
\begin{eqnarray}
S_{z||s,j}^{(t)}=\frac{c}{8\pi}\mathrm{Re}\left\{ \left(\overline{H_{y||s,j}^{(t)}},\,-\overline{H_{x||s,j}^{(t)}}\right)\hat{{\cal Q}}_{s,j}^{(3)}\left(\begin{array}{c}
H_{x||s,j}^{(t)}\\
H_{y||s,j}^{(t)}
\end{array}\right)\right\} \nonumber \\
\label{eq:St-z-mat}\\
=-\frac{c}{8\pi}\mathrm{Re}\left\{ \left(\begin{array}{c}
H_{x||s,j}^{(t)}\\
H_{y||s,j}^{(t)}
\end{array}\right)^{\dagger}i\hat{\sigma}_{y}\hat{{\cal Q}}_{s,j}^{(3)}\left(\begin{array}{c}
H_{x||s,j}^{(t)}\\
H_{y||s,j}^{(t)}
\end{array}\right)\right\} .\nonumber 
\end{eqnarray}
In the air, the $z$ component of the Poynting vector of the reflected
wave's $(s,j)$-harmonic {[}using Eqs. (\ref{eq:Ex1-vacuum}) and (\ref{eq:Ey1-vacuum}){]}
is written as
\begin{eqnarray}
S_{z||s,j}^{(r)} & = & -\frac{c^{2}}{8\pi\omega}\mathrm{Re}\left\{ \overline{H_{y||s,j}^{(r)}}\left[p_{s,j}^{(1)}H_{y||s,j}^{(r)}\right.\right.\label{eq:Sr-z-mat}\\
 &  & \left.+jg\frac{sgH_{x||s,j}^{(r)}+jgH_{y||s,j}^{(r)}}{p_{s,j}^{(1)}}\right]+\overline{H_{x||s,j}^{(r)}}\nonumber \\
 &  & \left.\times\left[p_{s,j}^{(1)}H_{x||s,j}^{(r)}+sg\frac{sgH_{x||s,j}^{(r)}+jgH_{y||s,j}^{(r)}}{p_{s,j}^{(1)}}\right]\right\} \nonumber \\
 & = & \frac{c}{8\pi}\mathrm{Re}\left\{ \left(\begin{array}{c}
H_{x||s,j}^{(r)}\\
H_{y||s,j}^{(r)}
\end{array}\right)^{\dagger}i\hat{\sigma}_{y}\hat{{\cal Q}}_{s,j}^{(1)}\left(\begin{array}{c}
H_{x||s,j}^{(r)}\\
H_{y||s,j}^{(r)}
\end{array}\right)\right\} .\nonumber 
\end{eqnarray}
The difference of sign in Eqs. (\ref{eq:St-z}) and (\ref{eq:Sr-z-mat})
reflects the fact that the energy flux of the transmitted wave {[}see Eq.
(\ref{eq:St-z}){]} flows in the positive direction of $z$ axis,
while that of reflected wave {[}see Eq. (\ref{eq:Sr-z-mat}){]} flows
in the negative direction of $z$ axis. Using the same formalism for
the incident wave, we obtain the $z$ component of its Poynting vector in
the form:
\begin{eqnarray}
S_{z}^{(i)} & = & \frac{c}{8\pi}\left[\left|H_{x}^{(i)}\right|^{2}+\left|H_{y}^{(i)}\right|^{2}\right].\label{eq:Si-z}
\end{eqnarray}

The reflectance, $R$, and the trasmittance, $T$, are defined in the usual way: 
\begin{eqnarray}
R=-\frac{\sum_{s,j=-\infty}^{\infty}S_{z||s,j}^{(r)}}{S_{z}^{(i)}};\label{eq:R-gen}
\end{eqnarray}
\begin{eqnarray}
T=\frac{\sum_{s,j=-\infty}^{\infty}S_{z||s,j}^{(t)}}{S_{z}^{(i)}}.\label{eq:T-gen}
\end{eqnarray}
Substuting Eqs. (\ref{eq:St-z-mat})\textendash (\ref{eq:Si-z}) into
Eqs. (\ref{eq:R-gen}) and (\ref{eq:T-gen}), one obtains the expressions
for the reflectance and transmittace in the final form of Eqs. (\ref{eq:R-final})
and (\ref{eq:T-final}). 

\section{Transmittance of the graphene layer without periodical structure
and the dispersion relation of magnetoplasmons}

When the metal film is absent ($d=0$), then right-hand sides of Eqs.
(\ref{eq:cond-H1}) and (\ref{eq:cond-H3}) become equal {[}as well as
those of Eqs. (\ref{eq:cond-E1}) and (\ref{eq:cond-E3}){]}. As a
consequence, the equality of left-hand sides of Eqs. (\ref{eq:cond-H1})
and (\ref{eq:cond-H3}) gives 
\begin{eqnarray}
\left(\begin{array}{c}
H_{x}^{(i)}\\
H_{y}^{(i)}
\end{array}\right)+\sum_{s,j=-\infty}^{\infty}\exp\left[i\mathbf{k}_{s,j}\mathbf{r}\right]\left(\begin{array}{c}
H_{x||s,j}^{(r)}\\
H_{y||s,j}^{(r)}
\end{array}\right)\nonumber \\
=\sum_{s,j=-\infty}^{\infty}\exp\left[i\mathbf{k}_{s,j}\mathbf{r}\right]\left(\hat{{\cal I}}+\hat{{\cal G}}{\cal \hat{Q}}_{s,j}^{(3)}\right)\left(\begin{array}{c}
H_{x||s,j}^{(t)}\\
H_{y||s,j}^{(t)}
\end{array}\right),\label{eq:cond-per-H13}\\
i\hat{\sigma}_{y}\left(\begin{array}{c}
H_{x}^{(i)}\\
H_{y}^{(i)}
\end{array}\right)\nonumber \\
-\sum_{s,j=-\infty}^{\infty}\exp\left[i\mathbf{k}_{s,j}\mathbf{r}\right]{\cal \hat{Q}}_{s,j}^{(1)}\left(\begin{array}{c}
H_{x||s,j}^{(r)}\\
H_{y||s,j}^{(r)}
\end{array}\right)\nonumber \\
=\sum_{s,j=-\infty}^{\infty}{\cal \hat{Q}}_{s,j}^{(3)}\left(\begin{array}{c}
H_{x||s,j}^{(t)}\\
H_{y||s,j}^{(t)}
\end{array}\right)\exp\left[i\mathbf{k}_{s,j}\mathbf{r}\right].\label{eq:cond-per-E13}
\end{eqnarray}
As in Sec. \ref{sec:main-equations}, we multiply Eqs. (\ref{eq:cond-per-H13})
and (\ref{eq:cond-per-E13}) by $\exp\left[-i\mathbf{k}_{s^{\prime},j^{\prime}}\mathbf{r}\right]$
and integrate over the area of one period of the structure $lD\le x\le\left(l+1\right)D$,
$l^{\prime}D\le y\le\left(l^{\prime}+1\right)D$. We obtain:
\begin{eqnarray}
\left(\begin{array}{c}
H_{x}^{(i)}\\
H_{y}^{(i)}
\end{array}\right)\delta_{s^{\prime},0}\delta_{0,j^{\prime}}+\left(\begin{array}{c}
H_{x||s^{\prime},j^{\prime}}^{(r)}\\
H_{y||s^{\prime},j^{\prime}}^{(r)}
\end{array}\right)\nonumber \\
=\left(\hat{{\cal I}}+\hat{{\cal G}}{\cal \hat{Q}}_{s^{\prime},j^{\prime}}^{(3)}\right)\left(\begin{array}{c}
H_{x||s^{\prime},j^{\prime}}^{(t)}\\
H_{y||s^{\prime},j^{\prime}}^{(t)}
\end{array}\right),\label{eq:per-H13}\\
i\hat{\sigma}_{y}\left(\begin{array}{c}
H_{x}^{(i)}\\
H_{y}^{(i)}
\end{array}\right)\delta_{s^{\prime},0}\delta_{0,j^{\prime}}-{\cal \hat{Q}}_{s^{\prime},j^{\prime}}^{(1)}\left(\begin{array}{c}
H_{x||s^{\prime},j^{\prime}}^{(r)}\\
H_{y||s^{\prime},j^{\prime}}^{(r)}
\end{array}\right)\nonumber \\
={\cal \hat{Q}}_{s^{\prime},j^{\prime}}^{(3)}\left(\begin{array}{c}
H_{x||s^{\prime},j^{\prime}}^{(t)}\\
H_{y||s^{\prime},j^{\prime}}^{(t)}
\end{array}\right).\label{eq:per-E13}
\end{eqnarray}

\subsection{Dispersion relation of magnetoplasmons}

\label{subsec:dr-magnetoplasmons}When $s^{\prime}\ne0$ or $j^{\prime}\ne0$,
Eqs. (\ref{eq:per-H13}) and (\ref{eq:per-E13}) take the form
\begin{eqnarray}
\left(\begin{array}{c}
H_{x||s^{\prime},j^{\prime}}^{(r)}\\
H_{y||s^{\prime},j^{\prime}}^{(r)}
\end{array}\right)=\left(\hat{{\cal I}}+\hat{{\cal G}}{\cal \hat{Q}}_{s^{\prime},j^{\prime}}^{(3)}\right)\left(\begin{array}{c}
H_{x||s^{\prime},j^{\prime}}^{(t)}\\
H_{y||s^{\prime},j^{\prime}}^{(t)}
\end{array}\right),\label{eq:per-H13-ne0}\\
-{\cal \hat{Q}}_{s^{\prime},j^{\prime}}^{(1)}\left(\begin{array}{c}
H_{x||s^{\prime},j^{\prime}}^{(r)}\\
H_{y||s^{\prime},j^{\prime}}^{(r)}
\end{array}\right)={\cal \hat{Q}}_{s^{\prime},j^{\prime}}^{(3)}\left(\begin{array}{c}
H_{x||s^{\prime},j^{\prime}}^{(t)}\\
H_{y||s^{\prime},j^{\prime}}^{(t)}
\end{array}\right).\label{eq:per-E13-ne0}
\end{eqnarray}
As a result, the amplidues of the magnetic field components corresponding to the transmitted wave 
are governed by the equation 
\begin{eqnarray}
\left[\hat{{\cal I}}+\left\{ \hat{{\cal G}}+\left({\cal \hat{Q}}_{s^{\prime},j^{\prime}}^{(1)}\right)^{-1}\right\} {\cal \hat{Q}}_{s^{\prime},j^{\prime}}^{(3)}\right]\left(\begin{array}{c}
H_{x||s^{\prime},j^{\prime}}^{(t)}\\
H_{y||s^{\prime},j^{\prime}}^{(t)}
\end{array}\right)=0\nonumber \\
\label{eq:Ht-eigen}
\end{eqnarray}
with 
\begin{eqnarray*}
\left(\hat{{\cal Q}}_{s^{\prime},j^{\prime}}^{(1)}\right)^{-1}=\left(\begin{array}{cc}
-\frac{c}{\omega}\frac{s^{\prime}j^{\prime}g^{2}}{p_{s^{\prime},j^{\prime}}^{(1)}} & -\frac{c}{\omega}\frac{\left(p_{s^{\prime},0}^{(1)}\right)^{2}}{p_{s^{\prime},j^{\prime}}^{(1)}}\\
\frac{c}{\omega}\frac{\left(p_{0,j^{\prime}}^{(1)}\right)^{2}}{p_{s^{\prime},j^{\prime}}^{(1)}} & \frac{c}{\omega}\frac{s^{\prime}j^{\prime}g^{2}}{p_{s^{\prime},j^{\prime}}^{(1)}}
\end{array}\right)
\end{eqnarray*}
being the inverse of the matrix $\hat{{\cal Q}}_{s^{\prime},j^{\prime}}^{(1)}$.
If we project $H_{x||s^{\prime},j^{\prime}}^{(t)}$, $H_{y||s^{\prime},j^{\prime}}^{(t)}$
on the direction of the wavevector $\mathbf{k}_{s^{\prime},j^{\prime}}=\left(s^{\prime}g,\,j^{\prime}g\right)$,
we have
\begin{eqnarray*}
\left(\begin{array}{c}
H_{s^{\prime},j^{\prime}}^{\left(\parallel\right)}\\
H_{s^{\prime},j^{\prime}}^{\left(\perp\right)}
\end{array}\right)=\hat{{\cal V}}_{s^{\prime},j^{\prime}}\left(\begin{array}{c}
H_{x||s^{\prime},j^{\prime}}^{(t)}\\
H_{y||s^{\prime},j^{\prime}}^{(t)}
\end{array}\right),
\end{eqnarray*}
where 
\begin{eqnarray*}
\hat{{\cal V}}_{s^{\prime},j^{\prime}}=\frac{1}{\sqrt{s^{\prime2}+j^{\prime2}}}\left(\begin{array}{cc}
s^{\prime} & j^{\prime}\\
-j^{\prime} & s^{\prime}
\end{array}\right)
\end{eqnarray*}
is the transformation matrix, $H_{s^{\prime},j^{\prime}}^{\left(\parallel\right)}$
and $H_{s^{\prime},j^{\prime}}^{\left(\perp\right)}$ are the components
of the magnetic field in the substrate, parallel and perpendicular
to the harmonic wavevector $\mathbf{k}_{s^{\prime},j^{\prime}}$.
In this case Eq. (\ref{eq:Ht-eigen}) can be rewritten as
\begin{eqnarray}
\hat{{\cal Z}}_{s^{\prime},j^{\prime}}\left(\begin{array}{c}
H_{s^{\prime},j^{\prime}}^{\left(\parallel\right)}\\
H_{s^{\prime},j^{\prime}}^{\left(\perp\right)}
\end{array}\right)=0,\label{eq:Ht-eigen-mod}
\end{eqnarray}
and the matrix 
\begin{eqnarray*}
\hat{{\cal Z}}_{s^{\prime},j^{\prime}}=\hat{{\cal V}}_{s^{\prime},j^{\prime}}\left[\hat{I}+\left\{ \hat{{\cal G}}+\left({\cal \hat{Q}}_{s^{\prime},j^{\prime}}^{(1)}\right)^{-1}\right\} {\cal \hat{Q}}_{s^{\prime},j^{\prime}}^{(3)}\right]\left(\hat{{\cal V}}_{s^{\prime},j^{\prime}}\right)^{-1}
\end{eqnarray*}
after some algebra will have the form\begin{widetext}
\begin{eqnarray*}
\hat{{\cal Z}}_{s^{\prime},j^{\prime}}={\displaystyle {\displaystyle \left(\begin{array}{cc}
1+\frac{4\pi\omega}{c^{2}}\Sigma_{xx}\frac{1}{p_{s^{\prime},j^{\prime}}^{(3)}}+\frac{p_{s^{\prime},j^{\prime}}^{(1)}}{p_{s^{\prime},j^{\prime}}^{(3)}} & \frac{4\pi}{\omega\varepsilon}\Sigma_{xy}p_{s^{\prime},j^{\prime}}^{(3)}\\
-\frac{4\pi\omega}{c^{2}}\Sigma_{xy}\frac{1}{p_{s^{\prime},j^{\prime}}^{(3)}} & 1+\frac{p_{s^{\prime},j^{\prime}}^{(3)}}{\varepsilon p_{s^{\prime},j^{\prime}}^{(1)}}+\frac{4\pi}{\omega\varepsilon}\Sigma_{xx}p_{s^{\prime},j^{\prime}}^{(3)}
\end{array}\right)}}.
\end{eqnarray*}
\end{widetext}Here we took into account that $\Sigma_{yy}=\Sigma_{xx}$,
$\Sigma_{yx}=-\Sigma_{xy}$. The Eq. (\ref{eq:Ht-eigen-mod}) possesses
a solution, only when the determinant of matrix $\hat{{\cal Z}}_{s^{\prime},j^{\prime}}$
is equal to zero. Thus, the magnetoplasmon dispersion relation
is
\begin{eqnarray*}
\mathrm{det}\left|\hat{{\cal Z}}_{s^{\prime},j^{\prime}}\right| & = & \left[p_{s^{\prime},j^{\prime}}^{(3)}+\frac{4\pi\omega}{c^{2}}\Sigma_{xx}+p_{s^{\prime},j^{\prime}}^{(1)}\right]\\
 &  & \times\left[\frac{\varepsilon}{p_{s^{\prime},j^{\prime}}^{(3)}}+\frac{1}{p_{s^{\prime},j^{\prime}}^{(1)}}+\frac{4\pi}{\omega}\Sigma_{xx}\right]\\
 &  & +\left(\frac{4\pi}{c}\Sigma_{xy}\right)^{2}=0,
\end{eqnarray*}
which after the formal substitution $\mathbf{k}_{s^{\prime},j^{\prime}}=\mathbf{k}=\left(k_{x},\,k_{y}\right)$
can be transformed into Eq. (\ref{eq:dr-magnetoplasmons}). 

\subsection{The transmittance and the reflectance of the graphene layer on substrate}

\label{subsec:R0-T0}
When $s^{\prime}=j^{\prime}=0$, Eqs. (\ref{eq:per-H13})
and (\ref{eq:per-E13}) have the form
\begin{eqnarray}
\left(\begin{array}{c}
H_{x}^{(i)}\\
H_{y}^{(i)}
\end{array}\right)+\left(\begin{array}{c}
H_{x||0,0}^{(r)}\\
H_{y||0,0}^{(r)}
\end{array}\right)=\left(\hat{{\cal I}}+\hat{{\cal G}}{\cal \hat{Q}}_{0,0}^{(3)}\right)\left(\begin{array}{c}
H_{x||0,0}^{(t)}\\
H_{y||0,0}^{(t)}
\end{array}\right),\nonumber \\
\label{eq:per-H13-0}\\
i\hat{\sigma}_{y}\left(\begin{array}{c}
H_{x}^{(i)}\\
H_{y}^{(i)}
\end{array}\right)-\hat{Q}_{0,0}^{(1)}\left(\begin{array}{c}
H_{x||0,0}^{(r)}\\
H_{y||0,0}^{(r)}
\end{array}\right)={\cal \hat{Q}}_{0,0}^{(3)}\left(\begin{array}{c}
H_{x||0,0}^{(t)}\\
H_{y||0,0}^{(t)}
\end{array}\right).\nonumber \\
\label{eq:per-E13-0}
\end{eqnarray}
Taking into account that $p_{0,0}^{(1)}=\left(\omega/c\right)$, $p_{0,0}^{(3)}=\left(\omega/c\right)\sqrt{\varepsilon}$
, matrixes ${\cal \hat{Q}}_{0,0}^{(1)}$ ${\cal \hat{Q}}_{0,0}^{(3)}$
can be represented in the simple form,
\begin{eqnarray}
{\cal \hat{Q}}_{0,0}^{(3)}=\sqrt{\frac{1}{\varepsilon}}i\hat{\sigma}_{y},\qquad\hat{{\cal Q}}_{0,0}^{(1)}=i\hat{\sigma}_{y},\label{eq:mat-13-00}
\end{eqnarray}
and Eqs. (\ref{eq:per-H13-0}) and (\ref{eq:per-E13-0}) can be
rewritten as
\begin{eqnarray}
\left(\begin{array}{c}
H_{x}^{(i)}\\
H_{y}^{(i)}
\end{array}\right)+\left(\begin{array}{c}
H_{x||0,0}^{(r)}\\
H_{y||0,0}^{(r)}
\end{array}\right)=\left(\hat{{\cal I}}+\sqrt{\frac{1}{\varepsilon}}\hat{{\cal G}}i\hat{\sigma}_{y}\right)\left(\begin{array}{c}
H_{x||0,0}^{(t)}\\
H_{y||0,0}^{(t)}
\end{array}\right),\nonumber \\
\label{eq:per-H13-0-mod}\\
i\hat{\sigma}_{y}\left(\begin{array}{c}
H_{x}^{(i)}\\
H_{y}^{(i)}
\end{array}\right)-i\hat{\sigma}_{y}\left(\begin{array}{c}
H_{x||0,0}^{(r)}\\
H_{y||0,0}^{(r)}
\end{array}\right)=\sqrt{\frac{1}{\varepsilon}}i\hat{\sigma}_{y}\left(\begin{array}{c}
H_{x||0,0}^{(t)}\\
H_{y||0,0}^{(t)}
\end{array}\right).\nonumber \\
\label{eq:per-E13-0-mod}
\end{eqnarray}
Multiplying Eq. (\ref{eq:per-E13-0-mod}) by $-i\hat{\sigma}_{y}$
and summing with Eq. (\ref{eq:per-H13-0-mod}) we obtain:
\begin{eqnarray}
2\left(\begin{array}{c}
H_{x}^{(i)}\\
H_{y}^{(i)}
\end{array}\right)=\hat{{\cal F}}\left(\begin{array}{c}
H_{x||0,0}^{(t)}\\
H_{y||0,0}^{(t)}
\end{array}\right),\label{eq:mat-Ht0}
\end{eqnarray}
where the matrix
\begin{eqnarray*}
\hat{{\cal F}} & = & \left(1+\sqrt{\frac{1}{\varepsilon}}\right)\hat{{\cal I}}+\sqrt{\frac{1}{\varepsilon}}\hat{{\cal G}}i\hat{\sigma}_{y}\\
 & = & \left(\begin{array}{cc}
1+\frac{4\pi}{c\sqrt{\varepsilon}}\Sigma_{xx}+\sqrt{\frac{1}{\varepsilon}} & \frac{4\pi}{c\sqrt{\varepsilon}}\Sigma_{xy}\\
-\frac{4\pi}{c\sqrt{\varepsilon}}\Sigma_{xy} & 1+\frac{4\pi}{c\sqrt{\varepsilon}}\Sigma_{xx}+\sqrt{\frac{1}{\varepsilon}}
\end{array}\right),
\end{eqnarray*}
and we took into account that $\Sigma_{yy}=\Sigma_{xx}$, $\Sigma_{yx}=-\Sigma_{xy}$.
As a result, the solution of Eq. (\ref{eq:mat-Ht0}) can be represented
as
\begin{eqnarray}
H_{x||0,0}^{(t)}=2\sqrt{\varepsilon}\frac{\left(\sqrt{\varepsilon}+\frac{4\pi}{c}\Sigma_{xx}+1\right)H_{x}^{(i)}-\frac{4\pi}{c}\Sigma_{xy}H_{y}^{(i)}}{\left(\sqrt{\varepsilon}+\frac{4\pi}{c}\Sigma_{xx}+1\right)^{2}+\left(\frac{4\pi}{c}\Sigma_{xy}\right)^{2}},\nonumber \\
\label{eq:Htx-00}\\
H_{y||0,0}^{(t)}=2\sqrt{\varepsilon}\frac{\frac{4\pi}{c}\Sigma_{xy}H_{x}^{(i)}+\left(\sqrt{\varepsilon}+\frac{4\pi}{c}\Sigma_{xx}+1\right)H_{y}^{(i)}}{\left(\sqrt{\varepsilon}+\frac{4\pi}{c}\Sigma_{xx}+1\right)^{2}+\left(\frac{4\pi}{c}\Sigma_{xy}\right)^{2}}.\nonumber \\
\label{eq:Hty-00}
\end{eqnarray}
Similarly, combining Eqs. (\ref{eq:mat-Ht0}) and (\ref{eq:per-E13-0-mod}),
one can obtain the expression for the amplitude of the reflected wave
magnetic field:
\begin{eqnarray*}
\left(\begin{array}{c}
H_{x||0,0}^{(r)}\\
H_{y||0,0}^{(r)}
\end{array}\right)=\left[\hat{{\cal I}}-2\sqrt{\frac{1}{\varepsilon}}\hat{{\cal F}}^{-1}\right]\left(\begin{array}{c}
H_{x}^{(i)}\\
H_{y}^{(i)}
\end{array}\right),
\end{eqnarray*}
or, equivalently
\begin{eqnarray}
H_{x||0,0}^{(r)} & = & \frac{\left(\sqrt{\varepsilon}+\frac{4\pi}{c}\Sigma_{xx}\right)^{2}-1+\left(\frac{4\pi}{c}\Sigma_{xy}\right)^{2}}{\left(\sqrt{\varepsilon}+\frac{4\pi}{c}\Sigma_{xx}+1\right)^{2}+\left(\frac{4\pi}{c}\Sigma_{xy}\right)^{2}}H_{x}^{(i)}\nonumber \\
 &  & +\frac{\frac{8\pi}{c}\Sigma_{xy}}{\left(\sqrt{\varepsilon}+\frac{4\pi}{c}\Sigma_{xx}+1\right)^{2}+\left(\frac{4\pi}{c}\Sigma_{xy}\right)^{2}}H_{y}^{(i)}\nonumber \\
\label{eq:Hrx-00}\\
H_{y||0,0}^{(r)} & = & -\frac{\frac{8\pi}{c}\Sigma_{xy}}{\left(\sqrt{\varepsilon}+\frac{4\pi}{c}\Sigma_{xx}+1\right)^{2}+\left(\frac{4\pi}{c}\Sigma_{xy}\right)^{2}}H_{x}^{(i)}\nonumber \\
 &  & +\frac{\left(\sqrt{\varepsilon}+\frac{4\pi}{c}\Sigma_{xx}\right)^{2}-1+\left(\frac{4\pi}{c}\Sigma_{xy}\right)^{2}}{\left(\sqrt{\varepsilon}+\frac{4\pi}{c}\Sigma_{xx}+1\right)^{2}+\left(\frac{4\pi}{c}\Sigma_{xy}\right)^{2}}H_{y}^{(i)}.\nonumber \\
\label{eq:Hry-00}
\end{eqnarray}
Taking into account the particular form of the matrices ${\cal \hat{Q}}_{0,0}^{(3)}$
and ${\cal \hat{Q}}_{0,0}^{(1)}$ {[}see Eq. (\ref{eq:mat-13-00}){]},
the expressions (\ref{eq:St-z-mat}) and (\ref{eq:Sr-z-mat}) for transmitted and reflected wave energy flux density in $z$ direction are written as
\begin{eqnarray*}
S_{z||0,0}^{(t)}=\frac{c}{8\pi}\mathrm{Re}\left\{ \frac{1}{\sqrt{\varepsilon}}\left(\begin{array}{c}
H_{x||0,0}^{(t)}\\
H_{y||0,0}^{(t)}
\end{array}\right)^{\dagger}\left(\begin{array}{c}
H_{x||0,0}^{(t)}\\
H_{y||0,0}^{(t)}
\end{array}\right)\right\} ,\\
S_{z||0,0}^{(r)}=-\frac{c}{8\pi}\mathrm{Re}\left\{ \left(\begin{array}{c}
H_{x||0,0}^{(r)}\\
H_{y||0,0}^{(r)}
\end{array}\right)^{\dagger}\left(\begin{array}{c}
H_{x||0,0}^{(r)}\\
H_{y||0,0}^{(r)}
\end{array}\right)\right\} .
\end{eqnarray*}
Accordingly, from Eqs. (\ref{eq:R-gen}) and (\ref{eq:T-gen}) the  transmittance
and reflectance can be expressed as
\begin{eqnarray}
T_{0}=\frac{\mathrm{Re}\left\{ \sqrt{\frac{1}{\varepsilon}}\left(\left|H_{x||0,0}^{(t)}\right|^{2}+\left|H_{y||0,0}^{(t)}\right|^{2}\right)\right\} }{\left|H_{x}^{(i)}\right|^{2}+\left|H_{y}^{(i)}\right|^{2}},\label{eq:T-0-gen}\\
R_{0}=\frac{\left|H_{x||0,0}^{(r)}\right|^{2}+\left|H_{y||0,0}^{(r)}\right|^{2}}{\left|H_{x}^{(i)}\right|^{2}+\left|H_{y}^{(i)}\right|^{2}}.\label{eq:R-0-gen}
\end{eqnarray}
Finally, after substituting the expressions for the magnetic
field components, (\ref{eq:Htx-00})\textendash (\ref{eq:Hry-00}) into Eqs.
(\ref{eq:T-0-gen}) and (\ref{eq:R-0-gen}) one can obtain the transmittance and the reflectance of bare graphene in the following form: 
\begin{eqnarray}
T_{0}=\frac{4\mathrm{Re}\left(\sqrt{\frac{1}{\varepsilon}}\right)}{\left|H_{x}^{(i)}\right|^{2}+\left|H_{y}^{(i)}\right|^{2}}\nonumber \\
\times\left\{ \left|\sqrt{\varepsilon}\frac{\left(\sqrt{\varepsilon}+\frac{4\pi}{c}\Sigma_{xx}+1\right)H_{x}^{(i)}-\frac{4\pi}{c}\Sigma_{xy}H_{y}^{(i)}}{\left(\sqrt{\varepsilon}+\frac{4\pi}{c}\Sigma_{xx}+1\right)^{2}+\left(\frac{4\pi}{c}\Sigma_{xy}\right)^{2}}\right|^{2}\right.\nonumber \\
\left.+\left|\sqrt{\varepsilon}\frac{\frac{4\pi}{c}\Sigma_{xy}H_{x}^{(i)}+\left(\sqrt{\varepsilon}+\frac{4\pi}{c}\Sigma_{xx}+1\right)H_{y}^{(i)}}{\left(\sqrt{\varepsilon}+\frac{4\pi}{c}\Sigma_{xx}+1\right)^{2}+\left(\frac{4\pi}{c}\Sigma_{xy}\right)^{2}}\right|^{2}\right\} ,\nonumber \\
\label{eq:T-0}\\
R_{0}=\frac{1}{\left|H_{x}^{(i)}\right|^{2}+\left|H_{y}^{(i)}\right|^{2}}\nonumber \\
\times\left\{ \left|\frac{\left(\sqrt{\varepsilon}+\frac{4\pi}{c}\Sigma_{xx}\right)^{2}-1+\left(\frac{4\pi}{c}\Sigma_{xy}\right)^{2}}{\left(\sqrt{\varepsilon}+\frac{4\pi}{c}\Sigma_{xx}+1\right)^{2}+\left(\frac{4\pi}{c}\Sigma_{xy}\right)^{2}}H_{x}^{(i)}+\right.\right.\nonumber \\
\left.+\frac{\frac{8\pi}{c}\Sigma_{xy}}{\left(\sqrt{\varepsilon}+\frac{4\pi}{c}\Sigma_{xx}+1\right)^{2}+\left(\frac{4\pi}{c}\Sigma_{xy}\right)^{2}}H_{y}^{(i)}\right|^{2}\nonumber \\
+\left|\frac{\frac{8\pi}{c}\Sigma_{xy}}{\left(\sqrt{\varepsilon}+\frac{4\pi}{c}\Sigma_{xx}+1\right)^{2}+\left(\frac{4\pi}{c}\Sigma_{xy}\right)^{2}}H_{x}^{(i)}\right.\nonumber \\
\left.\left.-\frac{\left(\sqrt{\varepsilon}+\frac{4\pi}{c}\Sigma_{xx}\right)^{2}-1+\left(\frac{4\pi}{c}\Sigma_{xy}\right)^{2}}{\left(\sqrt{\varepsilon}+\frac{4\pi}{c}\Sigma_{xx}+1\right)^{2}+\left(\frac{4\pi}{c}\Sigma_{xy}\right)^{2}}H_{y}^{(i)}\right|^{2}\right\} .\nonumber \\
\label{eq:R-0}
\end{eqnarray}

\section{Coefficient of magnetic circular dichroism\label{sec:Coef-MCD}}

\begin{figure}
\includegraphics[width=8.5cm]{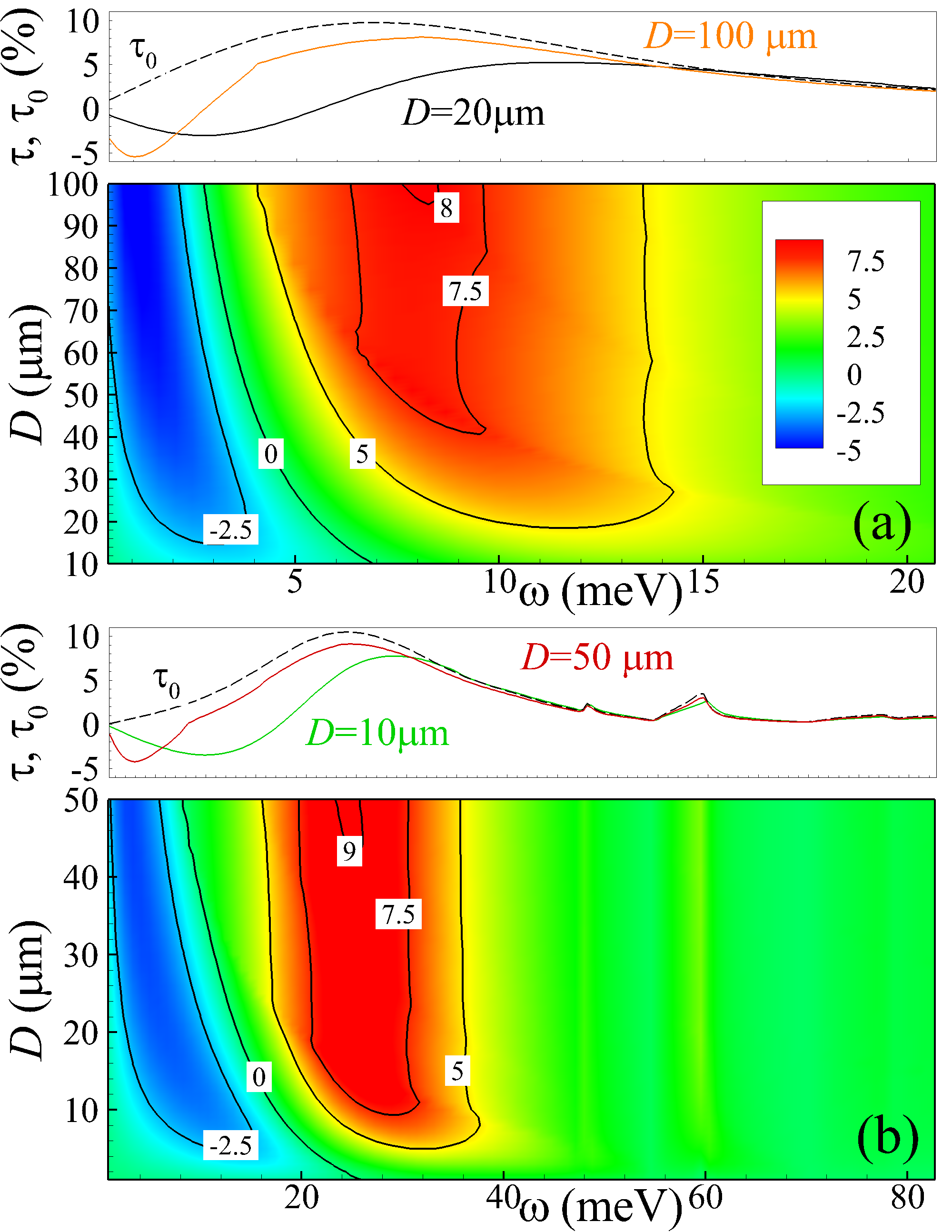}

\caption{Coefficient of circular dichroism $\tau$ (in percent, depicted by
color map) \emph{versus} frequency $\omega$ and hole array period
$D$ for two values of the magnetic field, $B=1.5\,$T {[}panel (a){]}
and $B=7\,$T {[}panel (b){]}. Other parameters of the structure are
the same as in Fig.\ref{fig:eigen-sp}. In the upper plots of panels
(a) and (b) we trace dependencies $\tau\left(\omega\right)$ from the 
lower plots at fixed values of the hole grating period $D=20\,\mu$m,
$D=100\,\mu$m {[}black and orange solid lines in panel (a){]}, and
$D=10\,\mu$m, $D=50\,\mu$m {[}green and red solid lines in panel
(b){]}. Also in the upper plots of panels (a) and (b) the coefficients
of circular dichroism $\tau_{0}$ of bare graphene are depicted by the dashed black lines. }
\label{fig:Isol_LR(w,D)_Al2O3}
\end{figure}

In order to characterize quantitatively the circular dichroism in
our magnetoactive graphene-based structure, we introduce the coefficient
of circular dichroism, 
\begin{eqnarray*}
\tau=\frac{T_{a}-T_{c}}{T_{a}+T_{c}},
\end{eqnarray*}
which is expressed in terms of the transmittance of the clockwise
and anticlockwise circularly polarized waves $T_{c}$ and $T_{a}$.
The dependence of the coefficient of the circular dichroism upon the frequency
$\omega$ and the period of the hole array $D$ is shown in Fig.\,\ref{fig:Isol_LR(w,D)_Al2O3}.
It can be seen that the coefficient of the optical dichroism $\tau$
achieves its maximum in the vicinity of $\omega\approx8\,$meV {[}for
weak magnetic field, see Fig.\,\ref{fig:Isol_LR(w,D)_Al2O3}(a){]}
or $\omega\approx25\,$meV {[}for strong magnetic field, see Fig.\,\ref{fig:Isol_LR(w,D)_Al2O3}(b){]}.
The maximal value of the optical dichoism coefficient grows monotonically
with the increase of the hole array period $D$. Notice that the
coefficient of optical dichroism of the perforated metal film deposited
on top of graphene monolayer $\tau$ is always smaller than that of
bare graphene, $\tau_{0}$
{[}compare dashed and solid lines in upper plots in Figs.\,\ref{fig:Isol_LR(w,D)_Al2O3}(a)
and \ref{fig:Isol_LR(w,D)_Al2O3}(b){]}. In other
words, the presence of the perforated metallic film on top of the graphene sheet 
reduces the relative contrast between the transmittances of clockwise- and
anticlockwise-polarized waves. The physical reason for this
is the diffraction of the incident electromagnetic wave on the graphene
with periodic structure, which gives rise to the partial conversion
of the incident wave energy into the energy of the excited magnetoplasmons,
thus reducing the transmittance of this structure for both circular polarizations (clockwise
and anticlockwise). However, the stronger decrease
of the anticlockwise circularly polarized wave transmittance (compared
to that of clockwise polarized wave) leads to the reduction of the
circular dichroism coefficient $\tau$. 

Moreover, the presence of the perforated metal film on
top of graphene results into the possibility to obtain a negative
circular dichroism coefficient (hence, $T_{a}<T_{c}$) at the low-frequency
region. This situation is in strong contrast with the case of bare graphene
without perforated metal film on top of it, where the coefficient
of circular dichroism $\tau_{0}$ is always possitive {[}$T_{a}>T_{c}$,
see upper plots in Figs.\,\ref{fig:Isol_LR(w,D)_Al2O3}(a) and \ref{fig:Isol_LR(w,D)_Al2O3}(b){]}.
It should be stressed that a decrease of the grating period
$D$ {[}as shown in Fig.\,\ref{fig:Isol_LR(w,D)_Al2O3}{]} and an increase
of the magnetic field $B$ {[}see Fig.\,\ref{fig:Isol_LR(w,B)_Al2O3}{]} lead to a broadening of the frequency region with negative
$\tau$.
\begin{figure}
\includegraphics[width=8.5cm]{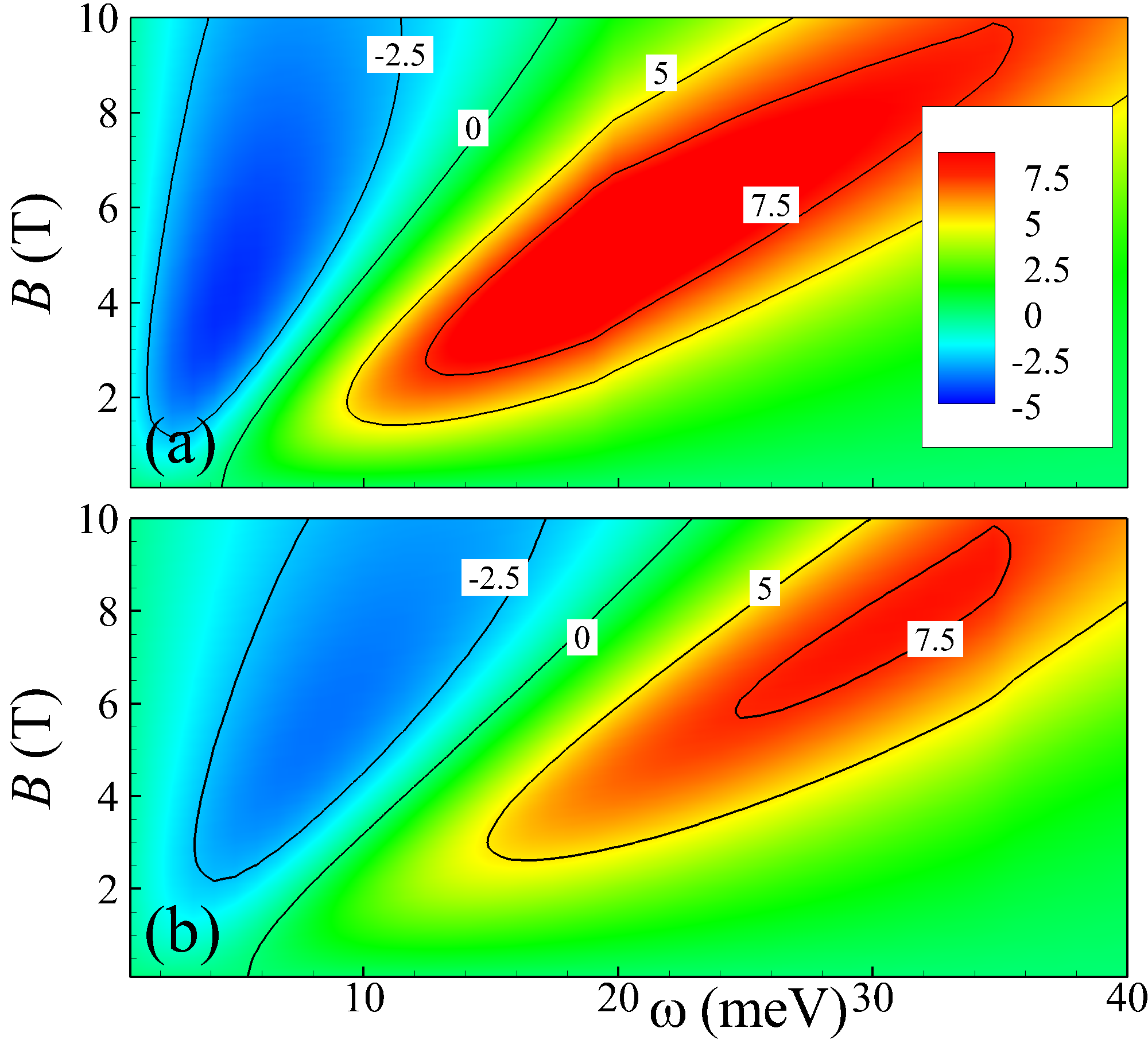}

\caption{Coefficient of circular dichroism $\tau$ (in percent, depicted by
color map) \emph{versus} frequency $\omega$ and magnetic field $B$
for two values of the period of the hole array, $D=20\,\mu$m {[}panel
(a){]} and $D=10\,\mu$m {[}panel (b){]}. Other parameters of the
structure are the same as in Fig.\ref{fig:eigen-sp}. }
\label{fig:Isol_LR(w,B)_Al2O3}
\end{figure}

\bibliographystyle{apsrev4-1}
\bibliography{difr_holes_mf_bib_submit}

\begin{thebibliography}{69}%
\makeatletter
\providecommand \@ifxundefined [1]{%
 \@ifx{#1\undefined}
}%
\providecommand \@ifnum [1]{%
 \ifnum #1\expandafter \@firstoftwo
 \else \expandafter \@secondoftwo
 \fi
}%
\providecommand \@ifx [1]{%
 \ifx #1\expandafter \@firstoftwo
 \else \expandafter \@secondoftwo
 \fi
}%
\providecommand \natexlab [1]{#1}%
\providecommand \enquote  [1]{``#1''}%
\providecommand \bibnamefont  [1]{#1}%
\providecommand \bibfnamefont [1]{#1}%
\providecommand \citenamefont [1]{#1}%
\providecommand \href@noop [0]{\@secondoftwo}%
\providecommand \href [0]{\begingroup \@sanitize@url \@href}%
\providecommand \@href[1]{\@@startlink{#1}\@@href}%
\providecommand \@@href[1]{\endgroup#1\@@endlink}%
\providecommand \@sanitize@url [0]{\catcode `\\12\catcode `\$12\catcode
  `\&12\catcode `\#12\catcode `\^12\catcode `\_12\catcode `\%12\relax}%
\providecommand \@@startlink[1]{}%
\providecommand \@@endlink[0]{}%
\providecommand \url  [0]{\begingroup\@sanitize@url \@url }%
\providecommand \@url [1]{\endgroup\@href {#1}{\urlprefix }}%
\providecommand \urlprefix  [0]{URL }%
\providecommand \Eprint [0]{\href }%
\providecommand \doibase [0]{http://dx.doi.org/}%
\providecommand \selectlanguage [0]{\@gobble}%
\providecommand \bibinfo  [0]{\@secondoftwo}%
\providecommand \bibfield  [0]{\@secondoftwo}%
\providecommand \translation [1]{[#1]}%
\providecommand \BibitemOpen [0]{}%
\providecommand \bibitemStop [0]{}%
\providecommand \bibitemNoStop [0]{.\EOS\space}%
\providecommand \EOS [0]{\spacefactor3000\relax}%
\providecommand \BibitemShut  [1]{\csname bibitem#1\endcsname}%
\let\auto@bib@innerbib\@empty
\bibitem [{\citenamefont {Faraday}(1846)}]{mo-Faraday1846-PTRSL}%
  \BibitemOpen
  \bibfield  {author} {\bibinfo {author} {\bibfnamefont {M.}~\bibnamefont
  {Faraday}},\ }\href {\doibase 10.1098/rstl.1846.0001} {\bibfield  {journal}
  {\bibinfo  {journal} {Philos. Trans. Roy. Soc. London}\ }\textbf {\bibinfo
  {volume} {136}},\ \bibinfo {pages} {1} (\bibinfo {year} {1846})}\BibitemShut
  {NoStop}%
\bibitem [{\citenamefont {Gabriel}\ and\ \citenamefont
  {Brodwin}(1965)}]{mo-nrps-Gabriel1965-mtt}%
  \BibitemOpen
  \bibfield  {author} {\bibinfo {author} {\bibfnamefont {G.}~\bibnamefont
  {Gabriel}}\ and\ \bibinfo {author} {\bibfnamefont {M.}~\bibnamefont
  {Brodwin}},\ }\href {\doibase 10.1109/TMTT.1965.1126001} {\bibfield
  {journal} {\bibinfo  {journal} {IEEE Trans. Microwave Theory Tech.}\ }\textbf
  {\bibinfo {volume} {13}},\ \bibinfo {pages} {364} (\bibinfo {year}
  {1965})}\BibitemShut {NoStop}%
\bibitem [{\citenamefont {Auracher}\ and\ \citenamefont
  {Witte}(1975)}]{mo-nrps-Auracher1975-OptComm}%
  \BibitemOpen
  \bibfield  {author} {\bibinfo {author} {\bibfnamefont {F.}~\bibnamefont
  {Auracher}}\ and\ \bibinfo {author} {\bibfnamefont {H.~H.}\ \bibnamefont
  {Witte}},\ }\href
  {http://apps.isiknowledge.com/InboundService.do?Func=Frame{\&}product=WOS{\&}action=retrieve{\&}SrcApp=Papers{\&}UT=A1975AA17600019{\&}SID=3EmklbC7p4hLNLm5ogD{\&}Init=Yes{\&}SrcAuth=mekentosj{\&}mode=FullRecord{\&}customersID=mekentosj{\&}DestFail=http://access.isiproducts.com/}
  {\bibfield  {journal} {\bibinfo  {journal} {Opt. Commun.}\ }\textbf {\bibinfo
  {volume} {13}},\ \bibinfo {pages} {435} (\bibinfo {year} {1975})}\BibitemShut
  {NoStop}%
\bibitem [{\citenamefont {D{\"{o}}tsch}\ \emph {et~al.}(2005)\citenamefont
  {D{\"{o}}tsch}, \citenamefont {Bahlmann}, \citenamefont {Zhuromskyy},
  \citenamefont {Hammer}, \citenamefont {Wilkens}, \citenamefont {Gerhardt},
  \citenamefont {Hertel},\ and\ \citenamefont
  {Popkov}}]{mo-rev-Dotsch2005-josab}%
  \BibitemOpen
  \bibfield  {author} {\bibinfo {author} {\bibfnamefont {H.}~\bibnamefont
  {D{\"{o}}tsch}}, \bibinfo {author} {\bibfnamefont {N.}~\bibnamefont
  {Bahlmann}}, \bibinfo {author} {\bibfnamefont {O.}~\bibnamefont
  {Zhuromskyy}}, \bibinfo {author} {\bibfnamefont {M.}~\bibnamefont {Hammer}},
  \bibinfo {author} {\bibfnamefont {L.}~\bibnamefont {Wilkens}}, \bibinfo
  {author} {\bibfnamefont {R.}~\bibnamefont {Gerhardt}}, \bibinfo {author}
  {\bibfnamefont {P.}~\bibnamefont {Hertel}}, \ and\ \bibinfo {author}
  {\bibfnamefont {A.~F.}\ \bibnamefont {Popkov}},\ }\href {\doibase
  10.1364/JOSAB.22.000240} {\bibfield  {journal} {\bibinfo  {journal} {Journal
  of the Optical Society of America B}\ }\textbf {\bibinfo {volume} {22}},\
  \bibinfo {pages} {240} (\bibinfo {year} {2005})}\BibitemShut {NoStop}%
\bibitem [{\citenamefont {Aplet}\ and\ \citenamefont
  {Carson}(1964)}]{mo-oi-Aplet1964-ao}%
  \BibitemOpen
  \bibfield  {author} {\bibinfo {author} {\bibfnamefont {L.~J.}\ \bibnamefont
  {Aplet}}\ and\ \bibinfo {author} {\bibfnamefont {J.~W.}\ \bibnamefont
  {Carson}},\ }\href {\doibase 10.1364/AO.3.000544} {\bibfield  {journal}
  {\bibinfo  {journal} {Appl. Opt.}\ }\textbf {\bibinfo {volume} {3}},\
  \bibinfo {pages} {544} (\bibinfo {year} {1964})}\BibitemShut {NoStop}%
\bibitem [{\citenamefont {Jalas}\ \emph {et~al.}(2013)\citenamefont {Jalas},
  \citenamefont {Petrov}, \citenamefont {Eich}, \citenamefont {Freude},
  \citenamefont {Fan}, \citenamefont {Yu}, \citenamefont {Baets}, \citenamefont
  {Popovi{\'{c}}}, \citenamefont {Melloni}, \citenamefont {Joannopoulos},
  \citenamefont {Vanwolleghem}, \citenamefont {Doerr},\ and\ \citenamefont
  {Renner}}]{mo-oi-Jalas2013-nature}%
  \BibitemOpen
  \bibfield  {author} {\bibinfo {author} {\bibfnamefont {D.}~\bibnamefont
  {Jalas}}, \bibinfo {author} {\bibfnamefont {A.}~\bibnamefont {Petrov}},
  \bibinfo {author} {\bibfnamefont {M.}~\bibnamefont {Eich}}, \bibinfo {author}
  {\bibfnamefont {W.}~\bibnamefont {Freude}}, \bibinfo {author} {\bibfnamefont
  {S.}~\bibnamefont {Fan}}, \bibinfo {author} {\bibfnamefont {Z.}~\bibnamefont
  {Yu}}, \bibinfo {author} {\bibfnamefont {R.}~\bibnamefont {Baets}}, \bibinfo
  {author} {\bibfnamefont {M.}~\bibnamefont {Popovi{\'{c}}}}, \bibinfo {author}
  {\bibfnamefont {A.}~\bibnamefont {Melloni}}, \bibinfo {author} {\bibfnamefont
  {J.~D.}\ \bibnamefont {Joannopoulos}}, \bibinfo {author} {\bibfnamefont
  {M.}~\bibnamefont {Vanwolleghem}}, \bibinfo {author} {\bibfnamefont {C.~R.}\
  \bibnamefont {Doerr}}, \ and\ \bibinfo {author} {\bibfnamefont
  {H.}~\bibnamefont {Renner}},\ }\href {\doibase 10.1038/nphoton.2013.185}
  {\bibfield  {journal} {\bibinfo  {journal} {Nat. Photonics}\ }\textbf
  {\bibinfo {volume} {7}},\ \bibinfo {pages} {579} (\bibinfo {year}
  {2013})}\BibitemShut {NoStop}%
\bibitem [{\citenamefont {Ribbens}(1965)}]{mo-circ-Ribbens1965-ao}%
  \BibitemOpen
  \bibfield  {author} {\bibinfo {author} {\bibfnamefont {W.~B.}\ \bibnamefont
  {Ribbens}},\ }\href {\doibase 10.1364/AO.4.001037} {\bibfield  {journal}
  {\bibinfo  {journal} {Appl. Opt.}\ }\textbf {\bibinfo {volume} {4}},\
  \bibinfo {pages} {1037} (\bibinfo {year} {1965})}\BibitemShut {NoStop}%
\bibitem [{\citenamefont {Takei}\ and\ \citenamefont
  {Mizumoto}(2010)}]{mo-circ-Takei2010-japj}%
  \BibitemOpen
  \bibfield  {author} {\bibinfo {author} {\bibfnamefont {R.}~\bibnamefont
  {Takei}}\ and\ \bibinfo {author} {\bibfnamefont {T.}~\bibnamefont
  {Mizumoto}},\ }\href {\doibase 10.1143/JJAP.49.052203} {\bibfield  {journal}
  {\bibinfo  {journal} {Japanese Journal of Applied Physics}\ }\textbf
  {\bibinfo {volume} {49}},\ \bibinfo {pages} {052203} (\bibinfo {year}
  {2010})}\BibitemShut {NoStop}%
\bibitem [{\citenamefont {Tien}\ \emph {et~al.}(1974)\citenamefont {Tien},
  \citenamefont {Schinke},\ and\ \citenamefont
  {Blank}}]{mo-switch-Tien1974-jap}%
  \BibitemOpen
  \bibfield  {author} {\bibinfo {author} {\bibfnamefont {P.~K.}\ \bibnamefont
  {Tien}}, \bibinfo {author} {\bibfnamefont {D.~P.}\ \bibnamefont {Schinke}}, \
  and\ \bibinfo {author} {\bibfnamefont {S.~L.}\ \bibnamefont {Blank}},\ }\href
  {\doibase 10.1063/1.1663724} {\bibfield  {journal} {\bibinfo  {journal}
  {Journal of Applied Physics}\ }\textbf {\bibinfo {volume} {45}},\ \bibinfo
  {pages} {3059} (\bibinfo {year} {1974})}\BibitemShut {NoStop}%
\bibitem [{\citenamefont {Ishida}\ \emph {et~al.}(2016)\citenamefont {Ishida},
  \citenamefont {Miura}, \citenamefont {Shoji}, \citenamefont {Mizumoto},
  \citenamefont {Nishiyama},\ and\ \citenamefont
  {Arai}}]{mo-switch-Ishida2016-jjap}%
  \BibitemOpen
  \bibfield  {author} {\bibinfo {author} {\bibfnamefont {E.}~\bibnamefont
  {Ishida}}, \bibinfo {author} {\bibfnamefont {K.}~\bibnamefont {Miura}},
  \bibinfo {author} {\bibfnamefont {Y.}~\bibnamefont {Shoji}}, \bibinfo
  {author} {\bibfnamefont {T.}~\bibnamefont {Mizumoto}}, \bibinfo {author}
  {\bibfnamefont {N.}~\bibnamefont {Nishiyama}}, \ and\ \bibinfo {author}
  {\bibfnamefont {S.}~\bibnamefont {Arai}},\ }\href {\doibase
  10.7567/JJAP.55.088002} {\bibfield  {journal} {\bibinfo  {journal} {Japanese
  Journal of Applied Physics}\ }\textbf {\bibinfo {volume} {55}},\ \bibinfo
  {pages} {088002} (\bibinfo {year} {2016})}\BibitemShut {NoStop}%
\bibitem [{\citenamefont {Shoji}\ and\ \citenamefont
  {Mizumoto}(2014)}]{mo-rev-Shoji2014-stam}%
  \BibitemOpen
  \bibfield  {author} {\bibinfo {author} {\bibfnamefont {Y.}~\bibnamefont
  {Shoji}}\ and\ \bibinfo {author} {\bibfnamefont {T.}~\bibnamefont
  {Mizumoto}},\ }\href {\doibase 10.1088/1468-6996/15/1/014602} {\bibfield
  {journal} {\bibinfo  {journal} {Sci. Technol. Adv. Mater.}\ }\textbf
  {\bibinfo {volume} {15}},\ \bibinfo {pages} {014602} (\bibinfo {year}
  {2014})}\BibitemShut {NoStop}%
\bibitem [{\citenamefont {Shoji}\ \emph {et~al.}(2016)\citenamefont {Shoji},
  \citenamefont {Miura},\ and\ \citenamefont {Mizumoto}}]{mo-rev-Shoji2016-jo}%
  \BibitemOpen
  \bibfield  {author} {\bibinfo {author} {\bibfnamefont {Y.}~\bibnamefont
  {Shoji}}, \bibinfo {author} {\bibfnamefont {K.}~\bibnamefont {Miura}}, \ and\
  \bibinfo {author} {\bibfnamefont {T.}~\bibnamefont {Mizumoto}},\ }\href
  {\doibase 10.1088/2040-8978/18/1/013001} {\bibfield  {journal} {\bibinfo
  {journal} {J. Opt.}\ }\textbf {\bibinfo {volume} {18}},\ \bibinfo {pages}
  {013001} (\bibinfo {year} {2016})}\BibitemShut {NoStop}%
\bibitem [{\citenamefont {Qiu}\ \emph {et~al.}(2011)\citenamefont {Qiu},
  \citenamefont {Wang},\ and\ \citenamefont
  {Solja{\v{c}}i{\'{c}}}}]{pc-Qiu2011-oe}%
  \BibitemOpen
  \bibfield  {author} {\bibinfo {author} {\bibfnamefont {W.}~\bibnamefont
  {Qiu}}, \bibinfo {author} {\bibfnamefont {Z.}~\bibnamefont {Wang}}, \ and\
  \bibinfo {author} {\bibfnamefont {M.}~\bibnamefont {Solja{\v{c}}i{\'{c}}}},\
  }\href {\doibase 10.1364/OE.19.022248} {\bibfield  {journal} {\bibinfo
  {journal} {Opt. Express}\ }\textbf {\bibinfo {volume} {19}},\ \bibinfo
  {pages} {22248} (\bibinfo {year} {2011})}\BibitemShut {NoStop}%
\bibitem [{\citenamefont {{\'{S}}migaj}\ \emph {et~al.}(2010)\citenamefont
  {{\'{S}}migaj}, \citenamefont {Romero-Vivas}, \citenamefont {Gralak},
  \citenamefont {Magdenko}, \citenamefont {Dagens},\ and\ \citenamefont
  {Vanwolleghem}}]{pc-circ-Smigaj2010-ol}%
  \BibitemOpen
  \bibfield  {author} {\bibinfo {author} {\bibfnamefont {W.}~\bibnamefont
  {{\'{S}}migaj}}, \bibinfo {author} {\bibfnamefont {J.}~\bibnamefont
  {Romero-Vivas}}, \bibinfo {author} {\bibfnamefont {B.}~\bibnamefont
  {Gralak}}, \bibinfo {author} {\bibfnamefont {L.}~\bibnamefont {Magdenko}},
  \bibinfo {author} {\bibfnamefont {B.}~\bibnamefont {Dagens}}, \ and\ \bibinfo
  {author} {\bibfnamefont {M.}~\bibnamefont {Vanwolleghem}},\ }\href {\doibase
  10.1364/OL.35.000568} {\bibfield  {journal} {\bibinfo  {journal} {Opt.
  Lett.}\ }\textbf {\bibinfo {volume} {35}},\ \bibinfo {pages} {568} (\bibinfo
  {year} {2010})}\BibitemShut {NoStop}%
\bibitem [{\citenamefont {Dmitriev}\ and\ \citenamefont
  {Kawakatsu}(2012)}]{pc-Dmitriev2012-ao}%
  \BibitemOpen
  \bibfield  {author} {\bibinfo {author} {\bibfnamefont {V.}~\bibnamefont
  {Dmitriev}}\ and\ \bibinfo {author} {\bibfnamefont {M.~N.}\ \bibnamefont
  {Kawakatsu}},\ }\href {\doibase 10.1364/AO.51.005917} {\bibfield  {journal}
  {\bibinfo  {journal} {Appl. Opt.}\ }\textbf {\bibinfo {volume} {51}},\
  \bibinfo {pages} {5917} (\bibinfo {year} {2012})}\BibitemShut {NoStop}%
\bibitem [{\citenamefont {Wang}\ and\ \citenamefont
  {Fan}(2005)}]{pc-circ-Wang2005-ol}%
  \BibitemOpen
  \bibfield  {author} {\bibinfo {author} {\bibfnamefont {Z.}~\bibnamefont
  {Wang}}\ and\ \bibinfo {author} {\bibfnamefont {S.}~\bibnamefont {Fan}},\
  }\href {\doibase 10.1364/OL.30.001989} {\bibfield  {journal} {\bibinfo
  {journal} {Opt. Lett.}\ }\textbf {\bibinfo {volume} {30}},\ \bibinfo {pages}
  {1989} (\bibinfo {year} {2005})}\BibitemShut {NoStop}%
\bibitem [{\citenamefont {Zhang}\ \emph {et~al.}(2013)\citenamefont {Zhang},
  \citenamefont {Yang}, \citenamefont {Chen}, \citenamefont {Li},\ and\
  \citenamefont {Xia}}]{pc-Zhang2013-olt}%
  \BibitemOpen
  \bibfield  {author} {\bibinfo {author} {\bibfnamefont {L.}~\bibnamefont
  {Zhang}}, \bibinfo {author} {\bibfnamefont {D.}~\bibnamefont {Yang}},
  \bibinfo {author} {\bibfnamefont {K.}~\bibnamefont {Chen}}, \bibinfo {author}
  {\bibfnamefont {T.}~\bibnamefont {Li}}, \ and\ \bibinfo {author}
  {\bibfnamefont {S.}~\bibnamefont {Xia}},\ }\href {\doibase
  10.1016/j.optlastec.2013.03.003} {\bibfield  {journal} {\bibinfo  {journal}
  {Optics and Laser Technology}\ }\textbf {\bibinfo {volume} {50}},\ \bibinfo
  {pages} {195} (\bibinfo {year} {2013})}\BibitemShut {NoStop}%
\bibitem [{\citenamefont {Fan}\ \emph {et~al.}(2011)\citenamefont {Fan},
  \citenamefont {Guo}, \citenamefont {Bai}, \citenamefont {Wang},\ and\
  \citenamefont {Chang}}]{pc-Fan2011-josab}%
  \BibitemOpen
  \bibfield  {author} {\bibinfo {author} {\bibfnamefont {F.}~\bibnamefont
  {Fan}}, \bibinfo {author} {\bibfnamefont {Z.}~\bibnamefont {Guo}}, \bibinfo
  {author} {\bibfnamefont {J.-J.}\ \bibnamefont {Bai}}, \bibinfo {author}
  {\bibfnamefont {X.-H.}\ \bibnamefont {Wang}}, \ and\ \bibinfo {author}
  {\bibfnamefont {S.-J.}\ \bibnamefont {Chang}},\ }\href {\doibase
  10.1364/JOSAB.28.000697} {\bibfield  {journal} {\bibinfo  {journal} {Journal
  of the Optical Society of America B}\ }\textbf {\bibinfo {volume} {28}},\
  \bibinfo {pages} {697} (\bibinfo {year} {2011})}\BibitemShut {NoStop}%
\bibitem [{\citenamefont {Wang}\ \emph {et~al.}(2009)\citenamefont {Wang},
  \citenamefont {Chong}, \citenamefont {Joannopoulos},\ and\ \citenamefont
  {Solja{\v{c}}i{\'{c}}}}]{pc-Wang2009-nature}%
  \BibitemOpen
  \bibfield  {author} {\bibinfo {author} {\bibfnamefont {Z.}~\bibnamefont
  {Wang}}, \bibinfo {author} {\bibfnamefont {Y.}~\bibnamefont {Chong}},
  \bibinfo {author} {\bibfnamefont {J.~D.}\ \bibnamefont {Joannopoulos}}, \
  and\ \bibinfo {author} {\bibfnamefont {M.}~\bibnamefont
  {Solja{\v{c}}i{\'{c}}}},\ }\href {\doibase 10.1038/nature08293} {\bibfield
  {journal} {\bibinfo  {journal} {Nature}\ }\textbf {\bibinfo {volume} {461}},\
  \bibinfo {pages} {772} (\bibinfo {year} {2009})}\BibitemShut {NoStop}%
\bibitem [{\citenamefont {Fan}\ \emph {et~al.}(2012)\citenamefont {Fan},
  \citenamefont {Chang}, \citenamefont {Niu}, \citenamefont {Hou},\ and\
  \citenamefont {Wang}}]{pc-circ-Fan2012-OptComm}%
  \BibitemOpen
  \bibfield  {author} {\bibinfo {author} {\bibfnamefont {F.}~\bibnamefont
  {Fan}}, \bibinfo {author} {\bibfnamefont {S.-J.}\ \bibnamefont {Chang}},
  \bibinfo {author} {\bibfnamefont {C.}~\bibnamefont {Niu}}, \bibinfo {author}
  {\bibfnamefont {Y.}~\bibnamefont {Hou}}, \ and\ \bibinfo {author}
  {\bibfnamefont {X.-H.}\ \bibnamefont {Wang}},\ }\href {\doibase
  10.1016/j.optcom.2012.05.044} {\bibfield  {journal} {\bibinfo  {journal}
  {Opt. Commun.}\ }\textbf {\bibinfo {volume} {285}},\ \bibinfo {pages} {3763}
  (\bibinfo {year} {2012})}\BibitemShut {NoStop}%
\bibitem [{\citenamefont {Chen}\ \emph
  {et~al.}(2015{\natexlab{a}})\citenamefont {Chen}, \citenamefont {Fan},
  \citenamefont {Wang}, \citenamefont {Wu}, \citenamefont {Zhang},\ and\
  \citenamefont {Chang}}]{mm-oi-Chen2015-oe}%
  \BibitemOpen
  \bibfield  {author} {\bibinfo {author} {\bibfnamefont {S.}~\bibnamefont
  {Chen}}, \bibinfo {author} {\bibfnamefont {F.}~\bibnamefont {Fan}}, \bibinfo
  {author} {\bibfnamefont {X.}~\bibnamefont {Wang}}, \bibinfo {author}
  {\bibfnamefont {P.}~\bibnamefont {Wu}}, \bibinfo {author} {\bibfnamefont
  {H.}~\bibnamefont {Zhang}}, \ and\ \bibinfo {author} {\bibfnamefont
  {S.}~\bibnamefont {Chang}},\ }\href {\doibase 10.1364/OE.23.001015}
  {\bibfield  {journal} {\bibinfo  {journal} {Optics Express}\ }\textbf
  {\bibinfo {volume} {23}},\ \bibinfo {pages} {1015} (\bibinfo {year}
  {2015}{\natexlab{a}})}\BibitemShut {NoStop}%
\bibitem [{\citenamefont {Chen}\ \emph
  {et~al.}(2015{\natexlab{b}})\citenamefont {Chen}, \citenamefont {Fan},
  \citenamefont {He}, \citenamefont {Chen},\ and\ \citenamefont
  {Chang}}]{mm-oi-Chen2015-ao}%
  \BibitemOpen
  \bibfield  {author} {\bibinfo {author} {\bibfnamefont {S.}~\bibnamefont
  {Chen}}, \bibinfo {author} {\bibfnamefont {F.}~\bibnamefont {Fan}}, \bibinfo
  {author} {\bibfnamefont {X.}~\bibnamefont {He}}, \bibinfo {author}
  {\bibfnamefont {M.}~\bibnamefont {Chen}}, \ and\ \bibinfo {author}
  {\bibfnamefont {S.}~\bibnamefont {Chang}},\ }\href {\doibase
  10.1364/AO.54.009177} {\bibfield  {journal} {\bibinfo  {journal} {Appl.
  Opt.}\ }\textbf {\bibinfo {volume} {54}},\ \bibinfo {pages} {9177} (\bibinfo
  {year} {2015}{\natexlab{b}})}\BibitemShut {NoStop}%
\bibitem [{\citenamefont {Ozbay}(2006)}]{spp-rev-Ozbay2006-science}%
  \BibitemOpen
  \bibfield  {author} {\bibinfo {author} {\bibfnamefont {E.}~\bibnamefont
  {Ozbay}},\ }\href {\doibase 10.1126/science.1114849} {\bibfield  {journal}
  {\bibinfo  {journal} {Science}\ }\textbf {\bibinfo {volume} {311}},\ \bibinfo
  {pages} {189} (\bibinfo {year} {2006})}\BibitemShut {NoStop}%
\bibitem [{\citenamefont {Stockman}(2011)}]{spp-rev-Stockman2011-pt}%
  \BibitemOpen
  \bibfield  {author} {\bibinfo {author} {\bibfnamefont {M.~I.}\ \bibnamefont
  {Stockman}},\ }\href {\doibase 10.1063/1.3554315} {\bibfield  {journal}
  {\bibinfo  {journal} {Phys. Today}\ }\textbf {\bibinfo {volume} {64}},\
  \bibinfo {pages} {39} (\bibinfo {year} {2011})}\BibitemShut {NoStop}%
\bibitem [{\citenamefont {Temnov}(2012)}]{spp-mf-rev-Temnov2012-NatPhot}%
  \BibitemOpen
  \bibfield  {author} {\bibinfo {author} {\bibfnamefont {V.~V.}\ \bibnamefont
  {Temnov}},\ }\href {\doibase 10.1038/nphoton.2012.220} {\bibfield  {journal}
  {\bibinfo  {journal} {Nat. Photonics}\ }\textbf {\bibinfo {volume} {6}},\
  \bibinfo {pages} {728} (\bibinfo {year} {2012})}\BibitemShut {NoStop}%
\bibitem [{\citenamefont {Armelles}\ \emph {et~al.}(2013)\citenamefont
  {Armelles}, \citenamefont {Cebollada}, \citenamefont
  {Garc{\'{i}}a-Mart{\'{i}}n},\ and\ \citenamefont
  {Gonz{\'{a}}lez}}]{spp-mf-rev-Armelles2013-aom}%
  \BibitemOpen
  \bibfield  {author} {\bibinfo {author} {\bibfnamefont {G.}~\bibnamefont
  {Armelles}}, \bibinfo {author} {\bibfnamefont {A.}~\bibnamefont {Cebollada}},
  \bibinfo {author} {\bibfnamefont {A.}~\bibnamefont
  {Garc{\'{i}}a-Mart{\'{i}}n}}, \ and\ \bibinfo {author} {\bibfnamefont
  {M.~U.}\ \bibnamefont {Gonz{\'{a}}lez}},\ }\href {\doibase
  10.1002/adom.201200011} {\bibfield  {journal} {\bibinfo  {journal} {Adv. Opt.
  Mater.}\ }\textbf {\bibinfo {volume} {1}},\ \bibinfo {pages} {10} (\bibinfo
  {year} {2013})}\BibitemShut {NoStop}%
\bibitem [{\citenamefont {Maksymov}(2015)}]{spp-mo-rev-Maksymov2015-nanomat}%
  \BibitemOpen
  \bibfield  {author} {\bibinfo {author} {\bibfnamefont {I.}~\bibnamefont
  {Maksymov}},\ }\href {\doibase 10.3390/nano5020577} {\bibfield  {journal}
  {\bibinfo  {journal} {Nanomaterials}\ }\textbf {\bibinfo {volume} {5}},\
  \bibinfo {pages} {577} (\bibinfo {year} {2015})}\BibitemShut {NoStop}%
\bibitem [{\citenamefont {Chiu}\ and\ \citenamefont
  {Quinn}(1972)}]{spp-Chiu1972-ncb}%
  \BibitemOpen
  \bibfield  {author} {\bibinfo {author} {\bibfnamefont {K.~W.}\ \bibnamefont
  {Chiu}}\ and\ \bibinfo {author} {\bibfnamefont {J.~J.}\ \bibnamefont
  {Quinn}},\ }\href {\doibase 10.1007/BF02911404} {\bibfield  {journal}
  {\bibinfo  {journal} {Il Nuovo Cimento B}\ }\textbf {\bibinfo {volume}
  {10}},\ \bibinfo {pages} {1} (\bibinfo {year} {1972})}\BibitemShut {NoStop}%
\bibitem [{\citenamefont {Temnov}\ \emph {et~al.}(2010)\citenamefont {Temnov},
  \citenamefont {Armelles}, \citenamefont {Woggon}, \citenamefont {Guzatov},
  \citenamefont {Cebollada}, \citenamefont {Garcia-Martin}, \citenamefont
  {Garcia-Martin}, \citenamefont {Thomay}, \citenamefont {Leitenstorfer},\ and\
  \citenamefont {Bratschitsch}}]{spp-mo-Temnov2010-NatPhot}%
  \BibitemOpen
  \bibfield  {author} {\bibinfo {author} {\bibfnamefont {V.~V.}\ \bibnamefont
  {Temnov}}, \bibinfo {author} {\bibfnamefont {G.}~\bibnamefont {Armelles}},
  \bibinfo {author} {\bibfnamefont {U.}~\bibnamefont {Woggon}}, \bibinfo
  {author} {\bibfnamefont {D.}~\bibnamefont {Guzatov}}, \bibinfo {author}
  {\bibfnamefont {A.}~\bibnamefont {Cebollada}}, \bibinfo {author}
  {\bibfnamefont {A.}~\bibnamefont {Garcia-Martin}}, \bibinfo {author}
  {\bibfnamefont {J.-M.}\ \bibnamefont {Garcia-Martin}}, \bibinfo {author}
  {\bibfnamefont {T.}~\bibnamefont {Thomay}}, \bibinfo {author} {\bibfnamefont
  {A.}~\bibnamefont {Leitenstorfer}}, \ and\ \bibinfo {author} {\bibfnamefont
  {R.}~\bibnamefont {Bratschitsch}},\ }\href {\doibase
  10.1038/nphoton.2009.265} {\bibfield  {journal} {\bibinfo  {journal} {Nat.
  Photonics}\ }\textbf {\bibinfo {volume} {4}},\ \bibinfo {pages} {107}
  (\bibinfo {year} {2010})}\BibitemShut {NoStop}%
\bibitem [{\citenamefont {Fan}\ \emph {et~al.}(2013)\citenamefont {Fan},
  \citenamefont {Chen}, \citenamefont {Lin}, \citenamefont {Miao},
  \citenamefont {Chang}, \citenamefont {Liu}, \citenamefont {Wang},\ and\
  \citenamefont {Lin}}]{spp-mf-Fei2013-apl}%
  \BibitemOpen
  \bibfield  {author} {\bibinfo {author} {\bibfnamefont {F.}~\bibnamefont
  {Fan}}, \bibinfo {author} {\bibfnamefont {S.}~\bibnamefont {Chen}}, \bibinfo
  {author} {\bibfnamefont {W.}~\bibnamefont {Lin}}, \bibinfo {author}
  {\bibfnamefont {Y.-P.}\ \bibnamefont {Miao}}, \bibinfo {author}
  {\bibfnamefont {S.-J.}\ \bibnamefont {Chang}}, \bibinfo {author}
  {\bibfnamefont {B.}~\bibnamefont {Liu}}, \bibinfo {author} {\bibfnamefont
  {X.-H.}\ \bibnamefont {Wang}}, \ and\ \bibinfo {author} {\bibfnamefont
  {L.}~\bibnamefont {Lin}},\ }\href {\doibase 10.1063/1.4826454} {\bibfield
  {journal} {\bibinfo  {journal} {Appl. Phys. Lett.}\ }\textbf {\bibinfo
  {volume} {103}},\ \bibinfo {pages} {161115} (\bibinfo {year}
  {2013})}\BibitemShut {NoStop}%
\bibitem [{\citenamefont {Fan}\ \emph {et~al.}(2016)\citenamefont {Fan},
  \citenamefont {Xu}, \citenamefont {Wang},\ and\ \citenamefont
  {Chang}}]{spp-mo-Fan2016-oe}%
  \BibitemOpen
  \bibfield  {author} {\bibinfo {author} {\bibfnamefont {F.}~\bibnamefont
  {Fan}}, \bibinfo {author} {\bibfnamefont {S.-T.}\ \bibnamefont {Xu}},
  \bibinfo {author} {\bibfnamefont {X.-H.}\ \bibnamefont {Wang}}, \ and\
  \bibinfo {author} {\bibfnamefont {S.-J.}\ \bibnamefont {Chang}},\ }\href
  {\doibase 10.1364/OE.24.026431} {\bibfield  {journal} {\bibinfo  {journal}
  {Opt. Express}\ }\textbf {\bibinfo {volume} {24}},\ \bibinfo {pages} {26431}
  (\bibinfo {year} {2016})}\BibitemShut {NoStop}%
\bibitem [{\citenamefont {Firby}\ and\ \citenamefont
  {Elezzabi}(2015)}]{spp-nrps-Firby2015-optica}%
  \BibitemOpen
  \bibfield  {author} {\bibinfo {author} {\bibfnamefont {C.~J.}\ \bibnamefont
  {Firby}}\ and\ \bibinfo {author} {\bibfnamefont {A.~Y.}\ \bibnamefont
  {Elezzabi}},\ }\href {\doibase 10.1364/OPTICA.2.000598} {\bibfield  {journal}
  {\bibinfo  {journal} {Optica}\ }\textbf {\bibinfo {volume} {2}},\ \bibinfo
  {pages} {598} (\bibinfo {year} {2015})}\BibitemShut {NoStop}%
\bibitem [{\citenamefont {Davoyan}\ and\ \citenamefont
  {Engheta}(2013)}]{spp-circ-Davoyan2013-njp}%
  \BibitemOpen
  \bibfield  {author} {\bibinfo {author} {\bibfnamefont {A.~R.}\ \bibnamefont
  {Davoyan}}\ and\ \bibinfo {author} {\bibfnamefont {N.}~\bibnamefont
  {Engheta}},\ }\href {\doibase 10.1088/1367-2630/15/8/083054} {\bibfield
  {journal} {\bibinfo  {journal} {New J. Phys.}\ }\textbf {\bibinfo {volume}
  {15}},\ \bibinfo {pages} {083054} (\bibinfo {year} {2013})}\BibitemShut
  {NoStop}%
\bibitem [{\citenamefont {Firby}\ and\ \citenamefont
  {Elezzabi}(2016)}]{spp-oi-Firby2016-ol}%
  \BibitemOpen
  \bibfield  {author} {\bibinfo {author} {\bibfnamefont {C.~J.}\ \bibnamefont
  {Firby}}\ and\ \bibinfo {author} {\bibfnamefont {A.~Y.}\ \bibnamefont
  {Elezzabi}},\ }\href {\doibase 10.1364/OL.41.000563} {\bibfield  {journal}
  {\bibinfo  {journal} {Opt. Lett.}\ }\textbf {\bibinfo {volume} {41}},\
  \bibinfo {pages} {563} (\bibinfo {year} {2016})}\BibitemShut {NoStop}%
\bibitem [{\citenamefont {Chin}\ \emph {et~al.}(2013)\citenamefont {Chin},
  \citenamefont {Steinle}, \citenamefont {Wehlus}, \citenamefont {Dregely},
  \citenamefont {Weiss}, \citenamefont {Belotelov}, \citenamefont {Stritzker},\
  and\ \citenamefont {Giessen}}]{spp-fr-Chin2013-NatComm}%
  \BibitemOpen
  \bibfield  {author} {\bibinfo {author} {\bibfnamefont {J.~Y.}\ \bibnamefont
  {Chin}}, \bibinfo {author} {\bibfnamefont {T.}~\bibnamefont {Steinle}},
  \bibinfo {author} {\bibfnamefont {T.}~\bibnamefont {Wehlus}}, \bibinfo
  {author} {\bibfnamefont {D.}~\bibnamefont {Dregely}}, \bibinfo {author}
  {\bibfnamefont {T.}~\bibnamefont {Weiss}}, \bibinfo {author} {\bibfnamefont
  {V.~I.}\ \bibnamefont {Belotelov}}, \bibinfo {author} {\bibfnamefont
  {B.}~\bibnamefont {Stritzker}}, \ and\ \bibinfo {author} {\bibfnamefont
  {H.}~\bibnamefont {Giessen}},\ }\href {\doibase 10.1038/ncomms2609}
  {\bibfield  {journal} {\bibinfo  {journal} {Nat. Commun.}\ }\textbf {\bibinfo
  {volume} {4}},\ \bibinfo {pages} {1599} (\bibinfo {year} {2013})}\BibitemShut
  {NoStop}%
\bibitem [{\citenamefont {Caballero}\ \emph {et~al.}(2015)\citenamefont
  {Caballero}, \citenamefont {Garc{\'{i}}a-Mart{\'{i}}n},\ and\ \citenamefont
  {Cuevas}}]{spp-fr-Caballero2015-oe}%
  \BibitemOpen
  \bibfield  {author} {\bibinfo {author} {\bibfnamefont {B.}~\bibnamefont
  {Caballero}}, \bibinfo {author} {\bibfnamefont {A.}~\bibnamefont
  {Garc{\'{i}}a-Mart{\'{i}}n}}, \ and\ \bibinfo {author} {\bibfnamefont
  {J.~C.}\ \bibnamefont {Cuevas}},\ }\href {\doibase 10.1364/OE.23.022238}
  {\bibfield  {journal} {\bibinfo  {journal} {Opt. Express}\ }\textbf {\bibinfo
  {volume} {23}},\ \bibinfo {pages} {22238} (\bibinfo {year}
  {2015})}\BibitemShut {NoStop}%
\bibitem [{\citenamefont {Belotelov}\ \emph {et~al.}(2011)\citenamefont
  {Belotelov}, \citenamefont {Akimov}, \citenamefont {Pohl}, \citenamefont
  {Kotov}, \citenamefont {Kasture}, \citenamefont {Vengurlekar}, \citenamefont
  {Gopal}, \citenamefont {Yakovlev}, \citenamefont {Zvezdin},\ and\
  \citenamefont {Bayer}}]{spp-mo-Belotelov2011-NatNanotech}%
  \BibitemOpen
  \bibfield  {author} {\bibinfo {author} {\bibfnamefont {V.~I.}\ \bibnamefont
  {Belotelov}}, \bibinfo {author} {\bibfnamefont {I.~A.}\ \bibnamefont
  {Akimov}}, \bibinfo {author} {\bibfnamefont {M.}~\bibnamefont {Pohl}},
  \bibinfo {author} {\bibfnamefont {V.~A.}\ \bibnamefont {Kotov}}, \bibinfo
  {author} {\bibfnamefont {S.}~\bibnamefont {Kasture}}, \bibinfo {author}
  {\bibfnamefont {A.~S.}\ \bibnamefont {Vengurlekar}}, \bibinfo {author}
  {\bibfnamefont {A.~V.}\ \bibnamefont {Gopal}}, \bibinfo {author}
  {\bibfnamefont {D.~R.}\ \bibnamefont {Yakovlev}}, \bibinfo {author}
  {\bibfnamefont {A.~K.}\ \bibnamefont {Zvezdin}}, \ and\ \bibinfo {author}
  {\bibfnamefont {M.}~\bibnamefont {Bayer}},\ }\href {\doibase
  10.1038/nnano.2011.54} {\bibfield  {journal} {\bibinfo  {journal} {Nat.
  Nanotechnol.}\ }\textbf {\bibinfo {volume} {6}},\ \bibinfo {pages} {370}
  (\bibinfo {year} {2011})}\BibitemShut {NoStop}%
\bibitem [{\citenamefont {Kreilkamp}\ \emph {et~al.}(2013)\citenamefont
  {Kreilkamp}, \citenamefont {Belotelov}, \citenamefont {Chin}, \citenamefont
  {Neutzner}, \citenamefont {Dregely}, \citenamefont {Wehlus}, \citenamefont
  {Akimov}, \citenamefont {Bayer}, \citenamefont {Stritzker},\ and\
  \citenamefont {Giessen}}]{spp-mf-Kleilkamp2013-prx}%
  \BibitemOpen
  \bibfield  {author} {\bibinfo {author} {\bibfnamefont {L.~E.}\ \bibnamefont
  {Kreilkamp}}, \bibinfo {author} {\bibfnamefont {V.~I.}\ \bibnamefont
  {Belotelov}}, \bibinfo {author} {\bibfnamefont {J.~Y.}\ \bibnamefont {Chin}},
  \bibinfo {author} {\bibfnamefont {S.}~\bibnamefont {Neutzner}}, \bibinfo
  {author} {\bibfnamefont {D.}~\bibnamefont {Dregely}}, \bibinfo {author}
  {\bibfnamefont {T.}~\bibnamefont {Wehlus}}, \bibinfo {author} {\bibfnamefont
  {I.~A.}\ \bibnamefont {Akimov}}, \bibinfo {author} {\bibfnamefont
  {M.}~\bibnamefont {Bayer}}, \bibinfo {author} {\bibfnamefont
  {B.}~\bibnamefont {Stritzker}}, \ and\ \bibinfo {author} {\bibfnamefont
  {H.}~\bibnamefont {Giessen}},\ }\href {\doibase 10.1103/PhysRevX.3.041019}
  {\bibfield  {journal} {\bibinfo  {journal} {Phys. Rev. X}\ }\textbf {\bibinfo
  {volume} {3}},\ \bibinfo {pages} {041019} (\bibinfo {year}
  {2013})}\BibitemShut {NoStop}%
\bibitem [{\citenamefont {Xiao}\ \emph {et~al.}(2016)\citenamefont {Xiao},
  \citenamefont {Zhu}, \citenamefont {Li},\ and\ \citenamefont
  {Mortensen}}]{spp-gr-rev-Xiao2016-fp}%
  \BibitemOpen
  \bibfield  {author} {\bibinfo {author} {\bibfnamefont {S.}~\bibnamefont
  {Xiao}}, \bibinfo {author} {\bibfnamefont {X.}~\bibnamefont {Zhu}}, \bibinfo
  {author} {\bibfnamefont {B.-H.}\ \bibnamefont {Li}}, \ and\ \bibinfo {author}
  {\bibfnamefont {N.~A.}\ \bibnamefont {Mortensen}},\ }\href {\doibase
  10.1007/s11467-016-0551-z} {\bibfield  {journal} {\bibinfo  {journal}
  {Frontiers of Physics}\ }\textbf {\bibinfo {volume} {11}},\ \bibinfo {pages}
  {117801} (\bibinfo {year} {2016})}\BibitemShut {NoStop}%
\bibitem [{\citenamefont {{Garc{\'{i}}a de
  Abajo}}(2014)}]{spp-gr-rev-deAbajo2014-ACSPhot}%
  \BibitemOpen
  \bibfield  {author} {\bibinfo {author} {\bibfnamefont {F.~J.}\ \bibnamefont
  {{Garc{\'{i}}a de Abajo}}},\ }\href {\doibase 10.1021/ph400147y} {\bibfield
  {journal} {\bibinfo  {journal} {ACS Photonics}\ }\textbf {\bibinfo {volume}
  {1}},\ \bibinfo {pages} {135} (\bibinfo {year} {2014})}\BibitemShut {NoStop}%
\bibitem [{\citenamefont {Chen}\ \emph {et~al.}(2017)\citenamefont {Chen},
  \citenamefont {Argyropoulos}, \citenamefont {Farhat},\ and\ \citenamefont
  {Gomez-Diaz}}]{spp-gr-rev-Chen2017-nanophot}%
  \BibitemOpen
  \bibfield  {author} {\bibinfo {author} {\bibfnamefont {P.-Y.}\ \bibnamefont
  {Chen}}, \bibinfo {author} {\bibfnamefont {C.}~\bibnamefont {Argyropoulos}},
  \bibinfo {author} {\bibfnamefont {M.}~\bibnamefont {Farhat}}, \ and\ \bibinfo
  {author} {\bibfnamefont {J.~S.}\ \bibnamefont {Gomez-Diaz}},\ }\href
  {\doibase 10.1515/nanoph-2016-0137} {\bibfield  {journal} {\bibinfo
  {journal} {Nanophotonics}\ }\textbf {\bibinfo {volume} {6}},\ \bibinfo
  {pages} {1239} (\bibinfo {year} {2017})}\BibitemShut {NoStop}%
\bibitem [{\citenamefont {Bludov}\ \emph {et~al.}(2013)\citenamefont {Bludov},
  \citenamefont {Ferreira}, \citenamefont {Peres},\ and\ \citenamefont
  {Vasilevskiy}}]{spp-gr-rev-Bludov2013-ijmfb}%
  \BibitemOpen
  \bibfield  {author} {\bibinfo {author} {\bibfnamefont {Y.~V.}\ \bibnamefont
  {Bludov}}, \bibinfo {author} {\bibfnamefont {A.}~\bibnamefont {Ferreira}},
  \bibinfo {author} {\bibfnamefont {N.~M.~R.}\ \bibnamefont {Peres}}, \ and\
  \bibinfo {author} {\bibfnamefont {M.~I.}\ \bibnamefont {Vasilevskiy}},\
  }\href {\doibase 10.1142/S0217979213410014} {\bibfield  {journal} {\bibinfo
  {journal} {International Journal of Modern Physics B}\ }\textbf {\bibinfo
  {volume} {27}},\ \bibinfo {pages} {1341001} (\bibinfo {year}
  {2013})}\BibitemShut {NoStop}%
\bibitem [{\citenamefont {Low}\ and\ \citenamefont
  {Avouris}(2014)}]{spp-gr-rev-Low2014-ACSNano}%
  \BibitemOpen
  \bibfield  {author} {\bibinfo {author} {\bibfnamefont {T.}~\bibnamefont
  {Low}}\ and\ \bibinfo {author} {\bibfnamefont {P.}~\bibnamefont {Avouris}},\
  }\href {\doibase 10.1021/nn406627u} {\bibfield  {journal} {\bibinfo
  {journal} {ACS Nano}\ }\textbf {\bibinfo {volume} {8}},\ \bibinfo {pages}
  {1086} (\bibinfo {year} {2014})},\ \Eprint {http://arxiv.org/abs/1403.2799}
  {1403.2799} \BibitemShut {NoStop}%
\bibitem [{\citenamefont {Koppens}\ \emph {et~al.}(2011)\citenamefont
  {Koppens}, \citenamefont {Chang},\ and\ \citenamefont {{Garc\'{i}a de
  Abajo}}}]{spp-gr-rev-Koppens2011-nl}%
  \BibitemOpen
  \bibfield  {author} {\bibinfo {author} {\bibfnamefont {F.~H.~L.}\
  \bibnamefont {Koppens}}, \bibinfo {author} {\bibfnamefont {D.~E.}\
  \bibnamefont {Chang}}, \ and\ \bibinfo {author} {\bibfnamefont {F.~J.}\
  \bibnamefont {{Garc\'{i}a de Abajo}}},\ }\href {\doibase 10.1021/nl201771h}
  {\bibfield  {journal} {\bibinfo  {journal} {Nano Lett.}\ }\textbf {\bibinfo
  {volume} {11}},\ \bibinfo {pages} {3370} (\bibinfo {year}
  {2011})}\BibitemShut {NoStop}%
\bibitem [{\citenamefont {Nikitin}\ \emph {et~al.}(2011)\citenamefont
  {Nikitin}, \citenamefont {Guinea}, \citenamefont {Garc{\'{i}}a-Vidal},\ and\
  \citenamefont {Mart{\'{i}}n-Moreno}}]{spp-gr-Nikitin2011-prb}%
  \BibitemOpen
  \bibfield  {author} {\bibinfo {author} {\bibfnamefont {A.~Y.}\ \bibnamefont
  {Nikitin}}, \bibinfo {author} {\bibfnamefont {F.}~\bibnamefont {Guinea}},
  \bibinfo {author} {\bibfnamefont {F.~J.}\ \bibnamefont {Garc{\'{i}}a-Vidal}},
  \ and\ \bibinfo {author} {\bibfnamefont {L.}~\bibnamefont
  {Mart{\'{i}}n-Moreno}},\ }\href {\doibase 10.1103/PhysRevB.84.195446}
  {\bibfield  {journal} {\bibinfo  {journal} {Phys. Rev. B}\ }\textbf {\bibinfo
  {volume} {84}},\ \bibinfo {pages} {195446} (\bibinfo {year}
  {2011})}\BibitemShut {NoStop}%
\bibitem [{\citenamefont {Novoselov}\ \emph {et~al.}(2005)\citenamefont
  {Novoselov}, \citenamefont {Geim}, \citenamefont {Morozov}, \citenamefont
  {Jiang}, \citenamefont {Katsnelson}, \citenamefont {Grigorieva},
  \citenamefont {Dubonos},\ and\ \citenamefont
  {Firsov}}]{gr-qhe-Novoselov2005-nature}%
  \BibitemOpen
  \bibfield  {author} {\bibinfo {author} {\bibfnamefont {K.~S.}\ \bibnamefont
  {Novoselov}}, \bibinfo {author} {\bibfnamefont {A.~K.}\ \bibnamefont {Geim}},
  \bibinfo {author} {\bibfnamefont {S.~V.}\ \bibnamefont {Morozov}}, \bibinfo
  {author} {\bibfnamefont {D.}~\bibnamefont {Jiang}}, \bibinfo {author}
  {\bibfnamefont {M.~I.}\ \bibnamefont {Katsnelson}}, \bibinfo {author}
  {\bibfnamefont {I.~V.}\ \bibnamefont {Grigorieva}}, \bibinfo {author}
  {\bibfnamefont {S.~V.}\ \bibnamefont {Dubonos}}, \ and\ \bibinfo {author}
  {\bibfnamefont {A.~A.}\ \bibnamefont {Firsov}},\ }\href {\doibase
  10.1038/nature04233} {\bibfield  {journal} {\bibinfo  {journal} {Nature}\
  }\textbf {\bibinfo {volume} {438}},\ \bibinfo {pages} {197} (\bibinfo {year}
  {2005})}\BibitemShut {NoStop}%
\bibitem [{\citenamefont {Zhang}\ \emph {et~al.}(2005)\citenamefont {Zhang},
  \citenamefont {Tan}, \citenamefont {Stormer},\ and\ \citenamefont
  {Kim}}]{gr-qhe-Zhang2005-nature}%
  \BibitemOpen
  \bibfield  {author} {\bibinfo {author} {\bibfnamefont {Y.}~\bibnamefont
  {Zhang}}, \bibinfo {author} {\bibfnamefont {Y.-W.}\ \bibnamefont {Tan}},
  \bibinfo {author} {\bibfnamefont {H.~L.}\ \bibnamefont {Stormer}}, \ and\
  \bibinfo {author} {\bibfnamefont {P.}~\bibnamefont {Kim}},\ }\href {\doibase
  10.1038/nature04235} {\bibfield  {journal} {\bibinfo  {journal} {Nature}\
  }\textbf {\bibinfo {volume} {438}},\ \bibinfo {pages} {201} (\bibinfo {year}
  {2005})}\BibitemShut {NoStop}%
\bibitem [{\citenamefont {Tamagnone}\ \emph {et~al.}(2014)\citenamefont
  {Tamagnone}, \citenamefont {Fallahi}, \citenamefont {Mosig},\ and\
  \citenamefont {Perruisseau-Carrier}}]{gr-mf-oi-Tamagnone2014-NatPhot}%
  \BibitemOpen
  \bibfield  {author} {\bibinfo {author} {\bibfnamefont {M.}~\bibnamefont
  {Tamagnone}}, \bibinfo {author} {\bibfnamefont {A.}~\bibnamefont {Fallahi}},
  \bibinfo {author} {\bibfnamefont {J.~R.}\ \bibnamefont {Mosig}}, \ and\
  \bibinfo {author} {\bibfnamefont {J.}~\bibnamefont {Perruisseau-Carrier}},\
  }\href {\doibase 10.1038/nphoton.2014.109} {\bibfield  {journal} {\bibinfo
  {journal} {Nat. Photonics}\ }\textbf {\bibinfo {volume} {8}},\ \bibinfo
  {pages} {556} (\bibinfo {year} {2014})}\BibitemShut {NoStop}%
\bibitem [{\citenamefont {Tamagnone}\ \emph {et~al.}(2016)\citenamefont
  {Tamagnone}, \citenamefont {Moldovan}, \citenamefont {Poumirol},
  \citenamefont {Kuzmenko}, \citenamefont {Ionescu}, \citenamefont {Mosig},\
  and\ \citenamefont {Perruisseau-Carrier}}]{gr-mf-oi-Tamagnone2016-NatComm}%
  \BibitemOpen
  \bibfield  {author} {\bibinfo {author} {\bibfnamefont {M.}~\bibnamefont
  {Tamagnone}}, \bibinfo {author} {\bibfnamefont {C.}~\bibnamefont {Moldovan}},
  \bibinfo {author} {\bibfnamefont {J.-M.}\ \bibnamefont {Poumirol}}, \bibinfo
  {author} {\bibfnamefont {A.~B.}\ \bibnamefont {Kuzmenko}}, \bibinfo {author}
  {\bibfnamefont {A.~M.}\ \bibnamefont {Ionescu}}, \bibinfo {author}
  {\bibfnamefont {J.~R.}\ \bibnamefont {Mosig}}, \ and\ \bibinfo {author}
  {\bibfnamefont {J.}~\bibnamefont {Perruisseau-Carrier}},\ }\href {\doibase
  10.1038/ncomms11216} {\bibfield  {journal} {\bibinfo  {journal} {Nat.
  Commun.}\ }\textbf {\bibinfo {volume} {7}},\ \bibinfo {pages} {11216}
  (\bibinfo {year} {2016})}\BibitemShut {NoStop}%
\bibitem [{\citenamefont {Shimano}\ \emph {et~al.}(2013)\citenamefont
  {Shimano}, \citenamefont {Yumoto}, \citenamefont {Yoo}, \citenamefont
  {Matsunaga}, \citenamefont {Tanabe}, \citenamefont {Hibino}, \citenamefont
  {Morimoto},\ and\ \citenamefont {Aoki}}]{gr-mf-Shimano2013-NatComm}%
  \BibitemOpen
  \bibfield  {author} {\bibinfo {author} {\bibfnamefont {R.}~\bibnamefont
  {Shimano}}, \bibinfo {author} {\bibfnamefont {G.}~\bibnamefont {Yumoto}},
  \bibinfo {author} {\bibfnamefont {J.~Y.}\ \bibnamefont {Yoo}}, \bibinfo
  {author} {\bibfnamefont {R.}~\bibnamefont {Matsunaga}}, \bibinfo {author}
  {\bibfnamefont {S.}~\bibnamefont {Tanabe}}, \bibinfo {author} {\bibfnamefont
  {H.}~\bibnamefont {Hibino}}, \bibinfo {author} {\bibfnamefont
  {T.}~\bibnamefont {Morimoto}}, \ and\ \bibinfo {author} {\bibfnamefont
  {H.}~\bibnamefont {Aoki}},\ }\href {\doibase 10.1038/ncomms2866} {\bibfield
  {journal} {\bibinfo  {journal} {Nat. Commun.}\ }\textbf {\bibinfo {volume}
  {4}},\ \bibinfo {pages} {1841} (\bibinfo {year} {2013})}\BibitemShut
  {NoStop}%
\bibitem [{\citenamefont {Crassee}\ \emph {et~al.}(2010)\citenamefont
  {Crassee}, \citenamefont {Levallois}, \citenamefont {Walter}, \citenamefont
  {Ostler}, \citenamefont {Bostwick}, \citenamefont {Rotenberg}, \citenamefont
  {Seyller}, \citenamefont {van~der Marel},\ and\ \citenamefont
  {Kuzmenko}}]{gr-mf-Crassee2010-NatPhys}%
  \BibitemOpen
  \bibfield  {author} {\bibinfo {author} {\bibfnamefont {I.}~\bibnamefont
  {Crassee}}, \bibinfo {author} {\bibfnamefont {J.}~\bibnamefont {Levallois}},
  \bibinfo {author} {\bibfnamefont {A.~L.}\ \bibnamefont {Walter}}, \bibinfo
  {author} {\bibfnamefont {M.}~\bibnamefont {Ostler}}, \bibinfo {author}
  {\bibfnamefont {A.}~\bibnamefont {Bostwick}}, \bibinfo {author}
  {\bibfnamefont {E.}~\bibnamefont {Rotenberg}}, \bibinfo {author}
  {\bibfnamefont {T.}~\bibnamefont {Seyller}}, \bibinfo {author} {\bibfnamefont
  {D.}~\bibnamefont {van~der Marel}}, \ and\ \bibinfo {author} {\bibfnamefont
  {A.~B.}\ \bibnamefont {Kuzmenko}},\ }\href {\doibase 10.1038/nphys1816}
  {\bibfield  {journal} {\bibinfo  {journal} {Nat. Phys.}\ }\textbf {\bibinfo
  {volume} {7}},\ \bibinfo {pages} {48} (\bibinfo {year} {2010})}\BibitemShut
  {NoStop}%
\bibitem [{\citenamefont {Sounas}\ \emph {et~al.}(2013)\citenamefont {Sounas},
  \citenamefont {Skulason}, \citenamefont {Nguyen}, \citenamefont {Guermoune},
  \citenamefont {Siaj}, \citenamefont {Szkopek},\ and\ \citenamefont
  {Caloz}}]{gr-fr-Sounas2013-apl}%
  \BibitemOpen
  \bibfield  {author} {\bibinfo {author} {\bibfnamefont {D.~L.}\ \bibnamefont
  {Sounas}}, \bibinfo {author} {\bibfnamefont {H.~S.}\ \bibnamefont
  {Skulason}}, \bibinfo {author} {\bibfnamefont {H.~V.}\ \bibnamefont
  {Nguyen}}, \bibinfo {author} {\bibfnamefont {A.}~\bibnamefont {Guermoune}},
  \bibinfo {author} {\bibfnamefont {M.}~\bibnamefont {Siaj}}, \bibinfo {author}
  {\bibfnamefont {T.}~\bibnamefont {Szkopek}}, \ and\ \bibinfo {author}
  {\bibfnamefont {C.}~\bibnamefont {Caloz}},\ }\href {\doibase
  10.1063/1.4804437} {\bibfield  {journal} {\bibinfo  {journal} {Appl. Phys.
  Lett.}\ }\textbf {\bibinfo {volume} {102}},\ \bibinfo {pages} {191901}
  (\bibinfo {year} {2013})}\BibitemShut {NoStop}%
\bibitem [{\citenamefont {Berman}\ \emph {et~al.}(2008)\citenamefont {Berman},
  \citenamefont {Gumbs},\ and\ \citenamefont
  {Lozovik}}]{spp-mf-gr-Berman2008-prb}%
  \BibitemOpen
  \bibfield  {author} {\bibinfo {author} {\bibfnamefont {O.~L.}\ \bibnamefont
  {Berman}}, \bibinfo {author} {\bibfnamefont {G.}~\bibnamefont {Gumbs}}, \
  and\ \bibinfo {author} {\bibfnamefont {Y.~E.}\ \bibnamefont {Lozovik}},\
  }\href {\doibase 10.1103/PhysRevB.78.085401} {\bibfield  {journal} {\bibinfo
  {journal} {Phys. Rev. B}\ }\textbf {\bibinfo {volume} {78}},\ \bibinfo
  {pages} {085401} (\bibinfo {year} {2008})}\BibitemShut {NoStop}%
\bibitem [{\citenamefont {Rold{\'{a}}n}\ \emph {et~al.}(2009)\citenamefont
  {Rold{\'{a}}n}, \citenamefont {Fuchs},\ and\ \citenamefont
  {Goerbig}}]{spp-mf-gr-Roldan2009-prb}%
  \BibitemOpen
  \bibfield  {author} {\bibinfo {author} {\bibfnamefont {R.}~\bibnamefont
  {Rold{\'{a}}n}}, \bibinfo {author} {\bibfnamefont {J.-N.}\ \bibnamefont
  {Fuchs}}, \ and\ \bibinfo {author} {\bibfnamefont {M.~O.}\ \bibnamefont
  {Goerbig}},\ }\href {\doibase 10.1103/PhysRevB.80.085408} {\bibfield
  {journal} {\bibinfo  {journal} {Phys. Rev. B}\ }\textbf {\bibinfo {volume}
  {80}},\ \bibinfo {pages} {085408} (\bibinfo {year} {2009})}\BibitemShut
  {NoStop}%
\bibitem [{\citenamefont {Crassee}\ \emph {et~al.}(2012)\citenamefont
  {Crassee}, \citenamefont {Orlita}, \citenamefont {Potemski}, \citenamefont
  {Walter}, \citenamefont {Ostler}, \citenamefont {Seyller}, \citenamefont
  {Gaponenko}, \citenamefont {Chen},\ and\ \citenamefont
  {Kuzmenko}}]{spp-mf-gr-Crassee2012-nl}%
  \BibitemOpen
  \bibfield  {author} {\bibinfo {author} {\bibfnamefont {I.}~\bibnamefont
  {Crassee}}, \bibinfo {author} {\bibfnamefont {M.}~\bibnamefont {Orlita}},
  \bibinfo {author} {\bibfnamefont {M.}~\bibnamefont {Potemski}}, \bibinfo
  {author} {\bibfnamefont {A.~L.}\ \bibnamefont {Walter}}, \bibinfo {author}
  {\bibfnamefont {M.}~\bibnamefont {Ostler}}, \bibinfo {author} {\bibfnamefont
  {T.}~\bibnamefont {Seyller}}, \bibinfo {author} {\bibfnamefont
  {I.}~\bibnamefont {Gaponenko}}, \bibinfo {author} {\bibfnamefont
  {J.}~\bibnamefont {Chen}}, \ and\ \bibinfo {author} {\bibfnamefont {A.~B.}\
  \bibnamefont {Kuzmenko}},\ }\href {\doibase 10.1021/nl300572y} {\bibfield
  {journal} {\bibinfo  {journal} {Nano Lett.}\ }\textbf {\bibinfo {volume}
  {12}},\ \bibinfo {pages} {2470} (\bibinfo {year} {2012})}\BibitemShut
  {NoStop}%
\bibitem [{\citenamefont {Yan}\ \emph {et~al.}(2012)\citenamefont {Yan},
  \citenamefont {Li}, \citenamefont {Li}, \citenamefont {Zhu}, \citenamefont
  {Avouris},\ and\ \citenamefont {Xia}}]{spp-mf-gr-Yan2012-nl}%
  \BibitemOpen
  \bibfield  {author} {\bibinfo {author} {\bibfnamefont {H.}~\bibnamefont
  {Yan}}, \bibinfo {author} {\bibfnamefont {Z.}~\bibnamefont {Li}}, \bibinfo
  {author} {\bibfnamefont {X.}~\bibnamefont {Li}}, \bibinfo {author}
  {\bibfnamefont {W.}~\bibnamefont {Zhu}}, \bibinfo {author} {\bibfnamefont
  {P.}~\bibnamefont {Avouris}}, \ and\ \bibinfo {author} {\bibfnamefont
  {F.}~\bibnamefont {Xia}},\ }\href {\doibase 10.1021/nl3016335} {\bibfield
  {journal} {\bibinfo  {journal} {Nano Lett.}\ }\textbf {\bibinfo {volume}
  {12}},\ \bibinfo {pages} {3766} (\bibinfo {year} {2012})}\BibitemShut
  {NoStop}%
\bibitem [{\citenamefont {Tymchenko}\ \emph {et~al.}(2013)\citenamefont
  {Tymchenko}, \citenamefont {Nikitin},\ and\ \citenamefont
  {Mart{\'{i}}n-Moreno}}]{spp-gr-mo-Tymchenko2013-ACSNano}%
  \BibitemOpen
  \bibfield  {author} {\bibinfo {author} {\bibfnamefont {M.}~\bibnamefont
  {Tymchenko}}, \bibinfo {author} {\bibfnamefont {A.~Y.}\ \bibnamefont
  {Nikitin}}, \ and\ \bibinfo {author} {\bibfnamefont {L.}~\bibnamefont
  {Mart{\'{i}}n-Moreno}},\ }\href {\doibase 10.1021/nn403282x} {\bibfield
  {journal} {\bibinfo  {journal} {ACS Nano}\ }\textbf {\bibinfo {volume} {7}},\
  \bibinfo {pages} {9780} (\bibinfo {year} {2013})}\BibitemShut {NoStop}%
\bibitem [{\citenamefont {Poumirol}\ \emph {et~al.}(2017)\citenamefont
  {Poumirol}, \citenamefont {Liu}, \citenamefont {Slipchenko}, \citenamefont
  {Nikitin}, \citenamefont {Martin-Moreno}, \citenamefont {Faist},\ and\
  \citenamefont {Kuzmenko}}]{spp-gr-mo-Poumirol2017-NatComm}%
  \BibitemOpen
  \bibfield  {author} {\bibinfo {author} {\bibfnamefont {J.-M.}\ \bibnamefont
  {Poumirol}}, \bibinfo {author} {\bibfnamefont {P.~Q.}\ \bibnamefont {Liu}},
  \bibinfo {author} {\bibfnamefont {T.~M.}\ \bibnamefont {Slipchenko}},
  \bibinfo {author} {\bibfnamefont {A.~Y.}\ \bibnamefont {Nikitin}}, \bibinfo
  {author} {\bibfnamefont {L.}~\bibnamefont {Martin-Moreno}}, \bibinfo {author}
  {\bibfnamefont {J.}~\bibnamefont {Faist}}, \ and\ \bibinfo {author}
  {\bibfnamefont {A.~B.}\ \bibnamefont {Kuzmenko}},\ }\href {\doibase
  10.1038/ncomms14626} {\bibfield  {journal} {\bibinfo  {journal} {Nat.
  Commun.}\ }\textbf {\bibinfo {volume} {8}},\ \bibinfo {pages} {14626}
  (\bibinfo {year} {2017})}\BibitemShut {NoStop}%
\bibitem [{\citenamefont {Kuzmin}\ \emph {et~al.}(2016)\citenamefont {Kuzmin},
  \citenamefont {Bychkov}, \citenamefont {Shavrov},\ and\ \citenamefont
  {Temnov}}]{spp-gr-mo-Kuzmin2016-nl}%
  \BibitemOpen
  \bibfield  {author} {\bibinfo {author} {\bibfnamefont {D.~A.}\ \bibnamefont
  {Kuzmin}}, \bibinfo {author} {\bibfnamefont {I.~V.}\ \bibnamefont {Bychkov}},
  \bibinfo {author} {\bibfnamefont {V.~G.}\ \bibnamefont {Shavrov}}, \ and\
  \bibinfo {author} {\bibfnamefont {V.~V.}\ \bibnamefont {Temnov}},\ }\href
  {\doibase 10.1021/acs.nanolett.6b01517} {\bibfield  {journal} {\bibinfo
  {journal} {Nano Lett.}\ }\textbf {\bibinfo {volume} {16}},\ \bibinfo {pages}
  {4391} (\bibinfo {year} {2016})}\BibitemShut {NoStop}%
\bibitem [{\citenamefont {Fallahi}\ and\ \citenamefont
  {Perruisseau-Carrier}(2012)}]{spp-gr-mo-Fallahi2012-apl}%
  \BibitemOpen
  \bibfield  {author} {\bibinfo {author} {\bibfnamefont {A.}~\bibnamefont
  {Fallahi}}\ and\ \bibinfo {author} {\bibfnamefont {J.}~\bibnamefont
  {Perruisseau-Carrier}},\ }\href {\doibase 10.1063/1.4769095} {\bibfield
  {journal} {\bibinfo  {journal} {Appl. Phys. Lett.}\ }\textbf {\bibinfo
  {volume} {101}},\ \bibinfo {pages} {231605} (\bibinfo {year}
  {2012})}\BibitemShut {NoStop}%
\bibitem [{\citenamefont {Ferreira}\ \emph {et~al.}(2011)\citenamefont
  {Ferreira}, \citenamefont {Bludov}, \citenamefont {Pereira}, \citenamefont
  {Peres},\ and\ \citenamefont {{Castro Neto}}}]{gr-mf-cond-Ferreira2011-pra}%
  \BibitemOpen
  \bibfield  {author} {\bibinfo {author} {\bibfnamefont {A.}~\bibnamefont
  {Ferreira}}, \bibinfo {author} {\bibfnamefont {Y.~V.}\ \bibnamefont
  {Bludov}}, \bibinfo {author} {\bibfnamefont {V.}~\bibnamefont {Pereira}},
  \bibinfo {author} {\bibfnamefont {N.~M.~R.}\ \bibnamefont {Peres}}, \ and\
  \bibinfo {author} {\bibfnamefont {A.~H.}\ \bibnamefont {{Castro Neto}}},\
  }\href {\doibase 10.1103/PhysRevB.84.235410} {\bibfield  {journal} {\bibinfo
  {journal} {Phys. Rev. B}\ }\textbf {\bibinfo {volume} {84}},\ \bibinfo
  {pages} {235410} (\bibinfo {year} {2011})}\BibitemShut {NoStop}%
\bibitem [{Note1()}]{Note1}%
  \BibitemOpen
  \bibinfo {note} {The substrate dielectric function includes optical phonon
  response, \begin {equation} \label {DF} \varepsilon _3 (\omega )=\varepsilon
  _{\infty }+ \DOTSB \sum@ \slimits@ _{n=1}^{4}\protect \frac {f_n\omega
  _{TO,n}^2} {\omega _{TO,n}^2-\omega ^2-i\omega \Gamma _{TO,n}}, \end
  {equation} where $\varepsilon _{\infty }=3.2$, $\omega _{TO,1}=47.7\protect
  \tmspace +\thinmuskip {.1667em}$meV, $\omega _{TO,2}=54.8\protect \tmspace
  +\thinmuskip {.1667em}$meV, $\omega _{TO,3}=70.5\protect \tmspace
  +\thinmuskip {.1667em}$meV, $\omega _{TO,4}=78.7\protect \tmspace
  +\thinmuskip {.1667em}$meV are phonon frequencies, $\Gamma
  _{TO,1}=0.72\protect \tmspace +\thinmuskip {.1667em}$meV, $\Gamma
  _{TO,2}=0.54\protect \tmspace +\thinmuskip {.1667em}$meV, $\Gamma
  _{TO,3}=1.41\protect \tmspace +\thinmuskip {.1667em}$meV, $\Gamma
  _{TO,4}=1.57\protect \tmspace +\thinmuskip {.1667em}$meV are the phonon
  dampings, and $f_1=0.3$, $f_2=2.7$, $f_3=3.0$, $f_4=0.3$ are the weighting
  coefficients.}\BibitemShut {Stop}%
\bibitem [{Note2()}]{Note2}%
  \BibitemOpen
  \bibinfo {note} {The expressions for $R_{0}$ and $T_{0}$ as well as the
  details of derivation can be found in Appendix \ref
  {subsec:R0-T0}.}\BibitemShut {Stop}%
\bibitem [{\citenamefont {Born}\ and\ \citenamefont {Wolf}(1999)}]{Born99}%
  \BibitemOpen
  \bibfield  {author} {\bibinfo {author} {\bibfnamefont {M.}~\bibnamefont
  {Born}}\ and\ \bibinfo {author} {\bibfnamefont {E.}~\bibnamefont {Wolf}},\
  }\href@noop {} {\emph {\bibinfo {title} {Principles of Optics (7th Ed)}}}\
  (\bibinfo  {publisher} {Cambridge University Press},\ \bibinfo {year}
  {1999})\BibitemShut {NoStop}%
\bibitem [{Note3()}]{Note3}%
  \BibitemOpen
  \bibinfo {note} {Excepting narrow frequency windows at $55\protect \tmspace
  +\thinmuskip {.1667em}\protect \mathrm {meV}\lesssim \omega \lesssim
  60\protect \tmspace +\thinmuskip {.1667em}\protect \mathrm {meV}$, and
  $72\protect \tmspace +\thinmuskip {.1667em}\protect \mathrm {meV}\lesssim
  \omega \lesssim 75\protect \tmspace +\thinmuskip {.1667em}\protect \mathrm
  {meV}$, where, nevertheless, an increase of the Faraday rotation angle is
  negligibly small.}\BibitemShut {Stop}%
\bibitem [{\citenamefont {Basu}(1997)}]{Basu}%
  \BibitemOpen
  \bibfield  {author} {\bibinfo {author} {\bibfnamefont {P.~K.}\ \bibnamefont
  {Basu}},\ }\href@noop {} {\emph {\bibinfo {title} {Theory of Optical
  Processes in Semiconductors}}}\ (\bibinfo  {publisher} {Clarendon Press,
  Oxford},\ \bibinfo {year} {1997})\BibitemShut {NoStop}%
\bibitem [{Note4()}]{Note4}%
  \BibitemOpen
  \bibinfo {note} {Alternatively, it can be quantified in terms of the
  coefficient of circular dichroism introduced in Appendix \ref
  {sec:Coef-MCD}.}\BibitemShut {Stop}%
\bibitem [{\citenamefont {Dias}\ and\ \citenamefont
  {Peres}(2017)}]{spp-gr-spoof-Dias2017-acsphotonics}%
  \BibitemOpen
  \bibfield  {author} {\bibinfo {author} {\bibfnamefont {E.~J.~C.}\
  \bibnamefont {Dias}}\ and\ \bibinfo {author} {\bibfnamefont {N.~M.~R.}\
  \bibnamefont {Peres}},\ }\href {\doibase 10.1021/acsphotonics.7b00629}
  {\bibfield  {journal} {\bibinfo  {journal} {ACS Photonics}\ } (\bibinfo
  {year} {2017}),\ 10.1021/acsphotonics.7b00629}\BibitemShut {NoStop}%
\bibitem [{\citenamefont {Ebbesen}\ \emph {et~al.}(1998)\citenamefont
  {Ebbesen}, \citenamefont {Lezec}, \citenamefont {Ghaemi}, \citenamefont
  {Thio},\ and\ \citenamefont {Wolff}}]{spp-eot-Ebessen1998-nat}%
  \BibitemOpen
  \bibfield  {author} {\bibinfo {author} {\bibfnamefont {T.~W.}\ \bibnamefont
  {Ebbesen}}, \bibinfo {author} {\bibfnamefont {H.~J.}\ \bibnamefont {Lezec}},
  \bibinfo {author} {\bibfnamefont {H.~F.}\ \bibnamefont {Ghaemi}}, \bibinfo
  {author} {\bibfnamefont {T.}~\bibnamefont {Thio}}, \ and\ \bibinfo {author}
  {\bibfnamefont {P.~A.}\ \bibnamefont {Wolff}},\ }\href {\doibase
  10.1038/35570} {\bibfield  {journal} {\bibinfo  {journal} {Nature}\ }\textbf
  {\bibinfo {volume} {391}},\ \bibinfo {pages} {667} (\bibinfo {year}
  {1998})}\BibitemShut {NoStop}%
\end{thebibliography}%

\end{document}